%% file: main.tex
\RequirePackage{fix-cm}

\documentclass[twocolumn,final]{svjour3} 
\usepackage{graphicx}
\usepackage{balance}  % for  \balance command ON LAST PAGE  (only there!)
\usepackage{xspace,multirow,multicol,colortbl,xcolor,tabularx}
\usepackage[ruled, vlined, linesnumbered]{algorithm2e}
\usepackage{subfig}
\usepackage[font=bf]{caption}
\usepackage{comment}
\usepackage{flushend}
\usepackage{enumitem}
\usepackage{amsmath}
\usepackage{amssymb}
\usepackage{bbold}
\usepackage[colorlinks = true,
            linkcolor = blue,
            urlcolor  = blue,
            citecolor = blue,
            anchorcolor = blue]{hyperref}
\input{cmd.tex}

\begin{document}

\title{PANE: Scalable and Effective Attributed Network Embedding}

% \author[1]{Renchi Yang}\email{renchi@nus.edu.sg}

% \author[2]{Jieming Shi}\email{jieming.shi@polyu.edu.hk}
% % \equalcont{These authors contributed equally to this work.}

% \author[1]{Xiaokui Xiao}\email{xkxiao@nus.edu.sg}
% % \equalcont{These authors contributed equally to this work.}

% \author[3]{Yin Yang}\email{yyang@hbku.edu.qa}

% \author[4]{Sourav S. Bhowmick}\email{assourav@ntu.edu.sg}

% \affil[1]{\orgdiv{School of Computing}, \orgname{National University of Singapore}}

% \affil[2]{\orgdiv{Department of Computing}, \orgname{Hong Kong Polytechnic
% University}}
% % , \orgaddress{\postcode{}, \country{HongKong SAR}}

% \affil[3]{\orgdiv{College of Science and Engineering}, \orgname{Hamad bin Khalifa University}}
% %, \orgaddress{\city{Doha}, \postcode{}, \country{Qatar}}

% \affil[4]{\orgdiv{School of Computer Science and Engineering}, \orgname{Nanyang Technological University}}
% % , \orgaddress{\postcode{}, \country{Singapore}}

\author{Renchi Yang\thanks{Work was done when the first author was a doctoral student at NTU, Singapore, and a research fellow at NUS.}         \and
        Jieming Shi          \and
        Xiaokui Xiao        \and
        Yin Yang            \and
        Sourav S. Bhowmick  \and
        Juncheng Liu
}


% \institute{
% National University of Singapore \and
% The Hong Kong Polytechnic University \and
% Hamad bin Khalifa University \and
% Nanyang Technological University
% }
\institute{Renchi Yang \at
              Hong Kong Baptist University \\
              \email{renchi@hkbu.edu.hk}           %  \\
          \and
            Jieming Shi \at
              The Hong Kong Polytechnic University \\
              \email{jieming.shi@polyu.edu.hk}           %  \\
          \and
            Xiaokui Xiao \at
              National University of Singapore \\
              \email{xkxiao@nus.edu.sg}           %  \\
          \and
            Yin Yang \at
              Hamad Bin Khalifa University \\
              \email{yyang@hbku.edu.qa}           %  \\
          \and
            Sourav S. Bhowmick \at
              Nanyang Technological University (NTU) \\
              \email{assourav@ntu.edu.sg}           %  \\
          \and
            Juncheng Liu \at 
            National University of Singapore \\
            \email{juncheng@comp.nus.edu.sg}
}

\date{Received: date / Accepted: date}

% \input{abstract}

% \keywords{network embedding, attributed graph, random walk, matrix factorization}


\maketitle

\begin{abstract}
Given a graph $G$ where each node is associated with a set of attributes, \textit{attributed network embedding} (\textit{ANE}) maps each node $v \in G$ to a compact vector $X_v$, which can be used in downstream machine learning tasks. Ideally, $X_v$ should capture node $v$'s \textit{affinity} to each attribute, which considers not only $v$'s own attribute associations, but also those of its connected nodes along edges in $G$. 
It is challenging to obtain high-utility embeddings that enable accurate predictions; scaling effective ANE computation to massive graphs with millions of nodes pushes the difficulty of the problem to a whole new level. Existing solutions largely fail on such graphs, leading to prohibitive costs, low-quality embeddings, or both.

This paper proposes \algo, an effective and scalable approach to ANE computation for massive graphs that achieves state-of-the-art result quality on multiple benchmark datasets, measured by the accuracy of three common prediction tasks: attribute inference, link prediction, and node classification. 
\algo obtains high scalability and effectiveness through three main algorithmic designs. First, it formulates the learning objective based on a novel random walk model for attributed networks.
The resulting optimization task is still challenging on large graphs. 
Second, \algo includes a highly efficient solver for the above optimization problem, whose key module is a carefully designed initialization of the embeddings, which drastically reduces the number of iterations required to converge. Finally, \algo utilizes multi-core CPUs through non-trivial parallelization of the above solver, which achieves scalability while retaining the high quality of the resulting embeddings.

The performance of \algo depends upon the number of attributes in the input network. To handle large networks with numerous attributes, we further extend \algo to \newalg, which employs an effective attribute clustering technique. 
%for efficient processing attributed networks with substantial attributes 
Extensive experiments, comparing 10 existing approaches on 8 real datasets, demonstrate that \algo and \newalg consistently outperform all existing methods in terms of result quality, while being orders of magnitude faster. %In particular, for the large {\em MAG} data with over 59 million nodes, 0.98 billion edges, and 2000 attributes, \algo is the only known viable solution that obtains effective embeddings on a single server, within 12 hours.

\end{abstract}

\input{intro.tex}

\input{backgd.tex}

\input{overview}
\input{algo.tex}
\input{parallel.tex}

\input{HDPANE}
\input{attrInferlinkPred}
\input{exp.tex}
\input{rw.tex}

\input{conclude.tex}

\bibliographystyle{spmpsci}
\bibliography{main}% common bib file

\end{document}

%% file: cmd.tex
\newcommand{\argmax}[1]{\underset{#1}{\operatorname{arg}\,\operatorname{max}}\;}

\makeatletter
\newcommand*\bigcdot{\mathpalette\bigcdot@{.5}}
\newcommand*\bigcdot@[2]{\mathbin{\vcenter{\hbox{\scalebox{#2}{$\m@th#1\bullet$}}}}}
\makeatother

\def\done{\hspace*{\fill} {$\square$}}
\def\header{\vspace{2mm} \noindent}

\newcommand{\tadw}{$\mathtt{TADW}$\xspace}
\newcommand{\hsca}{$\mathtt{HSCA}$\xspace}
\newcommand{\aane}{$\mathtt{AANE}$\xspace}
\newcommand{\bane}{$\mathtt{BANE}$\xspace}

\newcommand{\anrl}{$\mathtt{ANRL}$\xspace}
\newcommand{\dane}{$\mathtt{DANE}$\xspace}
\newcommand{\arga}{$\mathtt{ARGA}$\xspace}
\newcommand{\stne}{$\mathtt{STNE}$\xspace}
\newcommand{\can}{$\mathtt{CAN}$\xspace}
\newcommand{\graphsage}{$\mathtt{SAGE}$\xspace}
\newcommand{\asne}{$\mathtt{ASNE}$\xspace}

\newcommand{\sane}{$\mathtt{SANE}$\xspace}
\newcommand{\prre}{$\mathtt{PRRE}$\xspace}
\newcommand{\nethash}{$\mathtt{NetHash}$\xspace}

\newcommand{\netvae}{$\mathtt{NetVAE}$\xspace}

\newcommand{\progan}{$\mathtt{ProGAN}$\xspace}

\newcommand{\lqanr}{$\mathtt{LQANR}$\xspace}
\newcommand{\dgi}{$\mathtt{DGI}$\xspace}
\newcommand{\pge}{$\mathtt{PGE}$\xspace}
\newcommand{\gatne}{$\mathtt{GATNE}$\xspace}
\newcommand{\nrp}{$\mathtt{NRP}$\xspace}

\newcommand{\bla}{$\mathtt{BLA}$\xspace}

\newcommand{\appr}{$\mathtt{APMI}$\xspace}
\newcommand{\rpapr}{$\mathtt{PAPMI}$\xspace}
\newcommand{\algo}{$\mathtt{PANE}$\xspace}
\newcommand{\algopt}{$\mathtt{Seq\textrm{-}PANE}$\xspace}

\newcommand{\algoptp}{$\mathtt{Par\textrm{-}PANE}$\xspace}

\newcommand{\psvdccd}{$\mathtt{PSVDCCD}$\xspace}
\newcommand{\svdccd}{$\mathtt{SVDCCD}$\xspace}
\newcommand{\isvd}{$\mathtt{GInit}$\xspace}
\newcommand{\smsvd}{$\mathtt{SMGInit}$\xspace}

\newcommand{\newalg}{$\mathtt{PANE}^{++}$\xspace}
\newcommand{\newalgt}{$\boldsymbol{\mathtt{PANE}}^{++}$\xspace}

\newcommand{\CM}{\mathbf{C}\xspace}
\newcommand{\IM}{\mathbf{I}\xspace}
\newcommand{\XM}{\mathbf{X}\xspace}
\newcommand{\AM}{\mathbf{A}\xspace}
\newcommand{\DM}{\mathbf{D}\xspace}

\newcommand{\RM}{\mathbf{R}\xspace}
\newcommand{\PM}{\mathbf{P}\xspace}
\newcommand{\YM}{\mathbf{Y}\xspace}

\newcommand{\UM}{\mathbf{U}\xspace}
\newcommand{\VM}{\mathbf{V}\xspace}
\newcommand{\WM}{\mathbf{W}\xspace}
\newcommand{\MM}{\mathbf{M}\xspace}
\newcommand{\FM}{\mathbf{F}\xspace}

\newcommand{\BM}{\mathbf{B}\xspace}
\newcommand{\SM}{\mathbf{S}\xspace}

\newcommand{\XMf}{\mathbf{X}_f\xspace}
\newcommand{\XMb}{\mathbf{X}_b\xspace}

\newcommand{\ie}{{\it i.e.},\xspace}
\newcommand{\eg}{{\it e.g.},\xspace}

\def\parallelfor{\textbf{parallel} \For}

\newcommand{\secref}[1]{\textcolor{black}{Section} \ref{#1}}

\newcommand{\tblref}[1]{\textcolor{black}{Table} \ref{#1}}
\newcommand{\equref}[1]{\textcolor{black}{Equation \eqref{#1}}}
\newcommand{\iequref}[1]{\textcolor{black}{Inequality \eqref{#1}}}
\newcommand{\figref}[1]{\textcolor{black}{Figure} \ref{#1}}

\newcommand{\lemref}[1]{\textcolor{black}{Lemma} \ref{#1}}

\newcommand{\algoref}[1]{\textcolor{black}{Algorithm} \ref{#1}}

\newlength\lengtha 

\newcommand{\revise}[1]{{\color{black}{#1}}}

\usepackage{cite}
\makeatletter
\newcommand\footnoteref[1]{\protected@xdef\@thefnmark{\ref{#1}}\@footnotemark}
\makeatother

\let\oldnl\nl% Store \nl in \oldnl
\newcommand{\nonl}{\renewcommand{\nl}{\let\nl\oldnl}}% Remove line number for one line

\DeclareMathOperator{\Tr}{Tr}

%% file: intro.tex
\section{Introduction}\label{sec:intro}

Network embedding is a fundamental task for graph analytics, which has attracted much attention from both academia (\textit{e.g.},  \cite{perozzi2014deepwalk,node2vec2016,tang2015line}) and industry (\textit{e.g.}, \cite{pbg2019,zhu2019aligraph}). Given an input graph or network $G$, network embedding converts each node $v \in G$ to a compact, fixed-length vector $X_v$, which captures the topological features of the graph around node $v$. In practice, however, graph data often comes with \textit{attributes} associated to nodes. While we could treat graph topology and attributes as separate features, doing so loses the important information of \textit{node-attribute affinity} \cite{meng2019co}, \textit{i.e.}, attributes that can be reached by a node through one or more hops along the edges in $G$. For instance, consider a graph containing companies and board members. An important type of insights that can be gained from such a network is that one company (\textit{e.g.}, Tesla) can reach attributes of another related company (\textit{e.g.}, SpaceX) connected via a common board member (Elon Musk). To incorporate such information, \textit{attributed network embedding} (ANE) maps both topological and attribute information surrounding a node to an embedding vector, which facilitates accurate predictions, either through the embeddings themselves or in downstream machine learning tasks.

Effective ANE computation is a highly challenging task, especially for massive graphs, \textit{e.g.}, with millions of nodes and billions of edges. In particular, each node $v \in G$ could be associated with a large number of attributes, which correspond to a high dimensional space; further, each attribute of $v$ could influence not only $v$'s own embedding, but also those of $v$'s neighbors, neighbors' neighbors, and far-reaching connections via multiple hops along the edges in $G$. Existing ANE solutions are immensely expensive and largely fail on massive graphs. Specifically, as reviewed in Section \ref{sec:rw}, one class of previous methods \textit{e.g.}, \cite{yang2015network,zhang2016homophily,yang2018binarized,huang2017accelerated}, explicitly constructs and factorizes an $n\times n$ matrix, where $n$ is the number of nodes in $G$. For a graph with 50 million nodes, storing such a matrix of double-precision values would take over 20 petabytes of memory, which is clearly infeasible in practice.
%that combines proximities and attribute similatities between nodes.
Another category of methods, \textit{e.g.},  \cite{zhang2018anrl,gao2018deep,pan2018adversarially,liu2018content}, employs deep neural networks to extract higher-level features from nodes' connections and attributes. For a large dataset, training such a neural network incurs vast computational costs; further, the training process is usually done on GPUs with limited graphics memory, \textit{e.g.}, 80GB on Nvidia's flagship H100 cards. Thus, for massive graphs, currently the only practical option is to compute ANE leveraging a large cluster, \textit{e.g.}, \cite{zhu2019aligraph}, which is not only expensive but may also have significant environmental impact.

%compress the proximity and attribute
%matrices at the same time such that non-linear features are captured. %These methods either incur tremendous overheads in computation and factorization of the $n\times n$ dense matrix or rely on expensive training courses, failing to handle large graphs efficiently.
%Particularly, most of them are only capable of processing graphs with hundreds of thousands of nodes without the power of distributed computation system like AliGraph \cite{zhu2019aligraph}. 

In addition, to the best of our knowledge, all existing ANE solutions are designed for undirected graphs. In particular, it is unclear how to incorporate edge direction information (\textit{e.g.}, asymmetric transitivity \cite{zhou2017scalable}) into their resulting embeddings. In practice, many graphs are directed and existing methods yield suboptimal result quality on such graphs as shown later in our experimental study. \textit{Can we compute effective embeddings of a massive, attributed, directed graph on a single server?}

This paper provides an affirmative answer to the above question by presenting a novel framework coined \algo\footnote{\scriptsize The work reported here is an extended version of \cite{yang2020scaling,yang2022scaling}.}. It incorporates several variants: a single-thread version \algopt, a parallel version \algoptp optimized for multicore CPUs, and a version called \newalg designed to handle networks with numerous attributes. \algo significantly advances the state of the art in ANE computation. Specifically, as demonstrated in our experiments in Section~\ref{sec:exp}, the embeddings obtained by \algo simultaneously achieve the highest prediction accuracy compared to existing methods for three common graph analytics tasks, namely attribute inference, link prediction, and node classification, on common benchmark graph datasets. On the largest {\em Microsoft Academic Knowledge Graph} ({\em MAG}) dataset involving tens of millions of nodes, a billion edges, millions of distinct attributes (in the {\em MAG-SC} variant of the dataset), and over a billion node-attribute associations, \algo is the only viable solution on a single server (10 CPU cores, 1TB memory) whose resulting embeddings lead to superior prediction accuracy for all tasks.

%0.875 AUC for attribute inference, 0.96 AUC for link prediction, and 0.57 micro-F1\footnote{The micro-F1 score, ranging from $0$ to $1$, is the harmonic mean of the precision and recall, which are computed through micro averaging~\cite{yang1999evaluation}.} for node classification. \algo obtains such results using 10 CPU cores, 1TB memory, and 12 hours running time.

\algo achieves effective and scalable ANE computation through four main contributions: (i) a well-thought-out problem formulation based on a novel random walk model, (ii) a highly efficient solver, (iii) non-trivial parallelization of the algorithm (\textit{i.e.}, \algoptp), and (iv) an effective attribute clustering technique capable of handling networks with a large attribute set (\textit{i.e.}, \newalg). Specifically, as presented in Section \ref{sec:objective}, \algo formulates the ANE task as an optimization problem with the objective of approximating normalized multi-hop node-attribute affinity using node-attribute \linebreak co-projections \cite{meng2019co}, guided by a \textit{shifted} pairwise mutual information (SPMI) metric. The affinity between a given node-attribute pair is defined via a random walk model specifically adapted to attributed networks. Further, we incorporate edge direction information by defining separate {\it forward} and {\it backward} affinity, embeddings, and SPMI metrics. Solving this optimization problem is still immensely expensive with off-the-shelf algorithms as it involves the joint factorization of two $O(n \cdot d)$-sized matrices, where $n$ and $d$ are the numbers of nodes and attributes in the input data, respectively. Thus, \algo includes a novel solver with a key module that seeds the optimizer through a highly effective greedy algorithm, which drastically reduces the number of iterations till convergence. Further, we devise a non-trivial parallelization of the \algo algorithm that utilizes modern multi-core CPUs without significantly compromising the result utility. 

For networks with numerous attributes and/or node-attribute associations, the \newalg variant of the proposed solution exploits an effective attribute clustering algorithm that groups similar attributes into \textit{super attributes} to significantly reduce the computational overhead while retaining the high result quality of the obtained embeddings. Extensive experiments using 8 real datasets and comparing against 10 existing solutions demonstrate that \algo consistently obtains high-utility embeddings with superior prediction accuracy for attribute inference, link prediction and node classification, at a fraction of the cost compared to existing methods. 

%\renchi{However, our experimental results also show that the efficiency of \algo and its parallel version is less than satisfactory on the attributed networks with numerous distinct attributes, and they even fail to handle the datasets with substantial attributes, \eg an variant MAG dataset (\ie {\em MAG-SC}) that has 2.78M distinct attributes and 1.1B node attribute associations. To circumvent the drawback of \algo on such graphs, we propose \newalg, which employs an attribute clustering method to significantly reduce the overheads in \algo without sacrificing too much result quality, by grouping similar attributes as super attributes accurately. In particular, \newalg is the only survival ANE solution to handle the {\em MAG-SC} dataset that has 2.78M distinct attributes and 1.1B node attribute associations, while achieving 0.98 AUC, \renchi{0.983 AUC}, and 0.769 micro-F1 for attribute inference, link prediction, and node classification tasks, respectively.}

%  We summarize our main contributions in this work as follows:
In summary, our main contributions are as follows:
\vspace{-\topsep}
\begin{itemize}[leftmargin=*]
\item We formulate the ANE task as an optimization problem with the objective of approximating multi-hop node-attribute affinity.
\item We further consider edge direction in our objective by defining {\it forward} and {\it backward} affinity matrices using the SPMI metric.
\item We propose several techniques to efficiently solve the optimization problem, including efficient approximation of the affinity matrices, fast joint factorization of the affinity matrices, and a key module to greedily seed the optimizer, which drastically reduces the number of iterations till convergence.
\item We develop a non-trivial parallelization technique of \algo (\ie \algoptp) to further boost efficiency on multicore CPUs.
\item We augment \algo to \newalg with an effective attribute clustering technique that scales well to a massive attribute set associated with the input network.
\item We experimentally demonstrate the superior performance of \algo and \newalg, in terms of both effectiveness and efficiency, against 10 competitors on 8 real datasets.
\end{itemize}

The rest of the paper is organized as follows. We formally define the ANE problem addressed in this paper in Section~\ref{sec:back}. We introduce the sequential and parallel versions of \algo in Sections~\ref{sec:opt} and~\ref{sec:parallel}, respectively. Section~\ref{sec:parallel} presents the extension of \algo to effectively handle large attribute set associated with an input network. We discuss how the embeddings generated by our proposed techniques can be exploited by representative machine learning tasks in Section~\ref{sec:ane-app}. We report exhaustive performance study of our proposed techniques in Section~\ref{sec:exp}. Related work is discussed in Section~\ref{sec:rw}. The last section concludes the paper.

\vspace{-\topsep}

%% file: backgd.tex
\section{Problem Formulation}\label{sec:back}

In this section, we formally define the problem addressed in this paper. We begin by introducing the notations and terminology used in this work. Next, we introduce the notion of {\em node-attribute affinity}. Finally, we formally define the ANE problem by exploiting the notion of node-attribute affinity. 
\subsection{Notations and Terminology}
\begin{table}[!t]
\centering
\renewcommand{\arraystretch}{1.4}
\begin{small}
\caption{Frequently used notations.} \label{tbl:notations}
\resizebox{\columnwidth}{!}{
	\begin{tabular}{|p{1.0in}|p{2.3in}|}
		\hline
		{\bf Notation} &  {\bf Description}\\
		\hline
		$G$=$(V,E_{V},R,E_{R})$   & A graph $G$ with node set $V$, edge set $E_{V}$, attribute set $R$, and node-attribute association set $E_{R}$.\\
		\hline
		$n, m, d$   & The numbers of nodes, edges, and attributes in $G$, respectively.\\
% 		\hline
% 		$\din(v_i)$   & The in-degree of node $v_i$ \\
% 		\hline
% 		$\dout(v_i)$   & The out-degree of node $v_i$ \\
        \hline
		$k$   & The space budget of embedding vectors. \\
		\hline
		$\AM, \DM, \PM, \RM$   & The adjacency, out-degree, random walk and attribute matrices of $G$. \\
		\hline
		$\RM_r, \RM_c$ & The row-normalized and column-normalized attribute matrices. See \equref{eq:norm-r}. \\
		\hline
		$\FM,\BM$ & The forward and backward affinity matrices. See Equations \eqref{eq:fwd-prob} and \eqref{eq:bwd-prob}.\\
		\hline
		$\FM',\BM'$ & The approximate forward and backward affinity matrices. See Equation \eqref{equ:approxFB}.\\
% 		\hline 
% 		$\rho(v_i,r_j)$ & The Attributed PageRank of $r_j$ w.r.t. $v_i$. See Equation \eqref{eq:arp-prob}.\\
		\hline
		$\XMf, \XMb, \YM$ & The forward and backward embedding vectors, and attribute embedding vectors.\\
		\hline
% 		 $\SM$ & The target information matrix. See Equation \eqref{eq:sm}.\\
% 		 \hline
% 		 $\pi(v_i,r_j)$ & The RWR score of $v_j$ w.r.t $v_i$.\\
% 		 \hline
% 		 $\epsilon$ & The absolute error threshold, $\epsilon \in [0,1]$.\\
% 		 \hline
		 		$\alpha$ & The random walk stopping probability. \\
		\hline

		 $n_b$ & The number of threads.\\
% 		\hline
% 		$\sigma_{k+1}(\MM)$   & The $(k+1)$-th largest singular value of matrix $\MM$. \\
		\hline
% 		$\log(x)$   & The logarithm of $x$ to the base $2$. \\
% 		\hline
		 $\kappa$ & The number of super attributes.\\
		\hline
	\end{tabular}
}
\end{small}
\vspace{-2mm}
\end{table}

%\textit{An attributed network} consists of both graph topology and node attributes, and is formally defined as follows.

%Apart from the graph topology, attributed directed networks also include attributes associated  with nodes. The formal definition of {\em attributed directed network} is presented as follows.

%In addition to the topological structure of the network, there are often attributes accompanied with nodes in real-world graphs which are referred to as {\em attributed networks}. The formal definition of {\em attributed network} is as follows.
% \begin{definition}
% An {\em attributed network} $G$ is a graph with a node set $V$ containing $n$ nodes, an edge set $E_V$ containing $m$ edges, a set of $d$ attributes $R$ $(|R|=d)$ and a set of node-attribute associations $E_{R}$, in which each element $(v_i,r_j,w)$ represents that the attribute weight of node $v_i$'s attribute $r_j$ is $w$.
% \end{definition}

Let $G=(V, E_V, R, E_R)$ be an {\it attributed network}, consisting of (i) a node set $V$ with cardinality $n$, (ii) a set of edges $E_V$ with cardinality $m$, each connecting two nodes in $V$, (iii) a set of attributes $R$ with cardinality $d$, and (iv) a set of node-attribute associations $E_R$, where each element is a tuple $(v_i ,r_j, w_{i,j})$ signifying that node $v_i \in V$ is directly associated with attribute $r_j \in R$ with a weight $w_{i,j}$ (\textit{i.e.}, the attribute value). Note that for a categorical attribute such as marital status, we first apply a pre-processing step that transforms the attribute into a set of binary ones through one-hot encoding.
%where $v_i$ is a node in $V$, $r_j$ is an attribute in $R$, and $w$ is the weight that indicates the degree that $v_i$ is associated with $r_j$. 
Without loss of generality, we assume that $G$ is a directed graph; if $G$ is undirected, then we treat each edge $(v_i, v_j)$ in $G$ as a pair of directed edges with opposing directions: $(v_i, v_j)$ and $(v_j, v_i)$.

Given a space budget $k \ll n$, a \textit{node embedding} maps a node $v \in V$ to a length-$k$ vector. The general goal of attributed network embedding (ANE) is to compute such an embedding $X_v$ for each node $v$ in the input graph, such that $X_v$ captures the graph structure and attribute information surrounding node $v$.
%Given $k \ll n, d$, we aim to construct an {\em embedding} in $\mathbb{R}^k$ for each node $v \in V$, as a concise representation of $v$'s topological and attribute information. In particular, these embeddings alone should enable us to infer (i) the attribute values of each node and (ii) whether two nodes are connected in $G$. 
In addition, following previous work \cite{meng2019co}, we
also allocate a space budget $\frac{k}{2}$ (explained later in Section \ref{sec:obj2}) for each attribute $r \in R$, and aim to compute an {\em attribute embedding} vector for $r$ of length $\frac{k}{2}$.

%in $\mathbb{R}^{k/2}$ for each attribute $r \in R$, which could be used to facilitate certain learning tasks (\eg attribute inference \cite{gong2014joint,yang2017bi}). 

% given embedding vectors instead of $G$, can enable us to (i) infer 

% we can infer (i) the attribute associated $v$ based on its embedding, and (ii) w

% In this paper, we study the problem of attributed network co-embedding for both directed and undirected graphs as shown below.
% %learning low-dimensional vector representations for the {\em attributed network}, which consists of both $n$ nodes and $d$ attributes.
% \begin{definition}\label{def:ance}
% Given an attributed network $G=(V,E_V,R,E_R)$, where $|V|=n$ and $|R|=d$, {\em attributed network embedding} aims to learn a source-node embedding vector $\XMf[v_i]\in\mathbb{R}^{\frac{k}{2}}$, a target-node embedding vector $\XMb[v_i]\in\mathbb{R}^{\frac{k}{2}}$ ($k\ll n$) for each node $v_i\in V$, and an embedding vector $\YM[r_j]\in\mathbb{R}^{k}$ ($k\ll d$) for each attribute $r_j\in R$, such that they can preserve both the topological features of $v_i$ and attribute information of $r_j$ in $G$.
% \end{definition}
% % The input graph $G$ can be either directed or undirected. For simplicity, in the following we assume that $G$ is directed; 
% The input graph $G$ can be either directed or undirected. For an undirected graph, we regard each undirected edge $(v_i, v_j)$ as two directed edges with opposite direction, \ie  $(v_i, v_j)$ and $(v_j, v_i)$.

% \vspace{-1mm}
%\subsection{Notations}
\header
 We denote matrices in bold uppercase, \eg $\MM$. We use $\MM[v_i]$ to denote the $v_i$-th row vector of $\MM$, and $\MM[:,r_j]$ to denote the $r_j$-th column vector of $\MM$. In addition, we use $\MM[v_i,r_j]$ to denote the element at the $v_i$-th row and $r_j$-th column of $\MM$. Given an index set $S$, we let $\MM[S]$ (resp.\ $\MM[:,S]$) be the matrix block of $\MM$ that contains the row (resp.\ column) vectors of the indices in $S$. 

Let $\AM$ be the adjacency matrix of the input graph $G$, \ie $\AM[v_i, v_j] = 1$ if  $(v_i,v_j)\in E_V$, otherwise $\AM[v_i, v_j] = 0$. Let $\DM$ be the diagonal out-degree matrix of $G$, \ie $\DM[v_i,v_i] = \sum_{v_j\in V}{\AM[v_i,v_j]}$. We define the random walk matrix of $G$ as $\PM = \DM^{-1}\AM$. 
% Accordingly, $\PM^{k}[v_i,v_j]$ denotes the probability that a $k$-hop $(k \ge 1)$ random walk from node $v_i$ would end at $v_j$. 
% In addition, the RWR (\ie {\em random walk with restart} \cite{}) score of a node $v_j$ with repsect to a node $v_i$ is denoted as $\pi(v_i,r_j)$.
% \begin{definition}(Personalized PageRank)\label{def:ppr}
% Given a random walk decay factor $\alpha$, transition matrix $\PM$, the Personalized PageRank of $v_j$ with respect to $v_i$ is defined as
% \begin{equation}
% \textstyle \pi(v_i,v_j)=\sum_{\ell=0}^{\infty}{\alpha(1-\alpha)^{\ell}\PM^{\ell}[v_i,v_j]}.
% \end{equation} 
% \end{definition}
% PPR describes the importance of the target node from the perspective of the source node, which is shown very effective in capturing the node relevance over the input graphs and widely used in homogeneous network embedding (HNE) \cite{yang13homogeneous,yin2019scalable,tsitsulin2018verse,zhou2017scalable}. 

Furthermore, we define an attribute matrix $\RM \in \mathbb{R}^{n\times d}$, such that $\RM[v_i,r_j] = w_{i,j}$ is the weight associated with the entry ($v_i$, $r_j$, $w_{ij}$) $ \in E_R$. We refer to $\RM[v_i]$ as node $v_i$'s \textit{attribute vector}. Based on $\RM$, we derive a row-normalized (resp.\ column-normalized) attribute matrices $\RM_r$ (resp.\ $\RM_c$) as follows: 
% \begin{align}
% &\textstyle\RM_r[v_i,r_j]=\frac{\RM[v_i,r_j]}{\sum_{v_l\in V}{\RM[v_l,r_j]}},\label{eq:row-r}\\ &\textstyle\RM_c[v_i,r_j]=\frac{\RM[v_i,r_j]}{\sum_{r_l\in R}{\RM[v_i,r_l]}}\label{eq:col-r}.
% \end{align}
\begin{small}
\begin{equation}\label{eq:norm-r}
	\begin{split}
\RM_r[v_i,r_j]=\frac{\RM[v_i,r_j]}{\sum_{r_l\in R}{\RM[v_i,r_l]}},\\ \RM_c[v_i,r_j]=\frac{\RM[v_i,r_j]}{\sum_{v_l\in V}{\RM[v_l,r_j]}}.
	\end{split}
\end{equation}
\end{small}
\tblref{tbl:notations} lists the frequently used notations in our paper.

\begin{figure}[!t]
\centering
\hspace{-12mm}
\begin{minipage}{0.3\textwidth}
\centering
\begin{small}
\vspace{7mm}
\includegraphics[width=0.57\columnwidth]{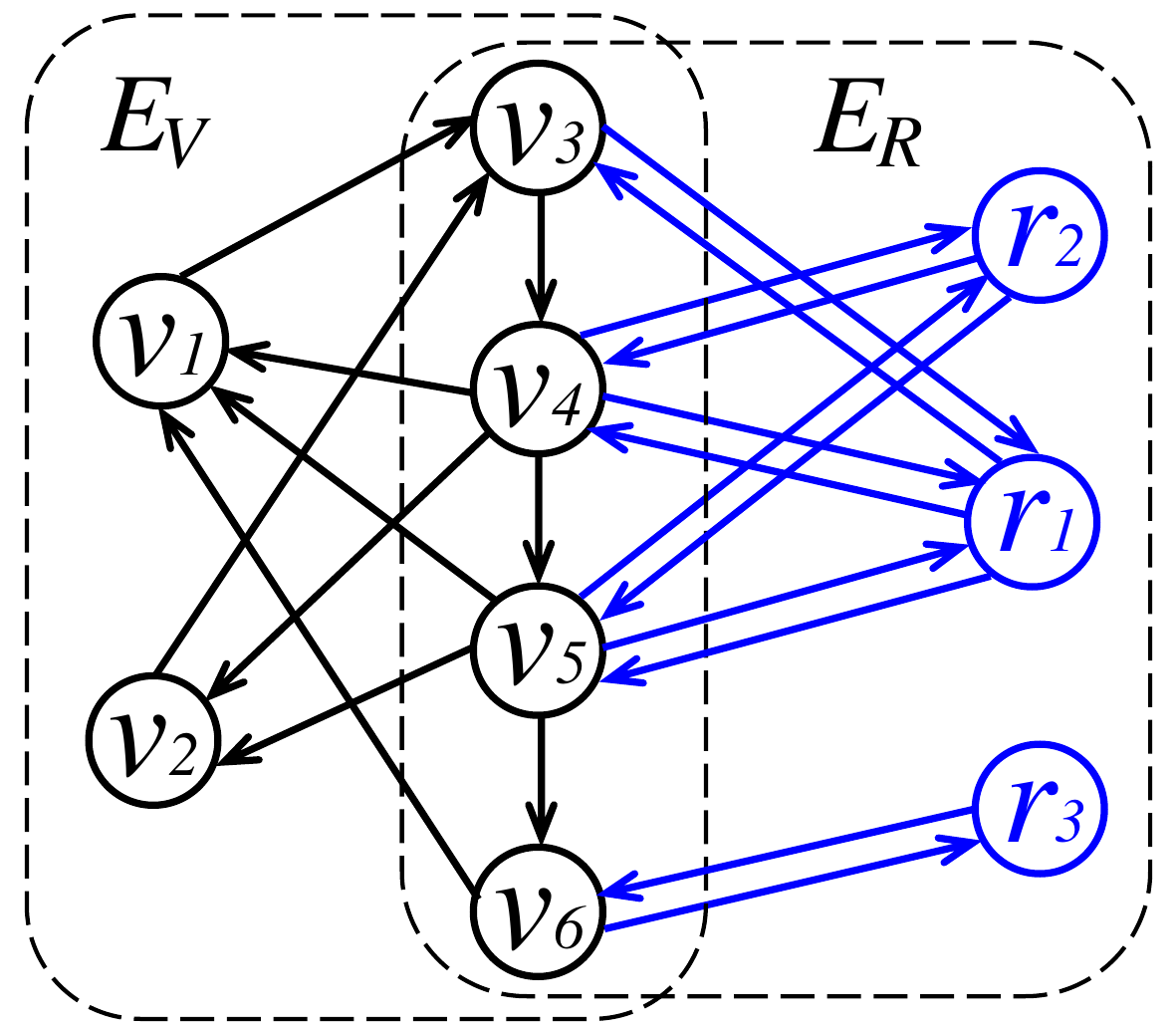}
\vspace{0mm}
\caption{Extended \\ graph $\mathcal{G}$}\label{fig:toy}
\end{small}
\end{minipage}
\hspace{-8mm}
\begin{minipage}{0.25\textwidth}
% \vspace{-2mm}
\centering
\captionsetup{type=table}
\renewcommand{\arraystretch}{1.1}
\begin{footnotesize}
\caption{Targets for $\XM[v_i]\cdot\YM[r_j]^{\top}$.}\label{tbl:toy}
\resizebox{\columnwidth}{!}{
\begin{tabular}{|c|c|c|c|}\hline
 & $\YM[r_1]$ & $\YM[r_2]$ & $\YM[r_3]$ \\ \hline
$\XMf[v_1]$ & 1.0 & 0.92 & 0.47 \\
$\XMb[v_1]$ & 0.93 & 0.88 & 1.17 \\ \hline
$\XMf[v_2]$ & 1.0 & 0.92 & 0.47 \\
$\XMb[v_2]$ & 1.11 & 1.08 & 0.8 \\ \hline
$\XMf[v_3]$ & 1.12 & 1.04 & 0.54 \\
$\XMb[v_3]$ & 1.06 & 0.95 & 0.99 \\ \hline
$\XMf[v_5]$ & 0.98 & 1.1 & 1.08 \\
$\XMb[v_5]$ & 1.09 & 1.22 & 0.61 \\ \hline
$\XMf[v_6]$ & 0.89 & 0.82 & 2.05 \\
$\XMb[v_6]$ & 0.53 & 0.61 & 1.6 \\ \hline
\end{tabular}
}
\end{footnotesize}
\end{minipage}
\vspace{-2mm}
\end{figure}

\header
%Given an input graph $G=(V, E_V, R, E_R)$, we define an \textit{extended graph} $\mathcal{G}$ based on $G$. 
Our solution utilizes an \textit{extended graph} $\mathcal{G}$ that incorporates additional nodes and edges into $G$. To illustrate, \figref{fig:toy} shows an example extended graph $\mathcal{G}$ constructed based on an input attributed network $G$ with 6 nodes $v_1$-$v_6$ and 3 attributes $r_1$-$r_3$. The left part of the figure (in black) shows the topology of $G$, \textit{i.e.}, the edge set $E_V$. The right part of the figure (in blue) shows the attribute associations $E_R$ in $G$. Specifically, for each attribute $r_j \in R$, we create an additional node in  $\mathcal{G}$; then, for each entry in $E_R$, \textit{e.g.}, ($v_3$, $r_1$, $w_{3, 1}$), we include in $\mathcal{G}$ a pair of edges with opposing directions connecting the node (\textit{e.g.}, $v_3$) with the corresponding attribute node (\textit{e.g.}, $r_1$), with an edge weight (\textit{e.g.}, $w_{3, 1}$). Note that in this example, nodes $v_1$ and $v_2$ are not associated with any attribute.

%% file: overview.tex
\subsection{Node-Attribute Affinity via Random Walks}\label{sec:objective}

As explained in Section \ref{sec:intro}, the resulting embedding of a node $v \in V$ should capture its \textit{affinity} with attributes in $R$, where the affinity definition should take into account both the attributes directly associated with $v$ in $E_R$, and the attributes of the nodes that $v$ can reach via edges in $E_V$. To effectively model node-attribute affinity via multiple hops in $\mathcal{G}$, we employ an adaptation of the \textit{random walks with restarts} (\textit{RWR})  \cite{jeh2003scaling,tong2006fast} technique to our setting with an extended graph $\mathcal{G}$. In the following, we refer to an RWR simply as a \textit{random walk}. Specifically, since $\mathcal{G}$ is directed, we distinguish two types of node-attribute affinity: \textit{forward affinity}, denoted as $\FM$, and \textit{backward affinity}, denoted as $\BM$. 

\header
\textbf{Forward affinity.} Given an attributed graph $G$, a node $v_i$, and random walk stopping probability $\alpha$ ($0<\alpha<1$), a \textit{forward random walk} on $\mathcal{G}$ starts from node $v_i$. At each step, assume that the walk is currently at node $v_l$. Then, the walk can either (i) with proabability $\alpha$, terminate at $v_l$ , or (ii)  with probability $1-\alpha$, follow an edge in $E_V$ to a random out-neighbor of $v_l$. After a random walk terminates at a node $v_l$, %if $v_l$ has any associated attribute in $E_R$, 
we randomly follow an edge in $E_R$ to an attribute $r_j$, with probability $\RM_r[v_l,r_j]$, \textit{i.e.}, a normalized edge weight defined in Equation \eqref{eq:norm-r}\footnote{In the degenerate case that $v_l$ is not associated with any attribute, \textit{e.g.}, $v_1$ in Figure \ref{fig:toy}, we simply restart the random walk from the source node $v_i$ and repeat the process.}. The forward random walk yields a \textit{node-to-attribute pair} $(v_i,r_j)$ and we add it to a collection $\mathcal{S}_f$. 

Suppose that we sample $n_r$ node-to-attribute pairs for each node $v_i$. The size of $\mathcal{S}_f$ is then $n_r\cdot n$, where $n$ is the number of nodes in $G$.
Denote $p_f(v_i,r_j)$ as the probability that a forward random walk starting from $v_i$ yields a node-to-attribute pair $(v_i, r_j)$.
Then, the \textit{forward affinity} $\FM[v_i,r_j]$ between note $v_i$ and attribute $r_j$ is defined as follows.
% \begin{small}
\begin{equation}\label{eq:fwd-prob}
	\FM[v_i,r_j] =  \log\left(\frac{n\cdot p_f(v_i,r_j)}{\sum_{v_h\in V}{p_f(v_h,r_j)}}+1\right)
\end{equation}
% \end{small}

To explain the intuition behind the above definition, note that in $\mathcal{S}_f$, the probabilities of observing node $v_i$, attribute $r_j$, and pair $(v_i, r_j)$ are $\mathbb{P}(v_i)=\frac{1}{n}$, $\mathbb{P}(r_j)=\frac{\sum_{v_h\in V}{\cdot p_f(v_h,r_j)}}{n}$, and $\mathbb{P}(v_i,r_j)=\frac{p_f(v_i,r_j)}{n}$, respectively. Thus, the above definition of forward affinity is a variant of the \textit{pointwise mutual information} (PMI) ~\cite{church1990word} between node $v_i$ and attribute $r_j$\footnote{The PMI quantifies how much more or less likely we are to see the two events co-occur, given their individual probabilities, and relative to the case where they are completely independent.}. In particular, given a collection of element pairs $\mathcal{S}$, the PMI of element pair $(x,y)\in \mathcal{S}$, denoted as $\textrm{PMI}(x,y)$, is defined as $\textrm{PMI}(x,y)=\log\left(\frac{\mathbb{P}(x,y)}{\mathbb{P}(x)\cdot\mathbb{P}(y)}\right)$, where $\mathbb{P}(x)$ (resp.\ $\mathbb{P}(y)$) is the probability of observing $x$ (resp.\ $y$) in $\mathcal{S}$ and $\mathbb{P}(x,y)$ is the  probability of observing pair $(x,y)$ in $\mathcal{S}$. The larger $\textrm{PMI}(x,y)$ is, the more likely  $x$ and $y$ co-occur in $\mathcal{S}$. Note that $\textrm{PMI}(x,y)$ can be negative. To avoid this, we use an alternative: shifted PMI, defined as $\textrm{SPMI}(x,y)=\log\left(\frac{\mathbb{P}(x,y)}{\mathbb{P}(x)\cdot\mathbb{P}(y)}+1\right)$, which is guaranteed to be non-negative, while retaining the original order of values of PMI. $\FM[v_i, r_j]$ in Equation \eqref{eq:fwd-prob} is then $\textrm{SPMI}(v_i,r_j)$.

Another way to understand Equation \eqref{eq:fwd-prob} is through an analogy to TF/IDF \cite{salton1986introduction} in natural language processing. Specifically, if we view all the forward random walks %starting from node $v_i$
as a ``document'', then $n \cdot p_f(v_i, r_j)$ is akin to the term frequency of $r_j$, whereas the denominator in Equation \eqref{eq:fwd-prob} is similar to the inverse document frequency of $r_j$. Thus, the normalization penalizes common attributes and compensates for rare attributes.

\header
\textbf{Backward affinity.} Next we define backward affinity in a similar fashion. Given an attributed network $G$, an attribute $r_j$ and stopping probability $\alpha$, a {\em backward random walk} starting from $r_j$ first randomly samples a node $v_l$ according to probability $\RM_c[v_l,r_j]$, defined in Equation \eqref{eq:norm-r}. Then, the walk starts from node $v_l$; at each step, the walk either terminates at the current node with $\alpha$ probability, or randomly jumps to an out-neighbor of current node with $1-\alpha$ probability. Suppose that the walk terminates at node $v_i$; then, it returns an \textit{attribute-to-node pair} $(r_j,v_i)$, which is added to a collection $\mathcal{S}_b$. After sampling $n_r$ attribute-to-node pairs for each attribute, the size of $\mathcal{S}_b$ becomes $n_r\cdot d$.
Let $p_b(v_i, r_j)$ be the probability that a backward random walk starting from attribute $r_j$ stops at node $v_i$. In collection $\mathcal{S}_b$, the probabilities of observing attribute $r_j$, node $v_i$ and pair $(r_j, v_i)$ are $\mathbb{P}(r_j)=\frac{1}{d}$, $\mathbb{P}(v_i)=\frac{\sum_{r_h\in R}{p_b(v_i,r_h)}}{d}$ and $\mathbb{P}(v_i,r_j)=\frac{p_b(v_i, r_j)}{d}$, respectively. By the definition of SPMI, we define backward affinity $\BM[v_i,r_j]$ as follows.
% \begin{small}
\begin{equation}\label{eq:bwd-prob}    
	\BM[v_i,r_j] = \log\left(\frac{d\cdot p_b(v_i,r_j)}{\sum_{r_h\in R}{p_b(v_i,r_h)}}+1\right).
\end{equation}
% \end{small}

\subsection{Objective Function} \label{sec:obj2}
Lastly, we define our objective function for ANE based on the notions of forward and backward node-attribute affinity defined in Equation \eqref{eq:fwd-prob} and Equation \eqref{eq:bwd-prob}, respectively.
%The association between nodes can be built upon node-to-attribute and attribute-to-node associations, explained later in Section \ref{sec:attr-app}.
Let $\FM[v_i,r_j]$ (resp.\ $\BM[v_i,r_j]$) be the forward affinity (resp.\ backward affinity) between node $v_i$ and attribute $r_j$. Given a space budget $k$, our objective is to learn (i) two embedding vectors for each node $v_i$, namely a \textit{forward embedding vector}, denoted as $\XMf[v_i]\in \mathbb{R}^{\frac{k}{2}}$ and a  \textit{backward embedding vector}, denoted as $\XMb[v_i]\in \mathbb{R}^{\frac{k}{2}}$, as well as (ii) an \textit{attribute embedding vector} $\YM[r_j]\in \mathbb{R}^{\frac{k}{2}}$ for each attribute $r_j$, such that the following objective is minimized:
% \begin{small}
\begin{align}
\mathcal{O}=\min_{\XMf,\YM,\XMb}&  \sum_{v_i\in V}\sum_{r_j\in R}\left(\FM[v_i,r_j]-\XMf[v_i]\cdot \YM[r_j]^{\top}\right)^2\nonumber\\
& \quad\quad +\left(\BM[v_i,r_j]-\XMb[v_i]\cdot \YM[r_j]^{\top}\right)^2\label{eq:obj1}
\end{align}
% \end{small}
Intuitively, in the above objective function, we approximate the forward node-attribute affinity $\FM[v_i,r_j]$ between node $v_i$ and attribute $r_j$ using the dot product of their respective embedding vectors, \textit{i.e.}, $\XMf[v_i]\cdot \YM[r_j]^{\top}$. Similarly, we also approximate the backward node-attribute affinity using $\XMb[v_i]\cdot \YM[r_j]^{\top}$. The objective is then to minimize the total squared error of such approximations, over all nodes and all attributes in the input data.

\header
{\bf Running Example.}
%We illustrate the effectiveness of our solution, using the attributed network in \figref{fig:toy}, where $v_1$-$v_6$ are nodes and $r_1$-$r_3$ are attributes. 
Assume that in the extended graph $\mathcal{G}$ shown in Figure \ref{fig:toy}, all attribute weights in $E_R$ are $1$, and the random walk stopping probability $\alpha$ is set to  $0.15$ \cite{jeh2003scaling,tong2006fast}. 
% \tblref{tbl:toy} lists the target values, \textit{i.e.}, the exact forward and backward affinity values. According to Equation \eqref{eq:obj1}, for the inner products of attribute embedding vectors of $r_1$-$r_3$ and that of  $v_1$-$v_6$.
\tblref{tbl:toy} lists the inner products of attribute embedding vectors of $r_1$-$r_3$ and embedding vectors of node $v_1$-$v_6$, which are the forward and backward affinity values preserved in these embedding vectors.
These values are calculated based on Equations \eqref{eq:fwd-prob} and \eqref{eq:bwd-prob}, using simulated random walks on $\mathcal{G}$ in \figref{fig:toy}. Observe, for example, that the node $v_1$ has high affinity values (both forward and backward) with attribute $r_1$, which agrees with the intuition that $v_1$ is connected to $r_1$ via many different intermediate nodes, \textit{i.e.}, $v_3, v_4, v_5$. 
Meanwhile, regarding node $v_5$, if we only consider forward affinity, then, observe that $v_5$ has a higher forward affinity value with $r_3$ than that with $r_1$. Such an affinity value fails to capture the fact that $v_5$ is associated with $r_1$ but not with $r_3$, which may lead to incorrect attribute inference. This issue is resolved when we consider both forward and backward affinity.

%% file: algo.tex
% \vspace{-1mm}
\section{\algopt: Single-Thread \algo}\label{sec:opt}
\begin{algorithm}[t]
	\begin{small}
		\caption{\algopt}
		\label{alg:mainopt}
		\BlankLine
		\KwIn{Attributed network $G$, space budget $k$, random walk stopping probability $\alpha$, error threshold $\epsilon$.}
		\KwOut{Forward and backward embedding vectors $\XMf$, $\XMb$ and attribute embedding vectors $\YM$.}
		$t\gets \frac{\log(\epsilon)}{\log(1-\alpha)}-1$\;
% 		\If{$\kappa \ge d$}{
    		$\FM^{\prime}, \BM^{\prime} \gets \mathtt{APMI}(\PM, \RM, \alpha, t)$\;
    		$\XMf,\YM, \XMb \gets \mathtt{SVDCCD}(\FM^{\prime},\BM^{\prime},k,t)$\;
% 		}\Else{
% 		    $\widetilde{\RM}\gets \mathtt{SCA}(\RM, \kappa, t)$\;
% 		    $\FM^{\prime}, \BM^{\prime} \gets \mathtt{APMI}(\PM, \widehat{\RM}, \alpha, t)$\;
% 		    $\XMf,\YM, \XMb \gets \mathtt{SVDCCD}(\FM^{\prime},\BM^{\prime},k,t)$\;
% 		}
    	\Return $\XMf,\YM,\XMb$\;
	\end{small}
\end{algorithm}

\begin{algorithm}[!t]
	\begin{small}
		\caption{\appr}
		\label{alg:appr}
		\KwIn{$\PM$, $\RM, \alpha, t$.}
		\KwOut{$\FM^{\prime},\BM^{\prime}$.}
		Compute $\RM_r$ and $\RM_c$ by \equref{eq:norm-r}\;
% 		Compute $\RM_c$ by Equation \equref{eq:col-r}\;
		$\PM_f^{(0)} \gets \RM_r, \ \PM_b^{(0)} \gets \RM_c$\;
		\For{$\ell \gets 1$ to $t$}{
			$\PM_f^{ (\ell)} \gets (1-\alpha)\cdot\PM \PM_f^{(\ell-1)} + \alpha\cdot\PM_f^{(0)}$\;
			$\PM_b^{(\ell)} \gets (1-\alpha)\cdot\PM^{\top} \PM_b^{(\ell-1)} + \alpha\cdot\PM_b^{(0)}$\;
		}
		Normalize $\PM_f^{ (t)}$ by columns to get $\widehat{\PM}_f^{ (t)}$\;
		Normalize $\PM_b^{ (t)}$ by rows to get $\widehat{\PM}_b^{ (t)}$\;
		$\FM^{\prime} \gets \log(n\cdot \widehat{\PM}_f^{ (t)}+1),\quad \BM^{\prime} \gets \log(d\cdot \widehat{\PM}_b^{ (t)}+1)$\;
% 		$\BM^{\prime} \gets \log(d\cdot \widehat{\PM}_b^{ (t)}+1)$\;
	    \Return $\FM^{\prime},\BM^{\prime}$\;
	\end{small}
\end{algorithm}

In this section, we describe a single-thread version of \algo, called \algopt. Further improved versions of \algo are presented in subsequent sections.
Observe that it is a challenging task to train embeddings of nodes and attributes that preserve our objective function in Equation \eqref{eq:obj1}, especially on massive attributed networks. First, node-attribute affinity values are defined by random walks, which are rather expensive to undertake in a large number from every node and attribute in order to accurately obtain the affinity values of all possible node-attribute pairs.
Second, our objective function preserves both forward and backward  affinity (\ie considering edge directions), which makes the training process hard to converge.
Further, jointly preserving both forward and backward affinity involves intensive computations, severely dragging down the performance.
To address these challenges, we propose \algopt that can efficiently handle large-scale data and produce high-quality ANE results.

At a high level, \algopt consists of two phases: (i) iteratively computing approximated versions $\FM'$ and $\BM'$ of the forward and backward affinity matrices with rigorous approximation error guarantees, without actually sampling random walks
(\secref{sec:apmi}), and (ii) initializing the  embedding vectors with a greedy algorithm for fast convergence, and 
%\textit{randomized SVD} \cite{musco2015randomized} and
then jointly factorizing $\FM'$ and $\BM'$ using {\em cyclic coordinate descent} \cite{wright2015coordinate} to efficiently obtain the output embedding vectors $\XMf,\XMb$, and $\YM$ (\secref{sec:svdccd}).
Given an attributed network $G$, space budget $k$, random walk stopping probability $\alpha$ and an error threshold $\epsilon$ as inputs, \algoref{alg:mainopt} outlines the proposed \algopt algorithm. We elaborate on these steps now.

\vspace{-2mm}
\subsection{Forward and Backward Affinity Approximation}\label{sec:apmi}
%In the first stage of \algopt, we compute forward and backward affinity values for each node-attribute pair $(v_i,r_j)$ in attributed network $G$.

In Section \ref{sec:objective}, node-attribute affinity values are defined using a large number of random walks, which are expensive to simulate on a massive graph.
%Recall that forward affinity value $\FM[v_i,r_j]$ and backward affinity value $\BM[v_j,r_j]$ are interpreted and defined by random walks in \secref{sec:attr-pr}.
%However, sampling numerous random walks over large-scale attributed networks to compute forward and backward affinity values for all possible node-attribute pairs in $G$ is rather inefficient.
Hence, we transform the forward and backward affinity in Equations \eqref{eq:fwd-prob} and \eqref{eq:bwd-prob} into their matrix forms and  propose \appr in Algorithm \ref{alg:appr}, which efficiently approximates forward and backward affinity matrices with error guarantee and in linear time complexity, without actually sampling random walks.

Observe that in Equations \eqref{eq:fwd-prob} and \eqref{eq:bwd-prob}, the key for forward and backward affinity computation is to obtain $p_f(v_i,r_j)$ and $p_b(v_i,r_j)$ for every pair $(v_i,r_j)\in V\times R$. Recall that $p_f(v_i,r_j)$ is the probability that a forward random walk starting from node $v_i$ picks attribute $r_j$, while $p_b(v_i,r_j)$ is the probability of a backward random walk from attribute $r_j$ stopping at node $v_i$. Given nodes $v_i$ and $v_l$, denote $\pi(v_i,v_l)$ as the probability that a random walk starting from $v_i$ stops at $v_l$, \ie the random walk score of $v_l$ with respect to $v_i$. By definition, $ p_f(v_i,r_j) =\sum_{v_l\in V}{\pi(v_i,v_l)\cdot{\RM_r}[v_l,r_j]}$, where ${\RM_r}[v_l,r_j]$ is the probability that node $v_l$ selects attribute $r_j$, according to Equation \eqref{eq:norm-r}. 
%from its nonzero attributes based on attribute weights
Similarly, $p_b(v_i,r_j)$ is formulated as $ p_b(v_i,r_j) =\sum_{v_l\in V}{\RM_c[v_l,r_j]\cdot \pi(v_l,v_i)}$, where $\RM_c[v_l,r_j]$ is the probability that attribute $r_j$ picks node $v_l$ from all nodes having $r_j$ based on their attribute weights. % (see \equref{eq:norm-r})
%In the first phase, we aim to compute the approximate forward and backward affinity matrices. Observe that the key to compute forward and backward affinity values is computing $p_f(v_i,r_j)$ and $p_b(v_i,r_j)$ for every pair $(v_i,r_j)\in V\times R$. 
%Recall that in \secref{sec:attr-pr}, $p_f(v_i,r_j)$ represents the probability that a forward attributed RWR starting from node $v_i$ picks attribute $r_j$ and $p_b(r_j,v_i)$ is the probability that a backward attributed RWR from attribute $r_j$ stops at node $v_i$, which by definition are $ p_f(v_i,r_j) =\sum_{v_l\in V}{\pi(v_i,v_l)\cdot{\RM_r}[v_l,r_j]}$ and $ p_b(v_i,r_j) =\sum_{v_l\in V}{\RM_c[v_l,r_j]\cdot \pi(v_l,v_i)},$
%where $\pi(v_i,v_l)$ (resp. $\pi(v_l,v_i)$) denotes the RWR score of $v_l$ (resp. $v_i$) w.r.t $v_i$ (resp. $v_l$), namely the probability that a RWR starting from node $v_i$ (resp. $v_l$) terminates at node $v_l$ (resp. $v_i$). 
 By the definition of random walk scores in \cite{jeh2003scaling,tong2006fast}, we derive the matrix form of $p_f$ and $p_b$ as follows.
% \begin{small}
\begin{equation*}\label{eq:fwd-bwd-prob-m}
\begin{split}
\PM_f &= \alpha\sum_{\ell=0}^{\infty}{(1-\alpha)^{\ell}\PM^{\ell}\RM_r},\\
\PM_b &= \alpha\sum_{\ell=0}^{\infty}{(1-\alpha)^{\ell}\PM^{\top \ell}\RM_c}.
\end{split}
\end{equation*}     
% \end{small}
%The computations of exact $\PM_f$ and $\PM_b$ are impossible, and thus we approximate them with $t$ iterations, \ie 
We only consider $t$ iterations to approximate $\PM_f$ and $\PM_b$ in \equref{eq:fwd-bwd-prob-m-t}, where $t$ is set to $\frac{\log(\epsilon)}{\log(1-\alpha)}-1$.
% \begin{equation}\label{eq:fwd-bwd-prob-m-t}
% \begin{split}
%  \PM^{(t)}_f &=  \alpha\sum_{\ell=0}^{t}{(1-\alpha)^{\ell}\PM^{\ell}\cdot\RM_r},\\
%  \PM^{(t)}_b &= \alpha\sum_{\ell=0}^{t}{(1-\alpha)^{\ell}\PM^{\top \ell}\cdot\RM_c}.
% \end{split}
% \end{equation}
% \begin{small}
\begin{equation}\label{eq:fwd-bwd-prob-m-t}
\begin{aligned}
\PM^{(t)}_f &=  \alpha\sum_{\ell=0}^{t}{(1-\alpha)^{\ell}\PM^{\ell}\RM_r},\\
\PM^{(t)}_b &= \alpha\sum_{\ell=0}^{t}{(1-\alpha)^{\ell}\PM^{\top \ell}\RM_c}.
\end{aligned}
\end{equation}    
% \end{small}
Then, we normalize $\PM_f^{ (t)}$ by columns and $\PM_b^{ (t)}$ by rows as follows.
\begin{small}
\begin{equation*}
\begin{aligned}
\widehat{\PM}_f^{ (t)}[v_i,r_j]&=\frac{\PM_f^{ (t)}[v_i,r_j]}{\sum_{v_l\in V}{\PM_f^{ (t)}[v_l,r_j]}},\\
\widehat{\PM}_b^{ (t)}[v_i,r_j]&=\frac{\PM_b^{ (t)}[v_i,r_j]}{\sum_{r_l\in R}{\PM_b^{ (t)}[v_i,r_l]}}.
\end{aligned}
\end{equation*}  
\end{small}

After normalization, we compute $\FM'$ and $\BM'$  according to the definitions of forward and backward affinity as follows.
\begin{equation}\label{equ:approxFB}
\FM^{\prime} = \log(n\cdot\widehat{\PM}_f^{ (t)}+1),\quad \BM^{\prime} = \log(d\cdot\widehat{\PM}_b^{ (t)}+1).    
\end{equation}

%After obtaining $\PM^{(t)}_f$ and $\PM^{(t)}_b$, we are able to compute approximate forward and backward affinity matrices $\FM^{\prime}$ and $\BM^{\prime}$.

\algoref{alg:appr} shows the pseudo-code of \appr to compute $\FM'$ and $\BM'$. Specifically, \appr takes as inputs random walk matrix $\PM$, attribute matrix $\RM$, random walk stopping probability $\alpha$ and the number of iterations $t$.  At Line 1, \appr begins by computing row-normalized attribute matrix $\RM_r$ and column-normalized attribute matrix $\RM_c$ according to \equref{eq:norm-r}. Then, \appr computes $\PM^{(t)}_f$ and $\PM^{(t)}_b$ based on \equref{eq:fwd-bwd-prob-m-t}. Note that $\PM$ is sparse and has $m$ non-zero entries. Thus, the computations of $\alpha\sum_{\ell=0}^{t}{(1-\alpha)^{\ell}\PM^{\ell}}$ and $\alpha\sum_{\ell=0}^{t}{(1-\alpha)^{\ell}\PM^{\top \ell}}$  in \equref{eq:fwd-bwd-prob-m-t} need $O(mnt)$ time, which is prohibitively expensive on large graphs. We avoid such expensive overheads and achieve a time cost of $O(mdt)$ for computing $\PM^{(t)}_f$ and $\PM^{(t)}_b$ by an iterative process as follows. 
Initially, we set $\PM_f^{ (0)}=\RM_r$ and $\PM_b^{ (0)}=\RM_c$ (Line 2). Then, we start an iterative process from Line 3 to 5 with $t$ iterations; at the $\ell$-th iteration, we compute $\PM_f^{ (\ell)}=(1-\alpha)\cdot\PM \PM_f^{ (\ell-1)} + \alpha\cdot\PM_f^{(0)}$ and $\PM_b^{ (\ell)}=(1-\alpha)\cdot\PM^{\top}\PM_b^{ (\ell-1)} + \alpha\cdot\PM_b^{(0)}$. After $t$ iterations, \appr normalizes $\PM_f^{ (t)}$ by column and $\PM_b^{ (t)}$ by row (Lines 6-7). At Line 8, \appr obtains $\FM^{\prime}$ and $\BM^{\prime}$ as the approximate forward and backward affinity matrices. The following lemma establishes the accuracy guarantee of \appr. 
% \appdref.

\begin{lemma}\label{lem:apa}
Given $\PM,{\RM_r},\alpha,\epsilon$ as inputs to \algoref{alg:appr}, the returned approximate forward and backward affinity matrices $\FM^{\prime}$, $\BM^{\prime}$ satisfy that, for every pair $(v_i,r_j)\in V\times R$,
% $\frac{2^{\FM^{\prime}[v_i,r_j]}-1}{2^{\FM[v_i,r_j]}-1}$ and $\frac{2^{\BM^{\prime}[v_i,r_j]}-1}{2^{\BM[v_i,r_j]}-1}$ is in
% \begin{align*}
% & \frac{2^{\FM^{\prime}[v_i,r_j]}-1}{2^{\FM[v_i,r_j]}-1}\in \left[\max\{0,\frac{\PM_f[v_i,r_j]-\epsilon}{\PM_f[v_i,r_j]}\}, \frac{\epsilon+\sum_{v_l\in V}{\max\{0,\PM_f[v_l,r_j]-\epsilon\}}}{\sum_{v_l\in V}{\max\{0,\PM_f[v_l,r_j]-\epsilon\}}}\right],\\
% & \frac{2^{\BM^{\prime}[v_i,r_j]}-1}{2^{\BM[v_i,r_j]}-1}\in \left[\max\{0,\frac{\PM_b[v_i,r_j]-\epsilon}{\PM_b[v_i,r_j]}\}, \frac{\epsilon+\sum_{r_l\in R}{\max\{0,\PM_b[v_i,r_l]-\epsilon\}}}{\sum_{r_l\in R}{\max\{0,\PM_b[v_i,r_l]-\epsilon\}}}\right].
% \end{align*}
% \begin{small}
\begin{align*}
	& \max\Big\{0,1-\frac{\epsilon}{\PM_f[v_i,r_j]}\Big\} \le \frac{2^{\FM^{\prime}[v_i,r_j]}-1}{2^{\FM[v_i,r_j]}-1}, \\
	& \max\Big\{0,1-\frac{\epsilon}{\PM_b[v_i,r_j]}\Big\} \le \frac{2^{\BM^{\prime}[v_i,r_j]}-1}{2^{\BM[v_i,r_j]}-1},
\end{align*} 
% \end{small}
and
% \begin{small}
\begin{align*}
	& \frac{2^{\FM^{\prime}[v_i,r_j]}-1}{2^{\FM[v_i,r_j]}-1}\le 1+\frac{\epsilon}{\sum_{v_l\in V}{\max\{0,\PM_f[v_l,r_j]-\epsilon\}}},\\
	& \frac{2^{\BM^{\prime}[v_i,r_j]}-1}{2^{\BM[v_i,r_j]}-1} \le 1+\frac{\epsilon}{\sum_{r_l\in R}{\max\{0,\PM_b[v_i,r_l]-\epsilon\}}}.
\end{align*}  
% \end{small}
\begin{proof}
First, with $t=\frac{\log(\epsilon)}{\log(1-\alpha)}-1$, we have
\begin{small}
\begin{equation}\label{eq:alpha-eps}
\sum_{\ell=t+1}^{\infty}{\alpha(1-\alpha)^{\ell}}=1-\sum_{\ell=0}^{t}{\alpha(1-\alpha)^{\ell}}=(1-\alpha)^{t+1}=\epsilon.
\end{equation}
\end{small}
By the definitions of $\PM_f,\PM^{(t)}_f$ and $\PM_b,\PM^{(t)}_b$ (\ie Equation \eqref{eq:fwd-bwd-prob-m-t}), for every pair $(v_i,r_j)\in V\times R$,
\begin{small}
\begin{align}
\PM_f[v_i,r_j]-\PM_f^{(t)}[v_i,r_j]&=\sum_{\ell=t+1}^{\infty}{\alpha(1-\alpha)^{\ell}\PM^{\ell}}[v_i]\cdot{\RM}^{\top}_r[r_j]\nonumber\\
&= \left(\sum_{\ell=t+1}^{\infty}{\alpha(1-\alpha)^{\ell}\PM^{\ell}}\right)[v_i]\cdot{\RM}^{\top}_r[r_j]\nonumber\\
&\le \sum_{\ell=t+1}^{\infty}{\alpha(1-\alpha)^{\ell}}=\epsilon\nonumber,
\end{align}
\end{small}
\begin{small}
\begin{align*}
\PM_b[v_i,r_j]-\PM_b^{(t)}[v_i,r_j] &=\sum_{\ell=t+1}^{\infty}{\alpha(1-\alpha)^{\ell}\PM^{\top\ell}[v_i]\cdot\RM^{\top}_c[r_j]}\nonumber\\
% =&\nonumber\\
&\le \sum_{v_l\in V}{\sum_{\ell=t+1}^{\infty}{\alpha(1-\alpha)^{\ell}}\cdot\RM_c[v_l,r_j]}\nonumber \\
&\le \sum_{v_l\in V}{\epsilon\cdot\RM_c[v_l,r_j]}=\epsilon\nonumber.
\end{align*}
\end{small}
Based on the above inequalities, $\forall{(v_i,r_j)\in V\times R}$,
\begin{align}
\max\{0, \PM_f[v_i,r_j]-\epsilon\}\le \PM^{(t)}_f[v_i,r_j] \le \PM_f[v_i,r_j],\label{eq:fwd-v-r}\\
\max\{0, \PM_b[v_i,r_j]-\epsilon\}\le \PM^{(t)}_b[v_i,r_j] \le \PM_b[v_i,r_j]\label{eq:bwd-v-r}.
\end{align}
According to Lines 6-9 in Algorithm \ref{alg:appr}, for every pair $(v_i,r_j)\in V\times R$,
\begin{small}
\begin{align*}
&\frac{2^{\FM^{\prime}[v_i,r_j]}-1}{2^{\FM[v_i,r_j]}-1}=\frac{\widehat{\PM}^{(t)}_f[v_i,r_j]}{\widehat{\PM}_f[v_i,r_j]}=\frac{\PM^{(t)}_f[v_i,r_j]\cdot\sum_{v_l\in V}{\PM_f[v_l,r_j]}}{\sum_{v_l\in V}{\PM^{(t)}_f[v_l,r_j]}\cdot\PM_f[v_i,r_j]}
% \label{eq:flog},
% &=\frac{\PM^{(t)}_f[v_i,r_j]}{\sum_{v_l\in V}{\PM^{(t)}_f[v_l,r_j]}}\times \frac{\sum_{v_l\in V}{\PM_f[v_l,r_j]}}{\PM_f[v_i,r_j]}\label{eq:flog},
\end{align*}
\end{small}
\begin{small}
\begin{align*}
&\frac{2^{\BM^{\prime}[v_i,r_j]}-1}{2^{\BM[v_i,r_j]}-1}
=\frac{\widehat{\PM}^{(t)}_f[v_i,r_j]}{\widehat{\PM}_f[v_i,r_j]}=\frac{\PM^{(t)}_b[v_i,r_j]\cdot\sum_{r_l\in R}{\PM_b[v_i,r_l]}}{\sum_{r_l\in R}{\PM^{(t)}_b[v_i,r_l]}\cdot\PM_b[v_i,r_j]}
% \label{eq:blog}.
% &=\frac{\PM^{(t)}_b[v_i,r_j]}{\sum_{r_l\in R}{\PM^{(t)}_b[v_i,r_l]}}\times \frac{\sum_{r_l\in R}{\PM_b[v_i,r_l]}}{\PM_b[v_i,r_j]}\label{eq:blog}.
\end{align*}
\end{small}
Plugging in the above equations completes our proof. \done
% Plugging Equations \eqref{eq:fwd-v-r} and \eqref{eq:bwd-v-r} into Equations \eqref{eq:flog} and \eqref{eq:blog} completes our proof. \done
% that, for every pair $(v_i,r_j)\in V\times R$,
% \begin{align*}
% \max\{0,1-\frac{\epsilon}{\PM_f[v_i,r_j]}\} \le \frac{2^{\FM^{\prime}[v_i,r_j]}-1}{2^{\FM[v_i,r_j]}-1}\le 1+\frac{\epsilon}{\sum_{v_l\in V}{\max\{0,\PM_f[v_l,r_j]-\epsilon\}}},\\
% \max\{0,1-\frac{\epsilon}{\PM_b[v_i,r_j]}\}\le
% \frac{2^{\BM^{\prime}[v_i,r_j]}-1}{2^{\BM[v_i,r_j]}-1}\le 1+\frac{\epsilon}{\sum_{r_l\in R}{\max\{0,\PM_b[v_i,r_l]-\epsilon\}}}.
% \end{align*}
\end{proof}
\end{lemma} 
% In next section, based on the approximate forward and backward affinity matrices, we show how to obtain the forward, backward and attribute embedding vectors.

\subsection{Joint Factorization of Affinity Matrices}\label{sec:svdccd}
This section presents the proposed algorithm \svdccd, outlined in \algoref{alg:svdccd}, which jointly factorizes the approximate forward and backward affinity matrices $\FM^{\prime}$ and $\BM^{\prime}$, in order to obtain the embedding vectors of all nodes and attributes, \ie $\XMf,\XMb$, and $\YM$. Specifically, the proposed \svdccd solver is based on the \textit{cyclic coordinate descent} (\textit{CCD}) framework, which iteratively updates each embedding value towards optimizing the objective function in Equation \eqref{eq:obj1}. Unfortunately, a direct application of CCD, starting from random initial values of the embeddings, requires numerous iterations to converge, leading to prohibitive overheads. 
Furthermore, CCD computation itself is expensive, especially on large-scale graphs.
To overcome these challenges, we firstly propose a greedy initialization method to facilitate fast convergence, and then design techniques for efficient refinement of initial embeddings, including dynamic maintenance and partial updates of intermediate results to avoid redundant computations in CCD.
% We address this problem with a greedy initilization method, as explained in the following.

\header
\textbf{Greedy initialization.} In many optimization problems, all we need for efficiency is a good initialization. Thus, a key component in the proposed \svdccd algorithm is such an initialization of embedding values based on {\em singular value decomposition} (\textit{SVD}) \cite{golub1971singular}. Note that unlike other matrix factorization problems, here SVD by itself cannot solve our problem because the objective function in Equation \eqref{eq:obj1} requires the joint factorization of both the forward and backward affinity matrices at the same time, which cannot be directly addressed with SVD.

%Therefore, we propose to initialize $\XMf,\XMb$ and $\YM$ by {\em singular value decomposition} (SVD) \cite{golub1971singular}, which gives a rough estimation for $\XMf,\XMb$ and $\YM$, and thus reduces the number of iterations required to converge.
%Specifically, at Line 1 in \algoref{alg:svdccd}, we invoke \isvd in \algoref{alg:isvd} to initialize $\XMf,\XMb$ and $\YM$. 

\algoref{alg:isvd} describes the \isvd module of \svdccd, which initializes embeddings $\XMf,\XMb$, and $\YM$.
Specifically, the algorithm first employs an efficient randomized SVD algorithm \cite{musco2015randomized} at Line 1 to decompose $\FM'$ into $\UM\in \mathbb{R}^{n\times \frac{k}{2}},\boldsymbol{\Sigma}\in \mathbb{R}^{\frac{k}{2}\times \frac{k}{2}}$, $\VM\in \mathbb{R}^{d\times \frac{k}{2}}$, and then initializes $\XMf=\UM\mathbf{\Sigma}$ and $\YM=\VM$ at Line 2, which satisfies $\XMf\cdot\YM^{\top}\approx \FM^{\prime}$. In other words, this initialization immediately gains a good approximation of the forward affinity matrix. 

Recall that our objective function in \equref{eq:obj1} also aims to find $\XMb$ such that $\XMb\YM^{\top}\approx \BM'$, \ie to approximate the backward affinity matrix well. We observe that the matrix $\VM$ (\ie $\YM$) returned by exact SVD is \textit{unitary}, \ie $\YM^{\top}\YM=\mathbf{I}$, which implies that $\XMb\approx\XMb\YM^{\top}\YM\approx\BM^{\prime}\YM$. Accordingly, we seed $\XMb$ with $\BM'\YM$ at Line 2 of \algoref{alg:isvd}. This initialization of $\XMb$ also leads to a relatively good approximation of the backward affinity matrix. Consequently, the number of iterations required by \svdccd is drastically reduced, as confirmed by our experiments in Section \ref{sec:exp}.

%The rationale of SVD-based initialization scheme is as follows. 
%Let $\UM,\mathbf{\Sigma},\VM^{\top}$ be the results returned by exact SVD over approximate forward affinity matrix $\FM^{\prime}$, \ie $\UM\mathbf{\Sigma}\VM^{\top}=\FM^{\prime}$. 
%Simply, we let $\XMf=\UM\mathbf{\Sigma}$ and $\YM=\VM$. 
%In addition, we aim to find $\XMb$ such that $\XMb\YM^{\top}=\BM^{\prime}$.
%Obviously, it satisfies $\XMb\YM^{\top}\YM=\BM^{\prime}\YM$ and $\XMb=\BM\YM(\YM^{\top}\YM)^{-1}$. Note that $\YM=\VM$ is returned by exact SVD and is unitary, \ie $\YM^{\top}\YM=\mathbf{I}$. Thus, its inverse also satisfies $(\YM^{\top}\YM)^{-1}=\mathbf{I}$. Then, we have $\XMb=\BM\YM$. However, performing exact SVD over $\FM$ is very time consuming and thus we choose to use randomized SVD to yield a rough result. Although, setting $\XMf=\UM\mathbf{\Sigma}$, $\YM=\VM$ and $\XMb=\BM^{\prime}\YM$ is unsatisfactory for producing accurate embedding vectors, but is sufficient for our initialization of CCD.

\header
\textbf{Efficient refinement of the initial embeddings.} In \algoref{alg:svdccd}, after initializing $\XMf,\XMb$ and $\YM$ at Line 1, we apply cyclic coordinate descent to refine the embedding vectors according to our objective function in \equref{eq:obj1} from Lines 2 to 14. The basic idea of CCD is to cyclically iterate through all entries in $\XMf,\XMb$ and $\YM$, one at a time, minimizing the objective function with respect to each entry (\ie coordinate direction). Specifically, in each iteration, CCD updates each entry of $\XMf, \XMb$ and $\YM$ according to the following rules:
% \jieming{Explain how CCD works intuitively; how below equations derived? Any reasoning?}
%we can directly apply {\em cyclic coordinate descent} (CCD) \cite{wright2015coordinate} to jointly optimize the objective function, \ie \equref{eq:obj1}. 
\begin{align}
 \XMf[v_i,l]\gets& \XMf[v_i,l]-\mu_f(v_i,l),\label{eq:xf-update}\\
 \XMb[v_i,l]\gets& \XMb[v_i,l]-\mu_b(v_i,l),\label{eq:xb-update}\\
 \YM[r_j,l]\gets& \YM[r_j,l]-\mu_y(r_j,l)\label{eq:y-update},
\end{align}
with $\mu_f(v_i,l),\mu_b(v_i,l)$ and $\mu_y(r_j,l)$  computed by:
%minimize objective function \equref{eq:obj1}, respectively. Thus, the entry updates are as follows:
\begin{small}
\begin{align}
 \mu_f(v_i,l)= \frac{\SM_f[v_i]\cdot\YM[:,l]}{\YM^{\top}[l]\cdot\YM[:,l]},\ \mu_b(v_i,l)= \frac{\SM_b[v_i]\cdot\YM[:,l]}{\YM^{\top}[l]\cdot\YM[:,l]},\label{eq:update-x-mu}\\
%  \end{align}
%  \begin{align}
 \mu_y(r_j,l)= \frac{\XMf^{\top}[l]\cdot\SM_f[:,r_j]+\XMb^{\top}[l]\cdot\SM_b[:,r_j]}{\XMf^{\top}[l]\cdot\XMf[:,l]+\XMb^{\top}[l]\cdot\XMb[:,l]},\label{eq:update-y-mu}\quad
\end{align}    
\end{small}
where $\SM_f=\XMf\YM^{\top}-\FM^{\prime}$ and $\SM_b=\XMb\YM^{\top}-\BM^{\prime}$ are obtained at Line 3 in \algoref{alg:isvd}. 

However, directly applying the above updating rules to learn $\XMf,\XMb$, and $\YM$ is inefficient, leading to many redundant matrix operations. 
%and iterations to achieve convergence of objective function \equref{eq:obj1}, which are expensive on large graphs. 
Lines 2-14 in \algoref{alg:svdccd} show how to efficiently apply the above updating rules by dynamically maintaining and partially updating intermediate results.
%we show how to reduce redundant computations and initializing the embedding vectors to speed up the learning course.
Specifically, each iteration in Lines 3-14 first fixes $\YM$ and updates each row of $\XMf$ and $\XMb$ (Lines 3-9), and then updates each column of $\YM$ with $\XMf$ and $\XMb$ fixed (Lines 10-14).
According to Equations \eqref{eq:update-x-mu} and \eqref{eq:update-y-mu}, 
$\mu_f(v_i,l)$, $\mu_b(v_i,l)$, and $\mu_y(r_j,l)$ are pertinent to $\SM_f[v_i]$, $\SM_b[v_i]$, and $\SM_f[:,r_j], \SM_b[:,r_j]$ respectively, where $\SM_f$ and $\SM_b$ further depend on embedding vectors $\XMf$, $\XMb$ and $\YM$. Therefore, whenever $\XMf[v_i,l], \XMb[v_i,l]$, and $\YM[r_j,l]$ are updated in the iteration (Lines 6-7 and Line 13), $\SM_f$ and $\SM_b$ need to be updated accordingly. It would be expensive if we directly recompute $\SM_f$ and $\SM_b$ by $\SM_f=\XMf\YM^{\top}-\FM^{\prime}$ and $\SM_b=\XMb\YM^{\top}-\BM^{\prime}$, whenever an entry in $\XMf,\XMb$ and $\YM$ is updated. %(in $O(ndk)$ time)

Instead, we dynamically maintain and partially update $\SM_f$ and $\SM_b$ according to Equations \eqref{eq:update-x-sf}, \eqref{eq:update-x-sb} and \eqref{eq:update-y-sf-sb}. 
Specifically, when $\XMf[v_i,l]$ and $\XMb[v_i,l]$ are updated (Lines 6-7), we update $\SM_f[v_i]$ and $\SM_b[v_i]$ respectively with $O(d)$ time at Lines 8-9 by 
\begin{align}
\SM_f[v_i]&\gets\SM_f[v_i]-\mu_f(v_i,l)\cdot \YM[:,l]^{\top},\label{eq:update-x-sf}\\
% \end{equation}
% When $\XMb[v_i,l]$ is updated at Line 7, $\SM_b[v_i]$ is updated in $O(d)$ time at Line 9 by 
% \begin{equation}
\SM_b[v_i]&\gets\SM_b[v_i]-\mu_b(v_i,l)\cdot \YM[:,l]^{\top},\label{eq:update-x-sb}
\end{align}
Whenever $\YM[r_j,l]$ is updated at Line 13, both $\SM_f[:,r_j]$ and $\SM_b[:,r_j]$ are updated in $O(n)$ time at Line 14 by
\begin{equation}\label{eq:update-y-sf-sb}
\begin{split}
\SM_f[:,r_j]\gets\SM_f[:,r_j]-\mu_y(r_j,l)\cdot \XMf[:,l],\\ \SM_b[:,r_j]\gets\SM_b[:,r_j]-\mu_y(r_j,l)\cdot \XMb[:,l].
\end{split}
\end{equation}

\begin{algorithm}[!t]
	\begin{small}
		\caption{\isvd}
		\label{alg:isvd}
		\KwIn{$\FM^{\prime},\BM^{\prime}, k, t$.}
		\KwOut{$\XMf,\XMb,\YM,\SM_f,\SM_b$.}
		$\UM, \boldsymbol{\Sigma}, \VM \gets \mathtt{RandSVD}(\FM^{\prime},\frac{k}{2},t)$\;
		$\YM\gets \VM,\ \XMf \gets \UM\boldsymbol{\Sigma},\ \XMb \gets \BM^{\prime}\cdot\YM$\;
		$\SM_f \gets \XMf\YM^{\top}-\FM^{\prime},\ \SM_b \gets \XMb\YM^{\top}-\BM^{\prime}$\;
		\Return $\XMf,\XMb,\YM,\SM_f,\SM_b$\;
	\end{small}
\end{algorithm}

\begin{algorithm}[t]
	\begin{small}
		\caption{\svdccd}
		\label{alg:svdccd}
		\BlankLine
		\KwIn{$\FM^{\prime}, \BM^{\prime}$, $k$, $t$.}
		\KwOut{$\XMf,\YM,\XMb$.}
		$\XMf,\XMb,\YM,\SM_f,\SM_b \gets \mathtt{GInit}(\FM^{\prime},\BM^{\prime}, k, t)$\;
		\For{$\ell\gets 1$ to $t$}{
		  %  $\WM_f \gets \SM_f\YM,\ \WM_b \gets \SM_b\YM,\ \widehat{\YM} \gets \YM^{\top}\YM$\;
		    \For{$v_i\in V$}{
		        \For{$l\gets 1$ to $\frac{k}{2}$}{
		          %  $y_l = \YM^{\top}[l]\cdot\YM[:,l]$\;
		          %  $\mu_f(v_i,l) \gets \frac{\SM_f[v_i]\cdot\YM[:,l]}{y_l}$\;
		          Compute $\mu_f(v_i,l),\mu_b(v_i,l)$ by \equref{eq:update-x-mu}\;
		            $\XMf[v_i,l]\gets \XMf[v_i,l]-\mu_f(v_i,l)$\;
		          %  $\mu_b(v_i,l) \gets \frac{\SM_b[v_i]\cdot\YM[:,l]}{y_l}$\;
		            $\XMb[v_i,l]\gets \XMb[v_i,l]-\mu_b(v_i,l)$\;
		            Update $\SM_f[v_i]$ by \equref{eq:update-x-sf}\;
		            Update $\SM_b[v_i]$ by \equref{eq:update-x-sb}\;
		          %  $\SM_f[v_i] \gets \SM_f[v_i]-\mu_f(v_i,l)\cdot\YM^{\top}[l]$\;
		          %  $\SM_b[v_i] \gets \SM_b[v_i]-\mu_b(v_i,l)\cdot\YM^{\top}[l]$\;
		        }
		    }
		  %  $\WM \gets \XMf^{\top}\SM_f+\XMb^{\top}\SM_b,\ \widehat{\XM} \gets \XMf^{\top}\XMf+ \XMb^{\top}\XMb$\;
		    \For{$r_j\in R$}{
		        \For{$l\gets 1$ to $\frac{k}{2}$}{
		          %  $\mu_y(r_j,l) \gets \frac{\XMf^{\top}[l]\cdot\SM_f[:,r_j]+\XMb^{\top}[l]\cdot\SM_b[:,r_j]}{\XMf^{\top}[l]\cdot\XMf[:,l]+\XMb^{\top}[l]\cdot\XMb[:,l]}$\;
		          Compute $\mu_y(r_j,l)$ by \equref{eq:update-y-mu}\;
		            $\YM[r_j,l]\gets \YM[r_j,l]-\mu_y(r_j,l)$\;
		            Update $\SM_f[:,r_j],\SM_b[:,r_j]$ by \equref{eq:update-y-sf-sb}\;
                    %  $\SM_f[:,r_j]\gets \SM_f[:,r_j]-\mu_y(r_j,l)\cdot \XMf[:,l]$\;
		          % $\SM_b[:,r_j]\gets \SM_b[:,r_j]-\mu_y(r_j,l)\cdot \XMb[:,l]$\;
		        }
		    }
		}
		\Return $\XMf,\YM,\XMb$\;
	\end{small}
\end{algorithm}

% \subsection{Theoretical Analysis}\label{sec:algo-als}
% \header
% {\bf Accuracy Analysis.}
% The following theorem shows that \algopt indeed produces forward and backward embeddings that well preserve FPMI and BPMI.
% \begin{theorem}\label{thm:main}
% When we set $t= \frac{\log(\epsilon)}{\log(1-\alpha)}-1$, the forward and backward embeddings produced by Algorithm \ref{alg:mainopt} satisfy, for every pair $(v,r)\in V\times R$ 
% \begin{align*}
% & \XMf[v]\cdot\YMf[r]^{\top}-\log(\FM[v,r]+1)\le(1+\epsilon)\cdot\sigma_{k+1}(\FM),\\
% &\XMf[v]\cdot\YMf[r]^{\top}-\log(\FM[v,r]+1)\\
% & \ge \log\left(\frac{\max\{\FM[v,r]-n\epsilon,0\}+1}{\FM[v,r]+1}\right)-(1+\epsilon)\cdot\sigma_{k+1}(\FM).
% \end{align*}
% and for every pair $(v,r)\in V\times R$,
% \begin{align*}
% & \XMb[v]\cdot\YMb[r]^{\top}-\log(\BM[v,r]+1)\le(1+\epsilon)\cdot\sigma_{k+1}(\BM),\\
% &\XMb[v]\cdot\YMb[r]^{\top}-\log(\BM[v,r]+1)\\
% & \ge \log\left(\frac{\max\{\BM[v,r]-n\epsilon,0\}+1}{\BM[v,r]+1}\right)-(1+\epsilon)\cdot\sigma_{k+1}(\BM).
% \end{align*}
% \vspace{-4mm}
% \begin{proof}
% See \textnormal{Appendix}.
% \end{proof}
% \end{theorem}

\subsection{Complexity Analysis}\label{sec:algo-als}
In the proposed algorithm \algopt (\algoref{alg:mainopt}), the maximum length of random walk is $ t=\frac{\log(\epsilon)}{\log(1-\alpha)}-1=\frac{\log(\frac{1}{\epsilon})}{\log(\frac{1}{1-\alpha})}-1$. According to \secref{sec:apmi}, \algoref{alg:appr} runs in time $ O\left(md\cdot t\right)=O\left(md\cdot\log\frac{1}{\epsilon}\right)$. Meanwhile, according to \cite{musco2015randomized}, given $\FM^{\prime}\in \mathbb{R}^{n\times d}$ as input, $\mathtt{RandSVD}$ in \algoref{alg:isvd} requires $O\left(ndkt\right)$ time, where $n$, $d$, $k$ are the number of nodes, number of attributes, and embedding space budget, respectively. The computation of $\SM_f,\SM_b$ costs $O(ndk)$ time. In addition, the $t$ iterations of CCD for updating $\XMf,\XMb$ and $\YM$ take $O(ndkt) = O(ndk\log\frac{1}{\epsilon})$ time. Therefore, the overall time complexity of \algoref{alg:mainopt} is $O\left((md+ndk)\cdot\log\left(\frac{1}{\epsilon}\right)\right).$ The memory consumption of intermediate results yielded in \algoref{alg:mainopt}, \ie $\FM^{\prime}$,$\BM^{\prime}$, $\UM,\boldsymbol{\Sigma},\VM$,$\SM_f$,$\SM_b$ are at most $O(nd)$. Hence, the space complexity of \algoref{alg:mainopt} is bounded by $O(nd+m)$.
% , where the $m$ term is due to the space required by the input graph.

%% file: parallel.tex
\vspace{-2mm}
\section{\algoptp: Parallel \algo}\label{sec:parallel}
\begin{algorithm}[t]
	\begin{small}
	\caption{\algoptp}
	\label{alg:mainoptp}
	\BlankLine
	\KwIn{Attributed network $G$, space budget $k$, random walk stopping probability $\alpha$, error threshold $\epsilon$, the number of threads $n_b$.}
	\KwOut{Forward and backward embedding vectors $\XMf$, $\XMb$ and attribute embedding vectors $\YM$.}
	Partition $V$ into $n_b$ subsets $\mathcal{V}\gets\{V_1,\cdots,V_{n_b}\}$ equally\;
	Partition $R$ into $n_b$ subsets $\mathcal{R}\gets\{R_1,\cdots,R_{n_b}\}$ equally\;
	$t\gets \frac{\log(\epsilon)}{\log(1-\alpha)}-1$\;
    $\FM^{\prime}, \BM^{\prime} \gets \mathtt{PAPMI}(\PM,\RM,\alpha,t,\mathcal{V},\mathcal{R})$\;
	$\XMf, \YM, \XMb \gets \mathtt{PSVDCCD}(\FM^{\prime}, \BM^{\prime},\mathcal{V},\mathcal{R},k,t)$\;
	\Return $\XMf,\YM,\XMb$\;
\end{small}
\end{algorithm}
Although single-thread \algo (\ie \algopt in \algoref{alg:mainopt}) runs in linear time to the size of the input attributed network, it still requires substantial time to handle large-scale attributed networks in practice.  
For instance, on {\em MAG} dataset that has $59.3$ million nodes, \algopt (single thread) takes about five days.
% To further boost efficiency, in this section we develop a parallel \algoptp (Algorithm \ref{alg:mainoptp}), and it takes only $11.9$ hours on {\em MAG} when using $10$ threads (\ie up to 10 times speedup).
Note that it is challenging to develop a parallel algorithm achieving such linear scalability to the number of threads on a multi-core CPU.
Specifically, \algopt involves various complex computational steps, including intensive matrix computation, factorization, and CCD updates.
% Therefore, it is non-trivial to assign computing tasks of both nodes and attributes to threads, to fully utilize the parallel power. 
Moreover, it is also challenging to maintain the intermediate result of each thread and combine them as the final result.
% To this end, we propose several parallelization techniques for  \algopt. 
%Parallel \algoptp significantly improves the efficiency of the training process, %without considerably compromising effectiveness, 
%as validated in our experiments in  \secref{sec:exp-effi}. For instance, parallel \algoptp is 9x faster than single-thread \algopt on a massive graph {\em MAG}, when using 10 cores.
%Parallel \algoptp also consists of two phases. 
To further boost efficiency, in this section we develop a parallel \algo (\algoptp in Algorithm \ref{alg:mainoptp}), which takes only $11.9$ hours on {\em MAG} when using $10$ threads (\ie up to 10 times speedup with respect to \algopt).
In the first phase, we adopt block matrix multiplication \cite{golub1996matrix} and propose \rpapr to compute forward and backward affinity matrices in a parallel manner (Section \ref{sec:papa}). In the second phase, we develop \psvdccd with a split-merge-based parallel SVD technique to efficiently decompose affinity matrices, and further propose a parallel CCD technique to refine the embeddings efficiently (Section \ref{sec:split-merge}).

\algoref{alg:mainoptp} illustrates the pseudo-code of parallel \algoptp. Compared to the single-thread version, parallel \algoptp takes as input an additional parameter, the number of threads $n_b$, and { randomly partitions the node set $V$, as well as the attribute set $R$, into $n_b$ subsets with equal size, denoted as $\mathcal{V}$ and $\mathcal{R}$, respectively (Lines 1-2).} \algoptp invokes \rpapr (\algoref{alg:rpapr}) at Line 4 to get $\FM'$ and $\BM'$, and then invokes \psvdccd (\algoref{alg:psvdccd}) to refine the embeddings. 

Note that the parallel version of \algopt does not return exactly the same outputs as the single-thread version, as some modules (\textit{e.g.}, the parallel version of SVD) introduce additional error. Nevertheless, as the experiments in Section \ref{sec:exp} demonstrates, the degradation of result utility in parallel \algopt is small but the speedup is significant.
%when running on a moderate number of CPU cores (10 in our experiments), which is well compensated by the running time reduction of parallel \algopt.

%Similar to \algopt, \algoptp runs in a two-phase framework with an additional input parameter, \ie the number of matrix blocks $n_b$. Before entering the two phases, \algoptp partitions node set $V$ and and attribute set $R$ into $n_b$ sub-sets equally, \ie $\mathcal{V}$ and $\mathcal{R}$, respectively (Line 1-2).
%After that, \algoptp sets the number of iterations $t=\frac{\log(\epsilon)}{\log(1-\alpha)}-1$ (Line 3). In the first phase, \algoptp invokes \rpapr, \ie parallelized \appr, with the following parameters: $\PM,\RM,\alpha,k,t$, partitioned node and attribute sub-sets $\mathcal{V},\mathcal{R}$, and returns $\FM^{\prime}$ as the approximate forward affinity matrix and $\BM^{\prime}$ as the approximate backward affinity matrix (Line 4). Second, \algoptp factorizes $\FM^{\prime}$ and $\BM^{\prime}$ to obtain attribute embedding vectors $\YM$, forward and backward embedding vectors $\XMf, \XMb$ by invoking \psvdccd, \ie parallelized \svdccd, with $\FM^{\prime},\BM^{\prime},k,t$, partitioned node and attribute sub-sets $\mathcal{V},\mathcal{R}$ as inputs (Line 5).
\vspace{-1mm}
\subsection{Parallel Forward and Backward Affinity Approximation}\label{sec:papa}
\begin{algorithm}[!t]
\begin{small}
\caption{\rpapr}
\label{alg:rpapr}
\BlankLine
\KwIn{$\PM,\RM,\alpha,t,\mathcal{V},\mathcal{R}$}
\KwOut{$\FM^{\prime}, \BM^{\prime}$}
	Compute $\RM_r$ and $\RM_c$ by \equref{eq:norm-r}\;
% {\nonl{{\color{blue}Line 1 is the same as Line 1 in Algorithm \ref{alg:appr}}}\;
% \setcounter{AlgoLine}{1}}
\parallelfor{$R_i\in \mathcal{R}$}{
    {${\PM_f}^{ (0)}_i \gets \RM_r[:,R_i], {\PM_b}^{ (0)}_i \gets \RM_c[:,R_i]$}\;
\For{$\ell \gets 1$ to $t$}{
	${\PM_f}^{ (\ell)}_i \gets (1-\alpha)\cdot\PM {\PM_f}^{ (\ell-1)}_i + \alpha\cdot{\PM_f}^{ (0)}_i$\;
	${\PM_b}^{ (\ell)}_i \gets (1-\alpha)\cdot\PM^{\top} {\PM_b}^{ (\ell-1)}_i + \alpha\cdot{\PM_b}^{ (0)}_i$\;
}
% \nonl{{\color{blue} Lines 4-6 (\ie computing ${\PM_f}^{ (t)}_i,{\PM_b}^{ (t)}_i$) are like Lines 3-5 in Algorithm \ref{alg:appr}}}\;
% Normalize ${\PM_f}^{ (t)}_i$ by columns to get ${\widehat{\PM}_{f_i}}^{ (t)}$\;
% $\mathbf{h}_i\gets \sum_{r_j\in R_i}{{\PM_b}^{ (t)}_i[:,r_j]}$\;
}
\setcounter{AlgoLine}{6}
${{\PM}_f}^{ (t)}\gets [{{\PM}_{f_1}}^{ (t)}\cdots {{\PM}_{f_{n_b}}}^{ (t)}]$\;
% $\mathbf{h}=\sum_{R_i\in \mathcal{R}}{\mathbf{h}_i}$\;
${{\PM}_b}^{ (t)}\gets [{\PM_b}^{ (t)}_1\cdots {\PM_b}^{ (t)}_{n_b}]$\;
{\nonl{Lines 9-10 are the same as Lines 6-7 in Algorithm \ref{alg:appr}}\;
\setcounter{AlgoLine}{10}}
\parallelfor{$V_i\in \mathcal{V}$}{
    $\FM^{\prime}[V_i] \gets \log(n\cdot {\widehat{\PM}_f}^{ (t)}[V_i] +1)$\;
    $\BM^{\prime}[V_i] \gets \log(d\cdot {\widehat{\PM}_b}^{ (t)}[V_i] +1)$\;
    % \nonl {\color{blue} \ie Lines 12-13 (computing $\FM^{\prime}[V_i],\BM^{\prime}[V_i]$) are like Lines 8-9 in Algorithm \ref{alg:appr}}\;
}
\setcounter{AlgoLine}{13}
\Return $\FM^{\prime}, \BM^{\prime}$
\end{small}
\end{algorithm}

%The sequential  \appr in \algoref{alg:appr}   incurs immense computational cost to compute forward and backward affinity matrices $\FM'$ and $\BM'$. 
%To utilize multiple  threads,
%address the deficiency,
We propose \rpapr in  Algorithm \ref{alg:rpapr} to estimate  $\FM'$ and $\BM'$  in parallel. After obtaining $\RM_r$ and $\RM_c$ based on Equation \eqref{eq:norm-r} at Line 1, \rpapr divides $\RM_r$ and $\RM_c$ into  matrix blocks according to two input parameters, the node subsets $\mathcal{V}=\{V_1,V_2,\cdots,V_{n_b}\}$ and attribute subsets $\mathcal{R}=\{R_1,R_2,\cdots,R_{n_b}\}$. 
Then, \rpapr parallelizes the matrix multiplications for computing $\PM^{(t)}_f$ and $\PM^{(t)}_b$ from Line 2 to 6, using $n_b$ threads in $t$ iterations. Specifically, the $i$-th thread initializes ${\PM_{f_i}}^{ (0)}$ by $\RM_r[:,R_i]$ and ${\PM_{b_i}}^{ (0)}$ by $\RM_c[:,R_i]$ (Line 3), and then computes ${\PM_f}^{ (\ell)}_i = (1-\alpha)\cdot\PM {\PM_f}^{ (\ell-1)}_i + \alpha\cdot{\PM_f}^{ (0)}_i$ and ${\PM_b}^{ (\ell)}_i = (1-\alpha)\cdot\PM^{\top} {\PM_b}^{ (\ell-1)}_i + \alpha\cdot{\PM_b}^{ (0)}_i$ (Lines 4-6). 
% Each thread finishes the computation by normalizing ${\PM_f}^{ (t)}_i$ by columns, \ie ${\widehat{\PM}_{f_i}}^{ (t)}[v_l,r_j]=\frac{{\PM_f}^{ (t)}_i[v_l,r_j]}{\sum_{v_h\in V}{{\PM_f}^{ (t)}_i[v_h,r_j]}}$, and summing up the columns of ${\PM_b}^{ (t)}_i$ to obtain a length-$n$ vector $\mathbf{h}_i$, in which the $v_l$-th element is $\mathbf{h}_i[v_l]=\sum_{r_j\in R}{{\PM_{b_i}}^{ (t)}[v_l,r_j]}$ (Lines 8-9).
Then, we use a main thread to aggregate the partial results of all threads at Lines 7-8. 
Specifically, $n_b$ matrix blocks ${{\PM}_{f_i}}^{ (t)}$ (resp. ${{\PM}_{b_i}}^{ (t)}$) are concatenated horizontally together as ${{\PM}_f}^{ (t)}$ (resp. ${{\PM}_b}^{ (t)}$) at Line 7 (resp. Line 8). 
At Lines 9-10, we normalize ${\widehat{\PM}_f}^{ (t)}$ and ${\widehat{\PM}_b}^{ (t)}$ in the same way as Lines 6-7 in Algorithm \ref{alg:appr}.
% After that, we sum up $n_b$ length-$n$ vectors $\mathbf{h}_i$ to $\mathbf{h}$, in which $\mathbf{h}[v_l]=\sum_{R_i\in \mathcal{R}}{\sum_{r_j\in R_i}{{\PM^{(t)}_{b_i}}[v_l,r_j]}}$ at Line 11, and concatenate $n_b$ matrix blocks, ${\PM_b}^{ (t)}_i[:,r_j]$ divided by $\mathbf{h}[r_j]$, horizontally together as ${\widehat{\PM}_b}^{(t)}$, in which $\widehat{\PM}^{(t)}_{b}[v_l,r_j]={{\PM^{ (t)}_{b_i}}[v_l,r_j]}/{\mathbf{h}[v_l]}$, at Line 12.
From Lines 11 to 13, \rpapr starts $n_b$ threads to compute $\FM'$ and $\BM'$ block by block in parallel, based on the definitions of forward and backward affinity. Specifically, the $i$-th thread computes $\FM^{\prime}[V_i]=\log(n\cdot {\widehat{\PM}_f}^{ (t)}[V_i] +1)$ and $\BM^{\prime}[V_i]=\log(d\cdot {\widehat{\PM}_b}^{ (t)}[V_i] +1)$.
Finally, \rpapr returns $\FM^{\prime}$ and $\BM^{\prime}$ as the approximate forward and backward affinity matrices (Line 14). \lemref{lem:papa} indicates the accuracy guarantee of \rpapr. 

\begin{lemma}\label{lem:papa}
Given same parameters $\PM,\RM,\alpha$ and $t$ as inputs to \algoref{alg:appr} and \algoref{alg:rpapr}, the two algorithms return the same approximate forward and backward affinity matrices $\FM^{\prime}$, $\BM^{\prime}$.
\begin{proof}
According to Line 3 in \algoref{alg:rpapr}, we have
\begin{equation*}
\RM_r=
\begin{bmatrix}
{\PM_f}^{(0)}_1 & {\PM_f}^{(0)}_2 & \cdots & {\PM_f}^{(0)}_{n_b}
\end{bmatrix},
\end{equation*}
where ${\PM_f}^{(0)}_1,\cdots,{\PM_f}^{(0)}_{n_b-1}\in \mathbb{R}^{n\times \frac{d}{n_b}}$ and ${\PM_f}^{(0)}_{n_b}\in \mathbb{R}^{n\times (d\%n_b)}$ ($d\%n_b$ denotes the remainder of integer $d$ divded by $n_b$), and
\begin{equation*}
\RM_c=
\begin{bmatrix}
{\PM_b}^{(0)}_1 & {\PM_b}^{(0)}_2 & \cdots & {\PM_b}^{(0)}_{n_b}
\end{bmatrix},
\end{equation*}
where ${\PM_b}^{(0)}_1,\cdots,{\PM_b}^{(0)}_{n_b-1}\in \mathbb{R}^{n\times \frac{d}{n_b}}$ and ${\PM_b}^{(0)}_{n_b}\in \mathbb{R}^{n\times (d\%n_b)}$. After $t$ iterations, by Lines 4-6 in \algoref{alg:rpapr}, we have
% \begin{small}
\begin{align*}
{\PM_f}^{(t)}_i&=\alpha\sum_{\ell=0}^{t}{(1-\alpha)^{\ell}\PM^{\ell}{\PM_f}^{(0)}_i}\ \textrm{and}\\
{\PM_b}^{(t)}_i&=\alpha\sum_{\ell=0}^{t}{(1-\alpha)^{\ell}\PM^{\top\ell}{\PM_b}^{(0)}_i}.
\end{align*}
% \end{small}
Thus, we can derive that
\begin{align*}
\PM^{(t)}_f=\begin{bmatrix} {\PM_f}^{(t)}_1 & \cdots & {\PM_f}^{(t)}_{n_b}
\end{bmatrix}=\alpha\sum_{\ell=0}^{t}{(1-\alpha)^{\ell}\PM^{\ell}\RM_r},\\
\PM^{(t)}_b=\begin{bmatrix}
{\PM_b}^{(t)}_1 & \cdots & {\PM_b}^{(t)}_{n_b}
\end{bmatrix}=\alpha\sum_{\ell=0}^{t}{(1-\alpha)^{\ell}\PM^{\ell}\RM_c}.
\end{align*}
% and
% \begin{align*}
% \mathbf{h}=\sum_{R_i\in \mathcal{R}}\sum_{r_j\in R_i}{\alpha(\sum_{\ell=0}^{t}{(1-\alpha)^{\ell}\PM^{\ell}{\PM_b}^{(0)}_i})}[:,r_j]=\sum_{r_j\in R}{\PM^{(t)}_b[:,r_j]}.\quad\quad\quad\quad\quad\quad
% \end{align*}
According to \iequref{eq:fwd-v-r} and \iequref{eq:bwd-v-r}, for every pair 
$(v_i,r_j)\in V\times R$,
\begin{align*}
\max\{0,\PM_f[v_i,r_j]-\epsilon\}\le \PM^{(t)}_f[v_i,r_j]\le \PM_f[v_i,r_j],\\ \max\{0,\PM_b[v_i,r_j]-\epsilon\}\le \PM^{(t)}_b[v_i,r_j]\le \PM_b[v_i,r_j].
\end{align*}
By Lines 9-10 in \algoref{alg:rpapr}, for $i$-th block and every pair $(v_l,r_j)\in V\times R_i$,
$$\small {\widehat{\PM}_{f}}^{(t)}[v_l,r_j]=\frac{{\PM_f}^{(t)}_i[v_l,r_j]}{\sum_{v_h\in V}{{\PM_f}^{(t)}_i[v_h,r_j]}}=\frac{\PM^{(t)}_f[v_l,r_j]}{\sum_{v_h\in V}{\PM^{(t)}_f[v_h,r_j]}},$$
\begin{small}
\begin{align*}
 {\widehat{\PM}_{b}}^{(t)}[v_l,r_j]&=\frac{{\PM_b}^{(t)}_i[v_l,r_j]}{\sum_{R_i\in \mathcal{R}}\sum_{r_h\in R_i}{{\PM_b}^{(t)}_i[v_l,r_h]}}\\
 &=\frac{\PM^{(t)}_b[v_l,r_j]}{\sum_{r_h\in R}{\PM^{(t)}_b[v_l,r_h]}}.  
\end{align*}
\end{small}
By Lines 11-13 in \algoref{alg:rpapr}, the results in in \lemref{lem:papa} are now at hand. \done
\end{proof}
\end{lemma}

\begin{algorithm}[!t]
\begin{small}
	\caption{\psvdccd}
	\label{alg:psvdccd}
	\KwIn{$\FM^{\prime}, \BM^{\prime}, \mathcal{V},\mathcal{R},k,t$.}
	\KwOut{$\XMf, \YM, \XMb$.}
	$\XMf, \XMb, \YM, \SM_f, \SM_b \gets \mathtt{SMGInit}(\FM^{\prime}, \BM^{\prime}, \mathcal{V}, k, t)$\;
	\For{$\ell\gets 1$ to $t$}{
	   % $\widehat{\YM} \gets \YM^{\top}\YM$\;
	    \parallelfor{$V_h\in \mathcal{V}$}{
	       % $\WM_f[V_i] \gets \SM_f[V_i]\cdot\YM, \WM_b[V_i] \gets \SM_b[V_i]\cdot\YM$\;
	        \For{$v_i \in V_h$}{
		      %  \For{$l\gets 1$ to $\frac{k}{2}$}{
		      %    Compute $\mu_f(v_i,l),\mu_b(v_i,l)$ by \equref{eq:update-x-mu}\;
		      %      $\XMf[v_i,l]\gets \XMf[v_i,l]-\mu_f(v_i,l)$\;
		      %      $\XMb[v_i,l]\gets \XMb[v_i,l]-\mu_b(v_i,l)$\;
		      %      Update $\SM_f[v_i]$ by \equref{eq:update-x-sf}\;
		      %      Update $\SM_b[v_i]$ by \equref{eq:update-x-sb}\;
		      %  }
		      \nonl {Lines 5-10 are the same as Lines 4-9 in Algorithm \ref{alg:svdccd}}\;
	        }
	    }
	    \setcounter{AlgoLine}{10}
	   % $\WM\gets \XMf^{\top}\SM_f+\XMb^{\top}\SM_b,\ \widehat{\XM}\gets \XMf^{\top}\XMf+\XMb^{\top}\XMb$\;
	   \parallelfor{$R_h\in \mathcal{R}$}{
	    \For{$r_j\in R_h$}{
	       % \For{$l\gets 1$ to $\frac{k}{2}$}{
		      %    Compute $\mu_y(r_j,l)$ by \equref{eq:update-y-mu}\;
		      %      $\YM[r_j,l]\gets \YM[r_j,l]-\mu_y(r_j,l)$\;
		      %      Update $\SM_f[:,r_j],\SM_b[:,r_j]$ by \equref{eq:update-y-sf-sb}\;
	       % }
	       \nonl {Lines 13-16 are the same as Lines 11-14 in Algorithm \ref{alg:svdccd}}\;
	    }
	    }
	}
	\setcounter{AlgoLine}{16}
	\Return $\XMf,\YM, \XMb$\;
\end{small}
\end{algorithm}

% \vspace{-1mm}
\subsection{Parallel Joint Factorization of Affinity Matrices}\label{sec:split-merge}
%The single-thread \svdccd in Algorithm \ref{alg:svdccd} involves expensive matrix factorization at Line 1 when invoking \isvd (\algoref{alg:isvd}) to initialize embedding vectors $\XM_f$, $\XM_b$, and $\YM$, and also needs intensive cyclic coordinate descent computations for training embedding vectors (Lines 2-14). 

This section presents the parallel algorithm \psvdccd in Algorithm \ref{alg:psvdccd} to further improve the efficiency of the joint affinity matrix factorization process. At Line 1 of the algorithm, we design a parallel initialization algorithm \smsvd with a split-and-merge-based parallel SVD technique for embedding vector initialization. 

Algorithm \ref{alg:split-merge} displays the pseudo-code of \smsvd, which takes as input $\FM^{\prime}$,  $\BM^{\prime}$, $\mathcal{V}$, and $k$.
Based on $\mathcal{V}$, \smsvd splits matrix $\FM'$ into $n_b$ blocks and launches $n_b$ threads. Then, the $i$-th thread applies $\mathtt{RandSVD}$ to block $\FM^{\prime}[V_i]$ generated by the rows of $\FM'$ based on node set $V_i\in \mathcal{V}$ (Line 1-3).
After obtaining $\VM_1,\cdots,\VM_{n_b}$, \smsvd merges them by concatenating  $\VM_1,\cdots,\VM_{n_b}$ into $\VM = [\VM_1\ \cdots\ \VM_{n_b}]^{\top}\in \mathbb{R}^{\frac{kn_b}{2}\times d}$, and then applies $\mathtt{RandSVD}$ over it to obtain $\WM\in \mathbb{R}^{\frac{kn_b}{2}\times \frac{k}{2}}$ and $\YM\in\mathbb{R}^{d\times \frac{k}{2}}$ (Lines 4-6).
At Line 7, \smsvd creates $n_b$ threads, and uses the $i$-th thread to handle node subset $V_i$ for initializing embedding vectors $\XMf[V_i]$ and $\XMb[V_i]$ at Lines 8-9, as well as computing $\SM_f$ and $\SM_b$ at Lines 10-11. 
% Specifically, the forward embedding vectors of node subset $V_i$ are initialized as $\XMf[V_i]=\UM_{i}\cdot \WM[(i-1)\cdot \frac{k}{2}:i\cdot \frac{k}{2}]$ at Line 8; the backward embedding vectors of $V_i$ are initialized as $\XMb[V_i] = \BM^{\prime}[V_i]\cdot\YM$ at Line 9; $\SM_f[V_i]$ and $\SM_b[V_i]$ for node subset $V_i$ are computed as  $\SM_f[V_i]=\XM_f[V_i]\cdot\YM^{\top}-\FM^{\prime}[V_i]$ at Line 10 and $\SM_b[V_i]=\XMb[V_i]\cdot\YM^{\top}-\BM^{\prime}[V_i]$ at Line 11. 
Finally, \smsvd returns initialized embedding vectors $\YM$, $\XMf$, and $\XMb$ as well as intermediate results $\SM_f,\SM_b$ at Line 12. \lemref{lem:smsvd} indicates that the initial embedding vectors produced by \smsvd and \isvd are close.
%\header
%{\bf Split-Merge-based Initialization.} Before entering the second phase of \algoptp, in other words,
%\psvdccd, we overcome the crucial difficulty of parallelizing \svdccd by introducing our split-and-merge-based SVD technique. In \algoref{alg:svdccd}, it initializes attribute embedding vectors $\YM$, forward and backward embedding vectors $\XMf,\XMb$ by performing SVD over the $n\times d$ dense approximate forward affinity matrix $\FM^{\prime}$, which is time comsuming on large graphs and hard to parallelize. 

\begin{algorithm}[!t]
\begin{small}
	\caption{\smsvd}
	\label{alg:split-merge}
	\KwIn{$\FM^{\prime}, \BM^{\prime}, \mathcal{V}, k, t$.}
	\KwOut{$\XMf, \XMb, \YM, \SM_f, \SM_b$.}
	\parallelfor{$V_i\in \mathcal{V}$}{
        $\boldsymbol{\Phi},  \boldsymbol{\Sigma}, \VM_i \gets \mathtt{RandSVD}(\FM^{\prime}[V_i],\frac{k}{2},t)$\;
        $\UM_i\gets \boldsymbol{\Phi}\boldsymbol{\Sigma}$\;
	}
	$\VM\gets\left[\VM_1\ \cdots\ \VM_{n_b}\right]^{\top}$\;
	$\boldsymbol{\Phi},  \boldsymbol{\Sigma}, \YM \gets \mathtt{RandSVD}(\VM,\frac{k}{2},t)$\;
	$\WM \gets \boldsymbol{\Phi}\boldsymbol{\Sigma}$\;
	\parallelfor{$V_i\in \mathcal{V}$}{
		$\XMf[V_i] \gets \UM_{i}\cdot \WM[(i-1)\cdot \frac{k}{2}:i\cdot \frac{k}{2}]$\;
		$\XMb[V_i] \gets \BM^{\prime}[V_i]\cdot\YM$\;
		$\SM_f[V_i] \gets \XM_f[V_i]\cdot\YM^{\top}-\FM^{\prime}[V_i]$\;
	    $\SM_b[V_i] \gets \BM^{\prime}[V_i]-\XMb[V_i]\cdot\YM^{\top}$\;
	}
	\Return $\XMf, \XMb, \YM, \SM_f, \SM_b$\;
\end{small}
\end{algorithm}

After obtaining $\XMf, \XMb$, and $\YM$ by \smsvd, Lines 2-16 in Algorithm \ref{alg:psvdccd} train embedding vectors by cyclic coordinate descent in parallel based on subsets  $\mathcal{V}$ and $\mathcal{R}$, in $t$ iterations. In each iteration, \psvdccd first fixes $\YM$ and launches $n_b$ threads to update $\XMf$ and $\XMb$ in parallel by blocks according to $\mathcal{V}$, and then updates $\YM$ using the $n_b$ threads in parallel by blocks according to $\mathcal{R}$, with $\XMf$ and $\XMb$ fixed. Specifically, 
Lines 5-10 are the same as Lines 4-9 of Algorithm \ref{alg:svdccd}, and Lines 13-16 are the same as Lines 11-14 of Algorithm \ref{alg:svdccd}.
% Specifically, from Lines 4 to 10, each thread is assigned a node subset $V_h$, and for each node $v_i$ in $V_h$, the thread updates $\XMf[v_i],\XMb[v_i]$ based on Equations \eqref{eq:xf-update} and \eqref{eq:xb-update}, and then updates $\SM_f[v_i]$ and $\SM_b[v_i]$ by Equations \eqref{eq:update-x-sf} and \eqref{eq:update-x-sb}. After updating $\XMf$ and $\XMb$, each thread is assigned an attribute subset $R_h$ to update $\YM[r_j]$ according to \equref{eq:y-update} and $\SM[:,r_j]$ by \equref{eq:update-y-sf-sb} for each attribute $r_j\in R_h$ (Lines 12-16).
Finally, \algoref{alg:psvdccd} returns embedding results at Line 17.%$\XMf$ and $\XMb$ as forward and backward embedding vectors, and $\YM$ as the attribute embedding vectors.

\begin{lemma}\label{lem:smsvd}
Given the same $\FM^{\prime},\BM^{\prime},k$ and $t$ as inputs to \algoref{alg:isvd} and \algoref{alg:split-merge}, the outputs $\XMf,\YM,\SM_f,\SM_b$ returned by both algorithms satisfy that $\XMf\cdot\YM^{\top}=\FM^{\prime}$,$\YM^{\top}\YM=\mathbf{I}$ and $\SM_f=\SM_b\YM=\mathbf{0}$, when $t=\infty$.
\begin{proof}
Let the output of \algoref{alg:isvd} be $\XMf,\XMb,\YM,\SM_f$ and $\SM_b$, and the results returned by \algoref{alg:split-merge} be $\widehat{\XM}_f$, $\widehat{\XM}_b$, $\widehat{\YM}$ and $\widehat{\SM}_f,\widehat{\SM}_b$. According to \cite{musco2015randomized}, $t=\infty$ implies that $\mathsf{RandSVD}$ produces the same factorized results as that returned by exact SVD. Therefore, $\XMf\cdot \YM^{\top}=\FM^{\prime}, \SM_f=\mathbf{0}, \XMb=\BM^{\prime}\YM$ and $\YM$ is unitary, \ie $\YM^{\top}\YM=\mathbf{I}$. This leads to $\SM_b\YM = (\XMb\YM^{\top}-\BM^{\prime})\YM=\mathbf{0}$. 

On the other hand, consider \algoref{alg:split-merge}. Based on Lines 2-3, we have $\UM_i\VM^{\top}_i=\FM^{\prime}[V_i]$,
\begin{small}
\begin{align*}
\FM^{\prime}=\begin{bmatrix}
\FM^{\prime}[V_1]\\
\FM^{\prime}[V_1]\\
\vdots\\
\FM^{\prime}[V_{n_b}]
\end{bmatrix}
&=
\begin{bmatrix}
\UM_1 & \mathbf{0} & \cdots & \mathbf{0} \\
\mathbf{0} & \UM_2 & \cdots & \mathbf{0} \\
\vdots & \vdots & \ddots & \vdots \\
\mathbf{0} & \mathbf{0} & \cdots & \UM_{n_b}
\end{bmatrix}
\cdot
\begin{bmatrix}
\VM^{\top}_1\\
\VM^{\top}_2\\
\vdots\\
\VM^{\top}_{n_b}
\end{bmatrix}.
\end{align*}
\end{small}
By Lines 5-6, $\WM\widehat{\YM}^{\top}=\VM$ and $\widehat{\YM}$ is a unitary matrix, \ie $\widehat{\YM}^{\top}\widehat{\YM}=\mathbf{I}$. Then by Line 8 and Line 10, we derive that
\begin{small}
\begin{align*}
\FM^{\prime} &= 
\begin{bmatrix}
\UM_1 & \mathbf{0} & \cdots & \mathbf{0} \\
\mathbf{0} & \UM_2 & \cdots & \mathbf{0} \\
\vdots & \vdots & \ddots & \vdots \\
\mathbf{0} & \mathbf{0} & \cdots & \UM_{n_b}
\end{bmatrix}
\cdot \begin{bmatrix}
\WM_1\\
\WM_2\\
\vdots\\
\WM_{n_b}
\end{bmatrix}
\cdot \widehat{\YM}^{\top}=
\begin{bmatrix}
\widehat{\XM}_f[V_1]\\
\widehat{\XM}_f[V_2]\\
\vdots\\
\widehat{\XM}_f[V_{n_b}]
\end{bmatrix}
\cdot \widehat{\YM}^{\top} \\
&= \widehat{\XM}_f\cdot \widehat{\YM}^{\top},
\end{align*}
\end{small}
and thus $\widehat{\SM}_f=\mathbf{0}$. In addition, according to Line 9 and Line 11, we have $\widehat{\XM}_b=\BM^{\prime}\widehat{\YM}$ and $\widehat{\SM}_b\widehat{\YM}=(\widehat{\XM}_b\widehat{\YM}^{\top}-\BM^{\prime})\widehat{\YM}=\mathbf{0}$. The proof is complete. \done
\end{proof}
\end{lemma}

\subsection{Complexity Analysis}\label{sec:algo-opt-als} Observe that the non-parallel parts of Algorithms \ref{alg:rpapr} (Lines 7-10) and \ref{alg:split-merge} (Lines 4-6) take $O(nd)$ time, as each of them performs a constant number of operations on $O(nd)$ matrix entries. Meanwhile, for the parallel parts of Algorithms~\ref{alg:rpapr} and \ref{alg:psvdccd}, each thread runs in $\textstyle O\left(\frac{md}{n_b}\cdot\log\left(\frac{1}{\epsilon}\right)\right)$ and $O(\frac{ndkt}{n_b})$ time, respectively, since we divide the workload evenly to $n_b$ threads.
%We assume that the workload between CPU cores are balanced, and thus the computational time complexity of \algoref{alg:mainoptp} is determined by the running time of a single thread.
Specifically, each thread in \algoref{alg:rpapr} runs in $\textstyle O\left(\tfrac{md}{n_b}\cdot\log\left(\tfrac{1}{\epsilon}\right)\right)$ time. \algoref{alg:psvdccd} first takes $O(\frac{n}{n_b}dkt)$ time for each thread to factorize a $\frac{n}{n_b}\times d$ matrix block of $\FM^{\prime}$ (Lines 1-3 in  \algoref{alg:split-merge}). In addition, Lines 4-6 in \algoref{alg:split-merge} requires $O(n_bdk)$ time. In merge course (\ie Lines 7-11 in \algoref{alg:split-merge}), the matrix multiplications take $O(\frac{n}{n_b}k^2)$ time. In the $t$ iterations of CCD (\ie Lines 2-16 in \algoref{alg:psvdccd}), each thread spends $O(\frac{ndkt}{n_b})$ time to update. Thus, the computational time complexity per thread in \algoref{alg:mainoptp} is $\textstyle O\left(\tfrac{md+ndk}{n_b}\cdot\log\left(\tfrac{1}{\epsilon}\right)\right).$ \algoref{alg:rpapr} and \algoref{alg:psvdccd} require $O(m+nd)$ and $O(nd)$ space, respectively. Therefore, the space complexity of \algoptp is $O(m+nd)$.

%% file: HDPANE.tex
\section{\newalgt: Scaling to Large Attribute Set}\label{sec:cluster}

The aforementioned algorithms of \algo run in time linear to the number $d$ of attributes in $G$, as shown in Sections~\ref{sec:algo-als} and \ref{sec:algo-opt-als}. Hence, when $d$ is large (\eg millions of attributes, as is the case in the \textit{MAG-SC} dataset in our experiments in Section \ref{sec:exp}), \algo (both \algopt and \algoptp) may still incur rather high computational costs.
%it suffers from severe efficiency issues on graphs when $\RM$ is high-dimensional, \ie $d$ is large (\eg $d\ge 10^4$). For instance, on the {\em MAG-SC} dataset that has $2.78$M distinct attributes ($d=2.78$M), neither \algopt nor \algoptp can handle the dataset on a single server. 
To overcome this problem,
%the drawback of \algo on these graphs, 
this section introduces \newalg, which significantly 
improves over \algo in terms of efficiency in the presence of a large attribute set, while retaining the high result quality of \algo.
% which is based on the idea of clustering similar attributes into {\em super attribute}
Specifically, \newalg first compresses the attribute matrix $\RM\in \mathbb{R}^{n\times d}$ into a lower-dimensional one $\widetilde{\RM}\in \mathbb{R}^{n\times \kappa}$. This is achieved by clustering $d$ attributes into $\kappa$ {\em super attributes}, where $\kappa \ll d$. Then, \newalg proceeds with the \algo algorithm,
%(\ie either \algopt or \algoptp)
with $\widetilde{\RM}$ replacing $\RM$. After obtaining the node embeddings $\XMf,\XMb$, and the super attribute embeddings $\widetilde{\YM}\in \mathbb{R}^{\kappa\times \frac{k}{2}}$ from \algo, \newalg reconstructs attribute embeddings $\YM\in \mathbb{R}^{d\times\frac{k}{2}}$, based on the cluster information obtained in the first step.
% To ensure the result quality, \newalg includes an effective clustering method for attributes. 

In the following, we present an effective attribute clustering technique in Section \ref{sec:cluster-idea}, and describe the complete \newalg algorithm and its analysis in  Sections~\ref{sec:cluster-algo-als} and \ref{sec:cluster-als}, respectively. 

% \begin{algorithm}[!t]
% 	\begin{small}
% 		\caption{\newalg}
% 		\label{alg:newalg}
% 		\KwIn{Attributed network $G$, space budget $k$, random walk stopping probability $\alpha$, error threshold $\epsilon$, integer $\kappa$.}
% 		\KwOut{Forward and backward embedding vectors $\XMf$, $\XMb$ and pseudo attribute embedding vectors $\YM$.}
% 		Generate a Gaussian random matrix $\TM \in \mathbb{R}^{d\times \kappa}\sim \mathcal{N}(0,1)$\;
% 		$\widetilde{\RM} \gets \frac{1}{\sqrt{\kappa}}\RM\TM$\;
% 		Invoke \algopt with $G$, $\widetilde{\RM}$, $k$, $\alpha$, and $\epsilon$\;
% 		Let $\XMf,\YM,\XMb$ be the output of \algopt\;
% 		\Return $\XMf,\YM,\XMb$\;
% 	\end{small}
% \end{algorithm}

\subsection{Attribute Clustering}\label{sec:cluster-idea}
% We propose \sca (short for {\em Spectral Compression of Attributes}). We borrow the idea from spectral clustering \cite{von2007tutorial}. Unlike it, we find clusters without applying k-means and we define the similarity matrix as follows.
As explained above, a key step in \newalg is to cluster the input $d$ attributes into $\kappa$ super attributes, denoted as $\{{c_1},{c_2},\cdots,{c_\kappa}\}$, where each super attribute $c_l$ corresponds to an attribute cluster $R_{c_l}\subset R$ consisting of multiple similar attributes. Our clustering ensures (i) that $R_{c_1}\cup R_{c_1}\cdots\cup R_{c_\kappa}=R$, and (ii) that no two attribute clusters have overlapping members. To preserve the information in the original attribute space, we need an appropriate attribute similarity measure, as well as an effective objective function for the attribute clustering, described in the following.

\header
{\bf Attribute Similarity.} Let $\SM_R[r_i,r_j]$ denote the similarity between two attributes $r_i,r_j\in R$. Recall that $\RM$ is the attribute matrix of the input graph $G$, where each row vector denotes an attribute vector of a node. Here, we focus on attributes (\ie columns of $\RM$) instead of nodes (rows of $\RM$). In particular, we view each node as a \textit{feature}; then, each column of $\RM$, say, $\RM[:,r_i]$ corresponding to attribute $r_i$, can be regarded as a feature vector of $r_i$. Accordingly, we define the similarity $\SM_R[r_i,r_j]$ between attributes $r_i,r_j$ as the cosine similarity of their feature vectors:
\begin{small}
\begin{align}
\SM_R[r_i,r_j]&=cosine(\RM[:,r_i]^{\top},\RM[:,r_j])=\frac{\RM[:,r_i]^{\top}\cdot\RM[:,r_j]}{\|\RM[:,r_i]\|\cdot \|\RM[:,r_j]\|}\notag\\
&=\RM_s[:,r_i]^{\top}\cdot\RM_s[:,r_j]\label{eq:srrij},
\end{align}   
\end{small}
\begin{small}
\begin{equation}\label{eq:r-l2norm}
{\textrm \em where} \;\; \RM_s[v_i,r_j]=\frac{\RM[v_i,r_j]}{\sqrt{\sum_{v_l\in V}{\RM[v_l,r_j]^2}}}=\frac{\RM[v_i,r_j]}{\|\RM[:,r_j]\|}.   
\end{equation}  
\end{small}
\equref{eq:srrij} indicates that we can simply calculate $\SM_R[r_i,r_j]$ using the dot product of the normalized feature vectors of attributes $r_i,r_j$ defined in \equref{eq:r-l2norm}.

\header
{\bf Objective Function.}
Let ${c_l}$ be a super attribute and $R_{c_l}$ is its corresponding attribute cluster. Intuitively, a good attribute cluster $R_{c_l}$ should satisfy that attributes within $R_{c_l}$ are similar to each other and dissimilar to those outside $R_{c_l}$. 
Inspired by the $\mathtt{RatioCut}$ algorithm \cite{von2007tutorial,hagen1992new}, we 
%a simple and direct way to
partition the attribute set $R$ into $\kappa$ disjoint subsets $R_{c_1},R_{c_2},\cdots,R_{c_\kappa}$ by solving the mincut problem, formulated as the following optimization problem:
\begin{equation}\label{eq:obj-phi}
\min_{R_{c_1},R_{c_2},\cdots,R_{c_\kappa}}\sum_{l=1}^{\kappa}{\Phi(R_{c_\kappa})},
\end{equation}
where $\Phi(R_{c_l})$ represents the {\em attribute cut} of $R_{c_l}$, defined as follows.
%as defined in \equref{eq:phi-single}.
% \begin{small}
\begin{equation}\label{eq:phi-single}
	\Phi(R_{c_l})=\sum_{r_i\in R_{c_l},r_j\in R\setminus R_{c_l}}{\frac{\SM_R[r_i,r_j]}{|R_{c_l}|}}
\end{equation}
% \end{small}

In the above formulation, $\Phi(R_{c_l})$ measures the averaged similarity between an attribute in $R_{c_l}$ and another outside $R_{c_l}$;
%$R\setminus R_{c_l}$;
intuitively, a good attribute cluster $R_{c_l}$ should have a low $\Phi(R_{c_l})$. As such, our objective in \equref{eq:obj-phi} is to find $\kappa$ partitions $R_{c_1},R_{c_2},\cdots,R_{c_\kappa}$ of $R$ such that the averaged similarities of attributes crossing different attribute clusters are minimized.

\begin{lemma}\label{lem:phi}
Given $\kappa$ disjoint subsets $\{R_{c_1},R_{c_2},\cdots,R_{c_\kappa}\}$ of attribute set $R$ and a clustering indicator matrix $\CM \in \mathbb{1}^{d\times \kappa}$ such that for each entry with index $r_j,c_l$,
% \begin{small}
\begin{equation}\label{eq:cluster}
	\textstyle \CM[r_j,c_l]=\begin{cases}
		\textstyle 1 \quad&\textstyle \text{$r_j \in R_{c_l}$,}
		\\
		\textstyle 0 \quad&\textstyle \text{$r_j \in R \setminus R_{c_l}$},
	\end{cases}
\end{equation}
% \end{small}
the objective in \equref{eq:obj-phi} is equivalent to minimizing the following:
\begin{small}
\begin{equation}
\sum_{l=1}^{\kappa}{\Phi(R_{c_l})}=\Tr\left(\sqrt{\CM^{\top}\CM}^{-1}\CM^{\top}(\IM-\SM_R)\CM\sqrt{\CM^{\top}\CM}^{-1}\right),
\end{equation}   
\end{small}
where $\Tr$ denotes the trace of a matrix.
\begin{proof}
According to the definitions of $\Phi(R_{c_l})$ and $\CM$ in \equref{eq:phi-single} and \equref{eq:cluster}, respectively, we have
% \begin{small}
\begin{align}
\Phi(R_{c_l})&=\sum_{r_i\in R_{c_l},r_j\in R\setminus R_{c_l}}{\frac{\SM_R[r_i,r_j]}{|R_{c_l}|}}\notag\\
&=\sum_{r_i,r_j\in R}{\SM_R[r_i,r_j]\cdot \left(\frac{\CM[r_i,c_l]}{\sqrt{|R_{c_l}|}}-\frac{\CM[r_j,c_l]}{\sqrt{|R_{c_l}|}}\right)^2}\notag\\
&=\frac{\CM[:,c_l]^{\top}}{\sqrt{|R_{c_l}|}}\cdot(\IM-\SM_R)\cdot\frac{\CM[:,c_l]}{\sqrt{|R_{c_l}|}}\notag.
\end{align}
% \end{small}
Note that $\sqrt{\CM^{\top}\CM}^{-1}$ is a $\kappa\times \kappa$ diagonal matrix, whose $(c_l,c_l)$ entry is equal to $\frac{1}{\sqrt{|R_{c_l}|}}$.
Therefore,
\begin{align*}
\sum_{l=1}^{\kappa}{\Phi(R_{c_l})}&=\sum_{l=1}^{\kappa}{\frac{\CM[:,c_l]^{\top}}{\sqrt{|R_{c_l}|}}\cdot(\IM-\SM_R)\cdot\frac{\CM[:,c_l]}{\sqrt{|R_{c_l}|}}}\\
&=\Tr\left(\sqrt{\CM^{\top}\CM}^{-1}\CM^{\top}\cdot(\IM-\SM_R)\cdot\CM\sqrt{\CM^{\top}\CM}^{-1}\right),
\end{align*}
which finishes the proof. \done
\end{proof}
\end{lemma}

Using \lemref{lem:phi}, the optimization objective in \equref{eq:obj-phi} can be transformed into the following:
% \begin{small}
\begin{equation}\label{eq:obj-trace}
	\max_{\CM\in \mathbb{1}^{d\times \kappa}}{\Tr\left(\sqrt{\CM^{\top}\CM}^{-1}\CM^{\top}\cdot\SM_R\cdot\CM\sqrt{\CM^{\top}\CM}^{-1}\right)}.
\end{equation}
% \end{small}
Therefore, the problem of finding $\kappa$ super attributes becomes computing a clustering indicator matrix (CIM) $\CM$ defined in \equref{eq:cluster} such that \equref{eq:obj-trace} is optimized. 

\header
{\bf Computing CIM $\CM$.}
Since CIM $\CM$ satisfies \equref{eq:cluster}, we have $\sqrt{\CM^{\top}\CM}^{-1}\CM^{\top}\cdot\CM\sqrt{\CM^{\top}\CM}^{-1}=\IM$. Further, according to \cite{sameh1982trace} and the Rayleigh-Ritz theorem (Section 5.5.2 of \cite{lutkepohl1997handbook}), 
\begin{small}
\begin{equation}\label{eq:trace-bound}
\Tr\left(\sqrt{\CM^{\top}\CM}^{-1}\CM^{\top}\cdot\SM_R\cdot\CM\sqrt{\CM^{\top}\CM}^{-1}\right)\le \Tr(\UM^{\top}\SM_R\UM),
\end{equation}  
\end{small} 
where the columns in the matrix $\UM\in \mathbb{R}^{d\times \kappa}$ are %the $\kappa$ largest eigenvectors of $\SM_R$, \ie 
the $\kappa$ eigenvectors corresponding to the $\kappa$ largest eigenvalues of $\SM_R$.
% Note that $\CM$ is a unit vector.
\iequref{eq:trace-bound} suggests that if we can find a CIM $\CM$ that minimizes the difference between $\CM\sqrt{\CM^{\top}\CM}^{-1}$ and $\UM$ as follows:
% \begin{small}
\begin{align}\label{eq:obj-diff}
 \| \CM\sqrt{\CM^{\top}\CM}^{-1}-\UM\|_F,
\end{align}    
% \end{small}
then, our objective in \equref{eq:obj-trace} can be roughly optimized. Recall that CIM $\CM$ satisfies \equref{eq:cluster}. Hence, for each attribute $r_i$ that belongs to subset $R_{c_l}$, we have $\textstyle\left(\CM\sqrt{\CM^{\top}\CM}^{-1}\right)[r_i,c_l]=\frac{1}{\sqrt{|R_{c_l}|}}$ and $$\textstyle\left(\CM\sqrt{\CM^{\top}\CM}^{-1}\right)[r_i,c_j]=0\ \forall{c_j\in \{c_1,\cdots,c_\kappa\}\setminus c_l}.$$ This implies that to minimize \equref{eq:obj-diff}, for each attribute $r_i$ and its 
corresponding super attribute ${c_l}$, we can simply ensure that $\UM[r_i,c_l]$ is the maximum entry in row vector $\UM[r_i]$. In other words, we choose the super attribute $c_l$ such that $c_l=\argmax{c_j\in \{c_1,\cdots,c_\kappa\}}{\UM[r_i,c_l]}$ and assign $r_i$ to the attribute cluster $R_{c_l}$. 

Now, the optimization problem in \equref{eq:obj-trace} is transformed to finding the top-$\kappa$ eigenvectors $\UM$ of $\SM_R$. However, by \equref{eq:srrij}, the construction of $\SM_R$ incurs $O(nd^2)$ time and $O(d^2)$ space, which is prohibitively expensive when $d$ is large. Observe that in \equref{eq:srrij}, $\SM_R$ is the dot product of $\RM_s$ and its transpose.
Suppose the exact SVD of $\RM^{\top}_s\in \mathbb{R}^{d\times n}$ is $\RM^{\top}_s=\widehat{\UM}\widehat{\boldsymbol{\Sigma}}\widehat{\VM}^{\top}$, where $\widehat{\UM}\in \mathbb{R}^{d\times d}$ contains the full left singular vectors of of $\RM^{\top}_s$ and the diagonal entries in $\widehat{\boldsymbol{\Sigma}}$ are the singular values of $\RM^{\top}_s$. According to \cite{strang1993introduction}, the columns in $\widehat{\UM}$ are the eigenvectors of matrix $\RM^{\top}_s\RM_s=\SM_R$ and the diagonal entries in $\widehat{\boldsymbol{\Sigma}}^2$ are the eigenvalues of $\SM_R$.
Since all singular values are non-negative, the $i$-th largest eigenvalue of $\SM_R$ is equal to the square of the $i$-th largest singular values of $\RM^{\top}_s$. Therefore, the $\kappa$ largest eigenvectors of $\SM_R$ are equal to the top-$\kappa$ left singular vectors of $\RM^{\top}_s$, \ie the $\kappa$ left singular vectors corresponding to the $\kappa$ largest singular values of $\RM^{\top}_s$.
Thus, the problem is transformed to computing the top-$\kappa$ left singular vectors of $\RM_s$, which eliminates the need to construct and materialize $\SM_R$ explicitly.

\subsection{Complete \newalg Algorithm}\label{sec:cluster-algo-als}

\begin{algorithm}[t]
	\begin{small}
		\caption{\newalg}
		\label{alg:paneplus}
		\BlankLine
		\KwIn{$G$, $\RM$, $k$, $\alpha$, $\kappa$}
		\KwOut{$\XM_f,{\YM},\XM_b$}
        Compute $\RM_s$ by \equref{eq:r-l2norm}\;
		Let $\UM$ be the approximate top-$\kappa$ left singular vectors returned by $\mathtt{RandSVD}(\RM^{\top}_s,\kappa,\frac{\log(\epsilon)}{\log(1-\alpha)}-1)$\;
% 		Normalize $\UM$ by \equref{eq:norm-u}\;
		Initialize $\CM \gets \mathbf{0}\in \mathbb{R}^{d\times \kappa}$\;
		\For{$r_i\in R$}{
		    $c_l \gets \argmax{c_j\in \{c_1,\cdots,c_\kappa\}}{\UM[r_i,c_j]}$\;
		    $\CM[r_i,c_l]\gets 1$
		}
		$\widetilde{\RM}\gets \RM\CM$\;
		Invoke \algopt with $\widetilde{\RM}$\;
		Let $\XM_f,\widetilde{\YM},\XM_b$ be the output of \algopt\;
		$\YM \gets \CM\widetilde{\YM}$\;
        \Return $\XM_f,\YM,\XM_b$\;
	\end{small}
\end{algorithm}

The pseudo-code of \newalg is displayed in Algorithm \ref{alg:paneplus}. Compared to \algopt, \newalg takes as input an additional parameter $\kappa$, \ie the number of super attributes. Overall, \newalg consists two phases: (i) constructing CIM $\CM$ and the super attribute matrix $\widetilde{\RM}$ (Lines 1-7); and (ii) invoking \algopt to obtain node embeddings $\XMf,\XMb$ and attribute embeddings $\YM$ (Lines 8-11). In the first phase, \newalg starts by normalizing attribute matrix $\RM$ as $\RM_s$ according to \equref{eq:r-l2norm} (Line 1). After that, \newalg obtains an approximate top-$\kappa$ left singular vectors $\UM$ by utilizing the efficient randomized SVD algorithm \cite{musco2015randomized} at Line 2, with dimensionality $\kappa$ and the number of iterations $\frac{\log(\epsilon)}{\log(1-\alpha)}-1$. Next, Algorithm \ref{alg:paneplus} proceeds to constructing CIM $\CM$ (Lines 3-6). More specifically, for each attribute $r_i\in R$, we find the super attribute $c_l$ such that $\UM[r_i,c_l]$ is maximized among all super attributes, and then set $\CM[r_i,c_l]=1$ (Lines 4-6). Accordingly, we obtain $\widetilde{\RM}=\RM\CM$ (Line 7). That is, for each node $v_i$, its attribute value on super attribute $c_l$ is computed by aggregating the values of all attributes in the attribute cluster $R_{c_l}$ that super attribute $c_l$ corresponds to, \ie
$\widetilde{\RM}[v_i,c_l]=\sum_{r_j\in R_{c_l}}{\RM[v_i,r_j]}$.
\newalg then invokes \algopt with $\widetilde{\RM}$ as the attribute matrix and obtains the returned forward embedding matrix $\XMf$, backward embedding matrix $\XMb$, as well as the embedding matrix $\widetilde{\YM}\in \mathbb{R}^{\kappa\times \frac{k}{2}}$ for $\kappa$ super attributes $\{c_1,c_2,\cdots,c_\kappa\}$ (Lines 8-9). Finally, \newalg computes $\YM=\CM\widetilde{\YM}$ as the attribute embeddings and return $\XMf,\XMb$, and $\YM$ as the output embeddings.

% \begin{equation}\label{eq:norm-u}
% \UM[r_i,l]=\frac{\UM[r_i,l]}{\sum_{r_j\in R}{\max\{\UM[r_j,l],0\}}}
% \end{equation}

\subsection{Complexity Analysis}\label{sec:cluster-als}
First, Line 1 in Algorithm \ref{alg:paneplus} needs to process every non-zero entry in $\RM$, and, thus, takes $O(|E_R|)$ time. Given $\RM_s$ as input, $\mathtt{RandSVD}$ \cite{musco2015randomized} at Line 2 requires $O\left((|E_R|+d\kappa)\cdot \kappa \log{(\frac{1}{\epsilon})}\right)$ time. Recall that in Lines 4-6, we need to find the largest value among $\kappa$ entries for each $r_i\in R$. This time cost can be bounded by $O(d\kappa)$. The sparse matrix multiplications at Line 7 and Line 10 can be implemented with $O(|E_R|\cdot \kappa)$ and $O(d\kappa)$ time, respectively. According to Section \ref{sec:algo-als}, the invocation of \algopt (Lines 8-9) in Algorithm \ref{alg:paneplus} takes $O((m+nk)\cdot\kappa\log{(\frac{1}{\epsilon})})$ time and $O(n\kappa+m)$ space. Overall, the total time complexity of \newalg is $O((m+nk+|E_R|+d\kappa)\cdot\kappa\log{(\frac{1}{\epsilon})})$. Regarding space complexity, $\RM$ and $\widetilde{\RM}$ require $O(|E_R|)$ and $O(d\kappa)$ space, respectively. Hence, the space overhead incurred by \newalg is $O(n\kappa+d\kappa+m+|E_R|)$.

\header
{\bf When to use \newalg.} Given a space budget $b$ (\eg total available RAM), we employ \newalg instead of \algopt / \algoptp when $d$ is very large (\eg $d\ge 10^4$) or $2nd + \frac{k}{2}\cdot (2n+d) \ge b$, where term $2nd$ represents the space overhead incurred by constructing $\FM$ and $\BM$, and $\frac{k}{2}\cdot (2n+d)$ is the total space cost for embedding vectors $\XMf,\XMb$, and $\YM$ in \algo. The rationale is that in the first condition ({\it i.e.}, $d$ is large), factorizing large $n\times d$ affinity matrices $\FM$ and $\BM$ will severely impede the efficacy of \algo. As for the second condition, $2nd + \frac{k}{2}\cdot (2n+d) \ge b$ means that other variants of \algo would run out of space due to the large size of the input network and the intermediate structures.

%% file: attrInferlinkPred.tex
\section{Usage of Embeddings}\label{sec:ane-app}
Given an input attributed network $G$, our ANE methods return two embedding vectors $\XMf[v_i]$ and $\XMb[v_i]$ for each node $v_i$, and an embedding vector $\YM[r_j]$ for each attribute $r_j$.
In this section, we explain how to use the embeddings to achieve high performance in downstream tasks, including node classification, attribute inference, and link prediction.

\header
\textbf{Node Classification.}
We  apply L2-normalization over the forward embedding vector $\XMf[v_i]$ and backward embedding vector $\XMb[v_i]$ for each node $v_i\in V$, and concatenate them as the feature representation of $v_i$, which is then used as input to train or evaluate node classifiers, such as a linear support-vector machine (SVM) classifier  \cite{cortes1995support} in Section \ref{sec:node-class}.

% According to the previous sections, our ANE solutions \algo and \newalg return two embedding vectors $\XMf[v_i]$ and $\XMb[v_i]$ for each node $v_i$, and an embedding vector $\YM[r_j]$ for each attribute $r_j$. Clearly, these can be simply fed to downstream machine learning operators as input features. In our experiments, node classification is implemented this way, as described in Section \ref{sec:node-class}. Besides this usage, the embedding vectors can be also be directly used to make predictions without additional training, for attribute inference and link prediction tasks, as follows.

\header
{\bf Attribute Inference.}
Attribute inference is a supervised task that aims to predict the existence of an attribute $r_j$ associated to a given node $v_i$.
We leverage the following three heuristics for attribute inference using the obtained embeddings. First, if the forward affinity $\FM[v_i,r_j]$ from node $v_i$ to attribute $r_j$ defined in  \equref{eq:fwd-prob} and the backward affinity $\BM[v_i,r_j]$ from attribute $r_j$ to $v_i$ defined in  \equref{eq:bwd-prob}
 are both high, then, intuitively, node $v_i$ is likely to be associated with attribute $r_j$.
Second, if an attribute $r_j$ appears frequently in different nodes of a training set, it tends to exist in the nodes of the test set. In other words, if the number of non-zero entries in column $\RM[:, r_j]$ is large, $r_j$ should be popular in many nodes, where $\RM$ is the attribute matrix.
Lastly, if a node $v_i$ has many attributes (\ie there are many non-zero entries in row $\RM[v_i]$), $v_i$ is more likely to be associated with attribute $r_j$.

Based on the above three heuristics, we derive the following indicator value for predicting whether node $v_i$ is associated with attribute $r_j$:
\begin{align}
&\FM[v_i,r_j] + \BM[v_i,r_j] + \log{(\gamma_{v_i}+1)}  +  \log{(\gamma_{r_j}+1)} \label{eq:attInferOrigin} \\
=&\log\left(\left(\tfrac{n\cdot p_f(v_i,r_j)}{\sum_{v_h\in V}{p_f(v_h,r_j)}}+1\right)\cdot \sqrt{(\gamma_{v_i}+1)\cdot (\gamma_{r_j}+1)}\right) \nonumber\\
&+\log\left(\left(\tfrac{d\cdot p_b(v_i,r_j)}{\sum_{r_h\in R}{p_b(v_i,r_h)}}+1\right)\cdot \sqrt{(\gamma_{v_i}+1)\cdot (\gamma_{r_j}+1)}\right).
\label{eq:attInferOrigin2}
\end{align}

Specifically, if \equref{eq:attInferOrigin} has a large value, then, node $v_i$ is likely to have attribute $r_j$.
Particularly, in \equref{eq:attInferOrigin}, in addition to the forward and backward affinities $\FM[v_i,r_j]$ and $\BM[v_i,r_j]$, $\gamma_{v_i}$ and $\gamma_{r_j}$ are the numbers of non-zero entries in $\RM[v_i]$ and $\RM[:,r_j]$ for node $v_i$ and attribute $r_j$ respectively, which serve as scaling factors to give more weights if $v_i$ and $r_j$ are popular in attribute matrix $\RM$.
Since forward and backward affinities $\FM[v_i,r_j]$ and $\BM[v_i,r_j]$ are based on shifted PMI (SPMI) explained in Section \ref{sec:objective}, we also apply logarithm operation over $\gamma_{v_i}$ and $\gamma_{r_j}$ shifted by 1, to make sure the factors are positive.
Then \equref{eq:attInferOrigin} is rewritten into \equref{eq:attInferOrigin2}, based on \equref{eq:fwd-prob} and \equref{eq:bwd-prob}.
\equref{eq:attInferOrigin2} provides a detailed interpretation about how the two scaling factors $\gamma_{v_i}$ and $\gamma_{r_j}$  work together as 
a factor $\sqrt{(\gamma_{v_i}+1)\cdot (\gamma_{r_j}+1)}$ to affect the forward and backward affinity computations.

% \begin{align}
% &\FM[v_i,r_j] + \BM[v_i,r_j] + \log{(\gamma_{v_i}+1)}  +  \log{(\gamma_{r_j}+1)} \label{eq:attInferOrigin} \\
% =&\log\left(\left(\tfrac{n\cdot p_f(v_i,r_j)}{\sum_{v_h\in V}{p_f(v_h,r_j)}}+1\right)\cdot \sqrt{(\gamma_{v_i}+1)\cdot (\gamma_{r_j}+1)}\right) \nonumber\\
% &+\log\left(\left(\tfrac{d\cdot p_b(v_i,r_j)}{\sum_{r_h\in R}{p_b(v_i,r_h)}}+1\right)\cdot \sqrt{(\gamma_{v_i}+1)\cdot (\gamma_{r_j}+1)}\right).
% \label{eq:attInferOrigin2}
% \end{align}
Recall that the embedding vectors are expected to satisfy that $\XMf[v_i]\cdot \YM[r_j]^{\top}$ preserves
$\FM[v_i,r_j]$, and  
$\XMb[v_i]\cdot \YM[r_j]^{\top}$ preserves
$\BM[v_i,r_j]$, according to objective function in \equref{eq:obj1}. Therefore, after obtaining the embeddings, we use $p(v_i,r_j)$ in \equref{eq:attr-infer} to infer if attribute $r_j$ is associated to node $v_i$.
% In attribute inference, we are given a node $v_i$, and we aim to predict its attribute value $r_j$ which is not included in the input data. Inferring the attributes of nodes can be done through the affinity between nodes and attributes, including both forward affinity and backward affinity. Based on objective function in \equref{eq:obj1}, given node $v_i$ and attribute $r_j$, $\XMf[v_i]\cdot \YM[r_j]^{\top}$ is expected to preserve forward affinity value $\FM[v_i,r_j]$, and $\XMb[v_i]\cdot \YM[r_j]^{\top}$ is expected to preserve backward affinity value $\BM[v_i,r_j]$. Thus, we  measure the attribute inference score between $v_i$ and $r_j$, denoted as 
% $p(v_i,r_j)$, by utilizing their embedding vectors as follows.
\begin{align}
p(v_i,r_j) =&  \XMf[v_i]\cdot\YM[r_j]^{\top}+\XMb[v_i]\cdot\YM[r_j]^{\top}\nonumber\\
& {} + \log{(\gamma_{v_i}+1)} + \log{(\gamma_{r_j}+1)}, \label{eq:attr-infer}
% &\approx \FM[v_i,r_j]+\BM[v_i,r_j] \nonumber\\
% & \quad + \log{(\gamma_{v_i}+1)} + \log{(\gamma_{r_j}+1)} \nonumber,
\end{align}
where $\gamma_{v_i}$ and $\gamma_{r_j}$ are the numbers of non-zero entries in $\RM[v_i]$ and $\RM[:,r_j]$, respectively.

% According to \equref{eq:fwd-prob} and \equref{eq:bwd-prob}, \equref{eq:attr-infer} is to approximate the following scaled affinity
% \begin{small}
% \begin{align*}
% &\FM[v_i,r_j] + \log{(\gamma_{v_i}+1)} + \BM[v_i,r_j] + \log{(\gamma_{r_j}+1)} \\
% &=\log\left(\left(\tfrac{n\cdot p_f(v_i,r_j)}{\sum_{v_h\in V}{p_f(v_h,r_j)}}+1\right)\cdot \sqrt{(\gamma_{v_i}+1)\cdot (\gamma_{r_j}+1)}\right) \\
% &\quad +\log\left(\left(\tfrac{d\cdot p_b(v_i,r_j)}{\sum_{r_h\in R}{p_b(v_i,r_h)}}+1\right)\cdot \sqrt{(\gamma_{v_i}+1)\cdot (\gamma_{r_j}+1)}\right).
% \end{align*}
% \end{small}
% The above score can then be normalized to the range of $r_j$, and output as the prediction.
% Note that in the above score, we scale forward affinity and backward affinity with node frequency $\gamma_{v_i}$ of $v_i$ and attribute frequency $\gamma_{r_j}$ of $r_j$, so as to incorporate the importance of node $v_i$ and attribute $r_j$ in the input node-attribute associations $E_R$. The intuition is that given a node (resp. an attribute) with high frequency in $E_R$, it is more likely to associated with other attributes (resp. nodes) compared to those with low frequencies. Their predicted scores should be weighted by their frequencies to reflect their importance over those with low frequencies.

\header
{\bf Link Prediction.}
Given two nodes $v_i$ and $v_j$ that are not directly connected, link prediction aims to predict if there will be an edge from $v_i$ to $v_j$.
Intuitively, if the affinity between $v_i$ to $v_j$ is strong, the probability of forming an edge from $v_i$ to $v_j$ is high.
We propose to evaluate the affinity  by combining forward affinities from $v_i$ and backward affinities to $v_j$ over the input attributed network, with the consideration of both graph topology and attributes.
Specifically, given nodes $v_i$ and $v_j$ and attribute $r_l$, $\FM[v_i,r_l]$ measures the affinity from $v_i$ to $r_l$, $\BM[v_j,r_l]$ evaluates the affinity from $r_l$ to $v_j$, and consequently $\FM[v_i,r_l]\times\BM[v_j,r_l]$ represents the affinity from node $v_i$ to node $v_j$ via attribute $r_l$ over the attributed network.
However, note that $\FM[v_i,r_l]\times\BM[v_j,r_l]$ does not consider the degrees of $v_i$ and $v_j$ for link prediction, while node degrees have been shown to be crucial in improving the performance of link prediction \cite{yang13homogeneous}.
Intuitively, if node $v_i$ (resp. $v_j$) has large out-edges (resp. in-edges), $v_i$ (resp. $v_j$) tends to connect to (resp. be connected to by) other nodes.
Therefore, we further use the out-degree $d_{out}(v_i)$ of $v_i$ and in-degree $d_{in}(v_j)$ of $v_j$ as weights for the node affinity values. \revise{In particular, the following equation is used to evaluate the weighted affinity between $v_i$ and $v_j$, by summing up all possible $\FM[v_i,r_l]\times \BM[v_j,r_l]$ for any $r_l\in R$,  with weights  $\sqrt{d_{out}(v_i)+1}$ and $\sqrt{d_{out}(v_j)+1}$:}
% \footnote{\revise{We increase the out-degree by 1 and apply a square root on the weights to avoid zero and excessively large weights, respectively, rendering them empirically superior to the weights in other forms.}}
% To perform link prediction, we need to evaluate the affinity between nodes, considering both graph topology and attributes. 
% The affinity between nodes can be obtained by concatenating forward affinity and backward affinity. 
% Given node $v_i$ and attribute $r_l$, recall that  $\FM[v_i,r_l]$ measures the affinity from $v_i$ to $r_l$ over the attributed network; similarly given node $v_j$ and attribute $r_l$, $\BM[v_j,r_l]$ measures the affinity from $r_l$ to $v_j$ over the  network. 
% Intuitively, $\FM[v_i,r_l]\cdot\BM[v_j,r_l]$ represents the affinity from node $v_i$ to node $v_j$ based on attribute $r_l$. 
%The affinity between nodes $v_i$ and $v_j$, denoted as $p^{\prime}(v_i,v_j)$, can be evaluated by summing up the affinity between the two nodes over all attributes in $R$, \ie
% \begin{small}
\begin{equation*}
\sum_{r_l\in R}{\sqrt{d_{out}(v_i)+1}\cdot\FM[v_i,r_l]\times \BM[v_j,r_l]\cdot\sqrt{d_{in}(v_j)+1}}.
\end{equation*}
% \end{small}

% However, $p^{\prime}(v_i,v_j)$ does not reflect the strength of connections (importance) of the nodes $v_i,v_j$ in the whole graph, which is shown critical in link prediction \cite{yang13homogeneous}. Intuitively, a node $v_i$ with massive out-going (resp. in-coming) edges is likely to connect to (resp. be connected by) other nodes. To incorporate the strength of connections from node $v_i$ and the strength of connections to node $v_j$ in $E_V$, we compute the weighted affinity $p(v_i,v_j)$ between nodes $v_i$ and $v_j$ as $\sqrt{d_{out}(v_i)+1}\cdot p^{\prime}(v_i,v_j) \cdot\sqrt{d_{in}(v_j)+1}$, where $d_{out}(v_i)$ and $d_{in}(v_j)$ represent the out-degree and in-degree of nodes $v_i$ and $v_j$, respectively.

As explained above, embedding vectors  $\XMf[v_i]\cdot \YM[r_l]^{\top}$ preserves $\FM[v_i,r_l]$, and $\XMb[v_j]\cdot \YM[r_l]^{\top}$ preserves $\BM[v_j,r_l]$.
Therefore, the node affinity $p(v_i,v_j)$ between $v_i$ and $v_j$ is estimated by \equref{eq:link-pred}.
For undirected graphs, we use $p(v_i,v_j)+p(v_j,v_i)$ as the score for predicting the edge between $v_i$ and $v_j$.
% \begin{small}
\begin{align}
p(v_i,v_j)&= \sum_{r_l\in R}\bigg(\left(\sqrt{d_{out}(v_i)+1}\cdot\XMf[v_i]\cdot\YM[r_l]^{\top}\right)\nonumber \\ 
&\quad \cdot\left(\sqrt{d_{in}(v_j)+1}\cdot\XMb[v_j]\cdot\YM[r_l]^{\top}\right)\bigg). \label{eq:link-pred} 
% \\
% & \approx \sum_{r_l\in R}{\sqrt{d_{out}(v_i)+1}\cdot\FM[v_i,r_l]\times \BM[v_j,r_l]\cdot\sqrt{d_{in}(v_j)+1}} \nonumber
% & = (\DM^{\frac{1}{2}}\FM)[v_i] \cdot (\BM^{\top}\DM_{in}^{\frac{1}{2}})[v_j]\nonumber.
\end{align}
% \end{small}
In the next section, we adopt the above methods to utilize the embedding results for node classification, link prediction, and attribute inference over real-world attributed networks.

%% file: exp.tex
% \vspace{-1mm}
\begin{table*}[t]
\centering
\renewcommand{\arraystretch}{1.4}
% \begin{small}
% \captionsetup{justification=centering}
\caption{Datasets. {\small (K=$10^3$, M=$10^6$)}}\label{tbl:exp-data}
% \resizebox{\columnwidth}{!}{
\begin{tabular}{|l|r|r|r|r|c|c|c|}
	\hline
	{\bf Name} & \multicolumn{1}{c|}{$\lvert V \rvert=n$} & \multicolumn{1}{c|}{$\lvert E_V \rvert=m$} & \multicolumn{1}{c|}{$\lvert R \rvert=d$}& \multicolumn{1}{c|}{$\lvert E_R \rvert$} & \multicolumn{1}{c|}{\bf $\lvert L \rvert$} & \multicolumn{1}{c|}{Type} & \multicolumn{1}{c|}{\bf  References}\\
	\hline
% 	{\em Wiki} & 2,405 & 17,981 & 4,973 &  19 \\
% 	\hline
% 	{\bf\em Cora} & 2.7K & 5.4K & 1.4K & 49.2K & 7  &  \cite{pan2018adversarially,zhou2018prre,liu2018content,yang2015network,meng2019co,yang2018binarized} \\
% 	\hline
	{\bf\em Citeseer} & 3.3K & 4.7K & 3.7K & 105.2K & 6 & directed & \cite{pan2018adversarially,zhou2018prre,liu2018content,yang2015network,meng2019co,yang2018binarized} \\
	\hline
	{\bf\em Pubmed} & 19.7K & 44.3K & 0.5K & 988K & 3  & directed & \cite{pan2018adversarially,zhou2018prre,meng2019co,zhang2018anrl} \\
	\hline
	{\bf\em Facebook} & 4K & 88.2K & 1.3K & 33.3K & 193 & undirected & \cite{leskovec2012learning,yang2013community,meng2019co,zhang2018anrl} \\
	\hline
	{\bf\em Flickr} & 7.6K & 479.5K & 12.1K & 182.5K & 9  & undirected & \cite{meng2019co} \\
	\hline
	{\bf\em Google+} & 107.6K & 13.7M & 15.9K & 300.6M &  468 & directed & \cite{leskovec2012learning,yang2013community} \\
	\hline
	{\bf\em TWeibo} & 2.3M & 50.7M & 1.7K & 16.8M & 8 & directed & - \\
	\hline
	{\bf\em MAG} & 59.3M & 978.2M & 2K & 434.4M & 100 & directed & - \\
	\hline
	{\bf\em MAG-SC} & 10.5M & 265.2M & 2.78M & 1.1B & 8 & directed & \cite{bojchevski2020scaling,yang2021effective} \\
	\hline
\end{tabular}
% }
% \end{small}
% \vspace{2mm}
\end{table*}

\section{Experiments}\label{sec:exp}
This section experimentally evaluates our proposed \algo and \newalg against 10 competitors on three tasks: node classification, link prediction, and attribute inference, over 8 real datasets. All experiments are conducted on a Linux machine powered by an %16 22-core 
Intel Xeon(R) E7-8880 v4@2.20GHz CPUs and 1TB RAM. The codes of all algorithms are collected from their respective authors, and all are implemented in Python, except \nrp, \tadw and \lqanr. For fair comparison of efficiency, we re-implement \tadw and \lqanr in Python.

% \vspace{-1mm}
\subsection{Experiments Setup}\label{sec:exp-set}

\header
{\bf Datasets.} \tblref{tbl:exp-data} lists the statistics of the datasets used in our experiments. {All graphs are directed except {\em Facebook} and {\em Flickr}}. $|V|$ and $|E_V|$ denote the number of nodes and edges in the graph, whereas $|R|$ and $|E_R|$ represent the number of attributes and the number of node-attribute associations (\ie the number of nonzero entries in attribute matrix $\RM$). In addition, $L$ is the set of \textit{node labels}, which are used in the node classification task. 
% {\em Cora}\footnote{\label{fn:linqs}{\url{http://linqs.soe.ucsc.edu/data}}},
% {\em Citeseer}\footnoteref{fn:linqs}, {\em Pubmed}\footnoteref{fn:linqs} 
{\em Citeseer}\footnote{\label{fn:linqs}{\url{http://linqs.soe.ucsc.edu/data}}} and {\em Flickr}\footnote{\url{https://github.com/mengzaiqiao/CAN} } are benchmark datasets used in prior work \cite{yang2015network,hamilton2017inductive,meng2019co,pan2018adversarially,zhou2018prre,liu2018content}. {\em Facebook}\footnote{\label{fn:snap}{\url{http://snap.stanford.edu/data}}} and {\em Google+}\footnoteref{fn:snap} are social networks used in \cite{leskovec2012learning}. For {\em Facebook} and {\em Google+}, we treat each ego-network as a label and extract attributes from their user profiles, which is consistent with the experiments in prior work \cite{meng2019co,yang2013community}. 

To evaluate the scalability of the proposed solutions, we also include three large datasets, namely \textit{TWeibo}\footnote{\url{https://www.kaggle.com/c/kddcup2012-track1}}, \textit{MAG}\footnote{\url{http://ma-graph.org/rdf-dumps/}} and {\em MAG-SC}\footnote{\url{https://figshare.com/articles/dataset/mag_scholar/12696653}}. These datasets have not been used in existing ANE work due to their massive size. Specifically, {\em TWeibo} \cite{kddcup2012tweibo} is a social network, in which each node represents a user and each directed edge represents a following relationship. We extract the $1657$ most popular tags and keywords from its user profile data as the node attributes. The labels are generated and categorized into eight types according to the ages of users. {\em MAG} dataset is extracted from the well-known {\em Microsoft Academic Knowledge Graph} \cite{sinha2015overview}, where each node represents a paper and each directed edge represents a citation. We extract frequently used distinct words from the abstract of all papers as the attribute set and regard the fields of study of each paper as its labels.
% which are able to cover about $80\%$ of most texts according to \cite{hyland2007there}. 
%We will make {\em TWeibo} and {\em MAG} datasets publicly available upon acceptance. 
{\em MAG-SC} is also a citation graph extracted from Microsoft Academic Knowledge Graph by \cite{bojchevski2020scaling}. In {\em MAG-SC}, the attributes of a node are the bag-of-words representation of the respective paper abstract. In total, there are  $2.78$ million distinct attributes in {\em MAG-SC}.

Note that \textit{Flickr}, \textit{Google+}, and \textit{MAG-SC} involve large values of $d$, \ie number of attributes, while the other datasets have relatively small $d$. Hence, \newalg is evaluated on these three datasets to validate its ability to handle large attribute sets effectively.

\header
{\bf Baselines and Parameter Settings.} We compare our  methods \algopt (single-thread \algo), \algoptp (parallel \algo), and \newalg against 10 state-of-the-art competitors: eight recent ANE methods including \bane \cite{yang2018binarized}, \can \cite{meng2019co}, \stne \cite{liu2018content}, \prre \cite{zhou2018prre}, \tadw \cite{yang2015network}, \arga \cite{pan2018adversarially}, \dgi \cite{velickovic2018deep} and \lqanr \cite{ijcai2019-low}, one state-of-the-art homogeneous network embedding method \nrp \cite{yang13homogeneous}, and one latest attributed heterogeneous network embedding algorithm \gatne \cite{cen2019representation}. All methods except \algoptp run on a single CPU core. Note that although \gatne itself is a parallel algorithm, its parallel version requires the proprietary AliGraph platform.

The parameters of all competitors are set as suggested in their respective papers. 
%Note that \nrp ignores node attributes and \gatne treats all nodes and edges as the same type. 
For \algopt, \algoptp, and \newalg, by default we set error threshold $\epsilon=0.015$ and random walk stopping probability $\alpha=0.5$, and we use $n_b=10$ threads for \algoptp and $\kappa=1024$ for \newalg. Unless otherwise specified, we set space budget $k=128$.

The efficiency evaluation results of all methods are presented in \secref{sec:exp-effi}. 
We report the evaluation results of all methods for link prediction, node classification, and attribute inference, in Sections \ref{sec:link-pred}, \ref{sec:node-class}, and \ref{sec:attr-infer} respectively. 
% Section \ref{sec:exp-param} analyses the parameter sensitivities of proposed methods. 
A method is excluded in our study if it cannot finish training within one week.

\begin{figure*}[!t]
\centering
\begin{small}
\begin{tabular}{cc}
\multicolumn{2}{c}{
\hspace{-2mm}\includegraphics[height=6mm]{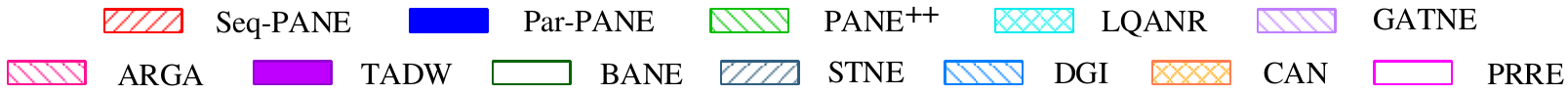}}\vspace{-2mm} \\
% \hspace{-5mm}\subfloat[{graphs with small $d$}.]{\includegraphics[width=0.32\linewidth]{./figure/time1-eps-converted-to.pdf}\label{fig:time-small}} &
% \hspace{-5mm}\subfloat[{graphs with small $d$}.]{\includegraphics[width=0.32\linewidth]{./figure/time2-eps-converted-to.pdf}\label{fig:time-large}} &
\hspace{0mm}\subfloat[\revise{graphs with small $d$}.]{\includegraphics[width=0.62\linewidth]{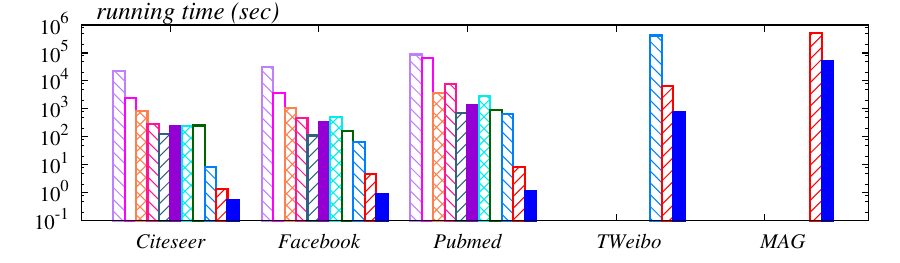}\label{fig:time-small}} &
\hspace{-4mm}\subfloat[\revise{graphs with large $d$}.]{\includegraphics[width=0.36\linewidth]{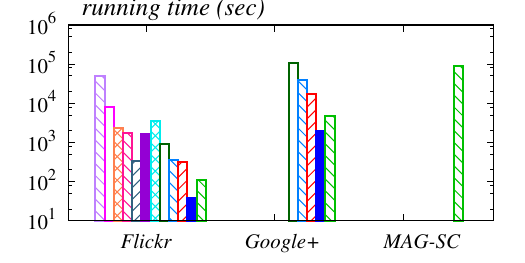}\label{fig:time-large}}
\end{tabular}
\end{small}
\vspace{-1mm}
\caption{\revise{Running time (best viewed in color).}} \label{fig:time-all}
\vspace{-2mm}
\end{figure*}

\subsection{Efficiency of ANE methods}\label{sec:exp-effi}
\figref{fig:time-small} and \figref{fig:time-large} report the running time required by each ANE method on datasets with small or large $d$, respectively. The $y$-axis is the running time (seconds) in $\log$-scale. The reported  time does not include the time for loading datasets and outputting embedding vectors. We omit any methods with  time exceeding one week. 
% For completeness, We report the running time of \nrp, but note that \nrp is designed only for homogeneous networks and not capable of handling attributed networks. %All methods are performed with a single thread except that \algoptp runs in parallel. 

%will not compare \algopt and \algoptp with it since \nrp is not applicable to attributed networks. 
%The first observation we can make is that both 
As shown in \figref{fig:time-small}, both  \algopt and \algoptp are significantly faster than all ANE competitors, often by orders of magnitude. For instance, on {\em Pubmed} in \figref{fig:time-small}, \algoptp takes 1.1 seconds and \algopt  requires 8.2 seconds, while the fastest ANE competitor \tadw consumes 405.3 seconds, demonstrating that \algopt  (resp. \algoptp) is $49\times$ (resp. $368\times$) faster. 
On large attributed networks including {\em TWeibo} and {\em MAG}, most existing ANE solutions cannot finish within a week, while \algopt  and \algoptp are able to handle such large-scale networks efficiently.
\algoptp is up to $9$ times faster than \algopt  over all datasets. For instance, on {\em MAG} dataset that has $59.3$ million nodes, when using $10$ threads, \algoptp requires $11.9$ hours while \algopt  takes about five days, which demonstrates the power of our parallel techniques in \secref{sec:parallel}. 
Note that \newalg is not reported in \figref{fig:time-small} since these datasets have small $d$ values whereas \newalg is designed for datasets with a large $d$.

As shown in \figref{fig:time-large}, on datasets with large $d$, \newalg is significantly faster than \algopt (both are single-threaded), validating the efficiency of techniques proposed in Section \ref{sec:cluster} for handling a large number of attributes.
\algopt is slower than \newalg on \textit{Flickr} and \textit{Google+} since it needs to construct, materialize, and decompose two high dimensional dense affinity matrices in $n\times d$ dimensions, while \newalg works on matrices in $n\times \kappa$ dimensions to obtain embeddings, where $\kappa \ll d$. All competitors are slower than our methods.
Moreover,  as reported in  \figref{fig:time-large}, \newalg is the only method that can efficiently handle \textit{MAG-SC} with 2.78M attributes, while all other methods including \algopt and \algoptp run out of memory or time.
Further, as we show shortly in Sections \ref{sec:link-pred}, \ref{sec:node-class}, and \ref{sec:attr-infer}, compared to \algopt, \newalg achieves comparable and sometimes even superior accuracy for link prediction, node classification, and attribute inference, which validates the efficiency and effectiveness of \newalg on attributed networks with numerous attributes.

% As for the attributed networks with large $d$, we can observe from \figref{fig:time-large} that \algopt  is only slightly faster than \dgi on {\em Flickr} and {\em Google+}. The reason is that on such datasets, \algopt needs to construct, materialize, and decompose two high dimensional dense affinity matrices $\FM$ and $\BM$, severely impairing the efficiency of \algopt. 
% In comparison, \newalg utilizes attribute clustering to mitigate the curse of dimensionality, and achieves over $3\times$ speedup over \algopt  using a single core. On the {\em MAG-SC} dataset with $d$ being up to $2.78$M, none of existing ANE solutions can process it using a single server and even \algoptp fails due to the immense time and memory overheads. In contrast, \newalg can finish training embeddings on such graph using only about one day. Recall that in Sections \ref{sec:attr-infer}, \ref{sec:link-pred}, and \ref{sec:node-class}, \newalg achieves comparable and even considerably superior result quality compared to \algopt. This verifies the effectiveness of attribute clustering introduced in \newalg for improved efficiency as well as attribute information preservation when processing attributed networks with large $d$.

% When $\kappa$ reaches $4096$ (still much less than $d$), \newalg and \algopt have similar performance. The reason is that \newalg decomposes $\RM$ before invoking \algopt and the super attribute matrix $\widetilde{\RM}$ is usually dense, both of which require non-negligible costs, especially when $d$ is not small.

\begin{figure*}[!t]
    \centering
    \captionsetup[subfloat]{captionskip=-0.5mm}
    \begin{small}
    \begin{tabular}{cccc}
    \multicolumn{4}{c}{\hspace{-4mm} \includegraphics[height=6mm]{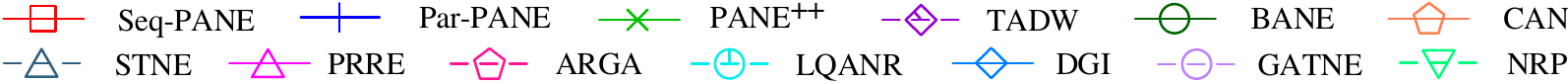}}\vspace{-3mm}  \\
    % \hspace{-4mm}\subfloat[{\em Wiki}]{\includegraphics[width=0.26\linewidth]{./figure/class/wiki-eps-converted-to.pdf}\label{fig:acc-class-wiki}} &
    % \hspace{-4mm}\subfloat[{\em Cora}]{\includegraphics[width=0.32\linewidth]{./figure/link/cora-eps-converted-to.pdf}\label{fig:link-cora}} &
    \hspace{-6mm}\subfloat[{\em Citeseer}]{\includegraphics[width=0.26\linewidth]{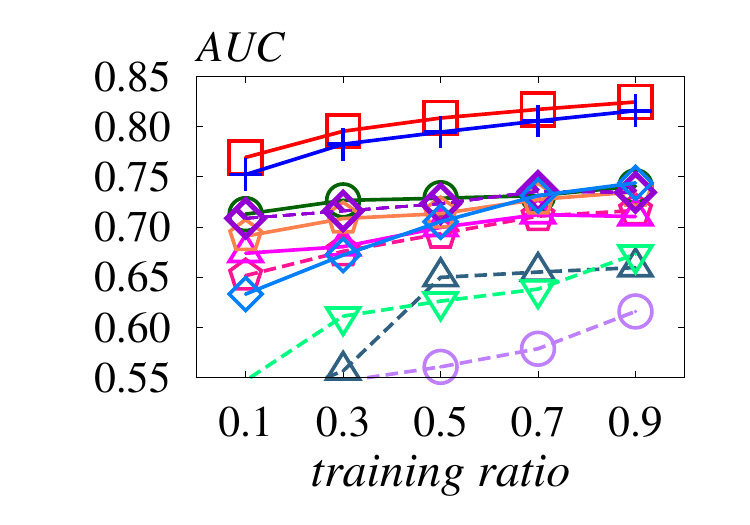}\label{fig:link-citeseer}} &
    \hspace{-5mm}\subfloat[{\em Pubmed}]{\includegraphics[width=0.26\linewidth]{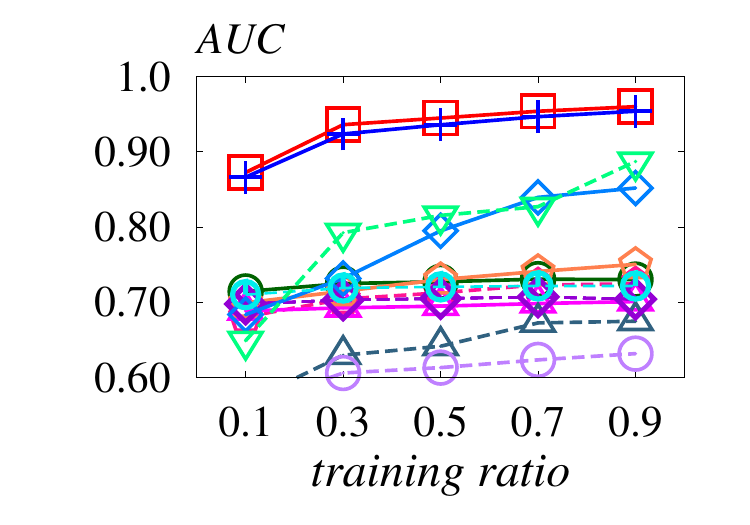}\label{fig:link-pubmed}} &
    \hspace{-5mm} \subfloat[{\em Facebook}]{\includegraphics[width=0.26\linewidth]{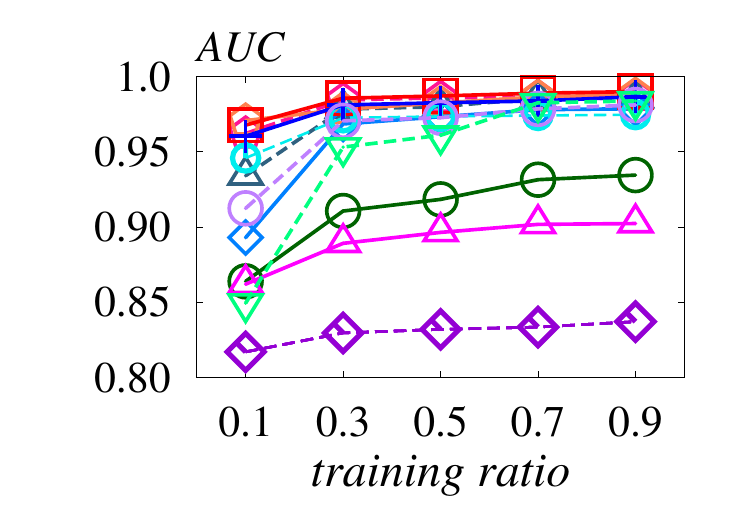}\label{fig:link-facebook}} &
    \hspace{-5mm}\subfloat[{\em Flickr}]{\includegraphics[width=0.26\linewidth]{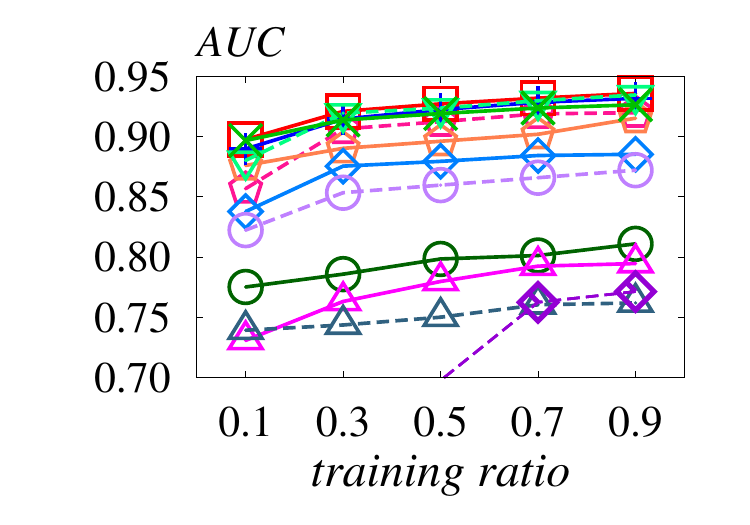}\label{fig:link-flickr}} \\
    \hspace{-6mm}\subfloat[{\em Google+}]{\includegraphics[width=0.26\linewidth]{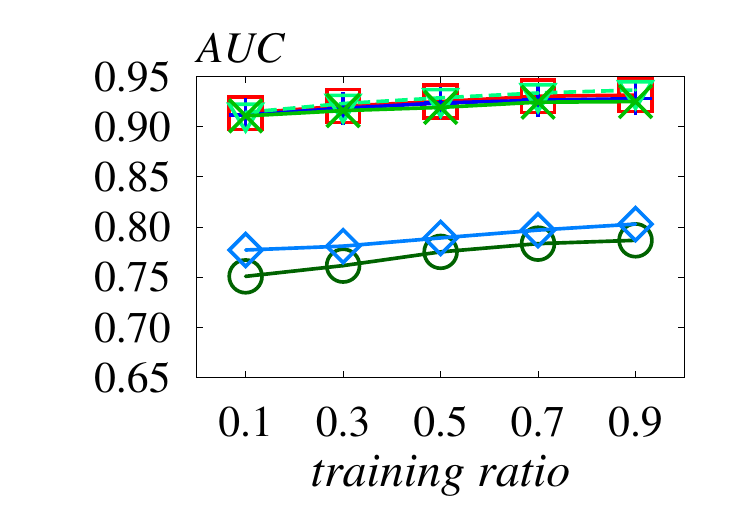}\label{fig:link-google}} &
    \hspace{-5mm}\subfloat[{\em TWeibo}]{\includegraphics[width=0.26\linewidth]{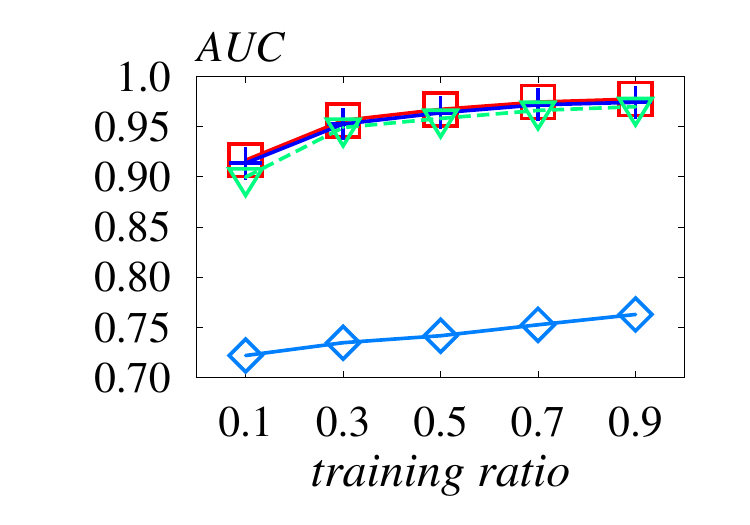}\label{fig:link-tweibo}} &
    \hspace{-5mm}\subfloat[{\em MAG}]{\includegraphics[width=0.26\linewidth]{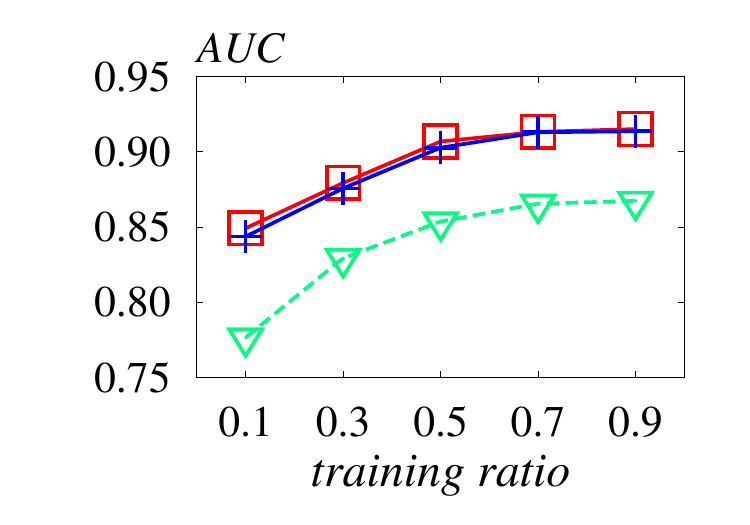}\label{fig:link-mag}} &
    \hspace{-5mm}\subfloat[{\em MAG-SC}]{\includegraphics[width=0.26\linewidth]{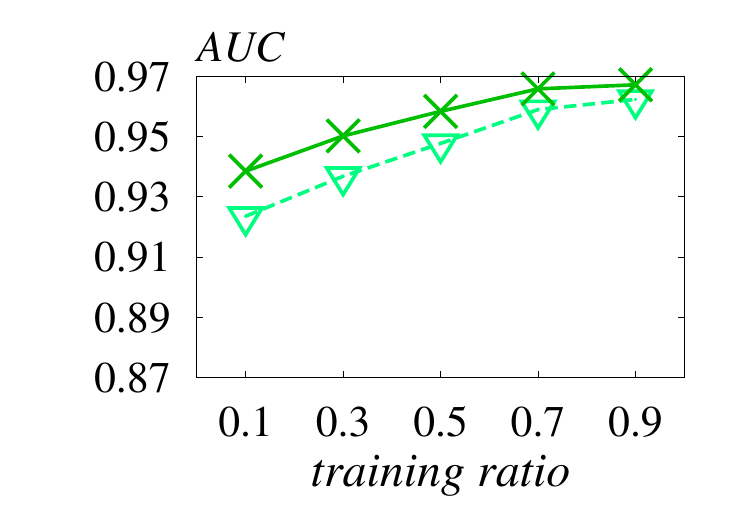}\label{fig:link-magsc}}
    \end{tabular}
    \vspace{-1mm}
    \caption{Link prediction results on graphs (best viewed in color).} \label{fig:link}
    \end{small}
    \vspace{-2mm}
\end{figure*}

\begin{figure*}[!t]
\centering
\captionsetup[subfloat]{captionskip=-0.5mm}
\begin{small}
\begin{tabular}{cccc}
\multicolumn{4}{c}{\hspace{-4mm} \includegraphics[height=6mm]{./figure/algo-legend-6.pdf}}\vspace{-2mm}  \\
% \hspace{-4mm}\subfloat[{\em Wiki}]{\includegraphics[width=0.26\linewidth]{./figure/class/wiki-eps-converted-to.pdf}\label{fig:acc-class-wiki}} &
% \hspace{-4mm}\subfloat[{\em Cora}]{\includegraphics[width=0.32\linewidth]{./figure/class/cora-eps-converted-to.pdf}\label{fig:acc-class-cora}} &
\hspace{-5mm}\subfloat[{\em Citeseer}]{\includegraphics[width=0.26\linewidth]{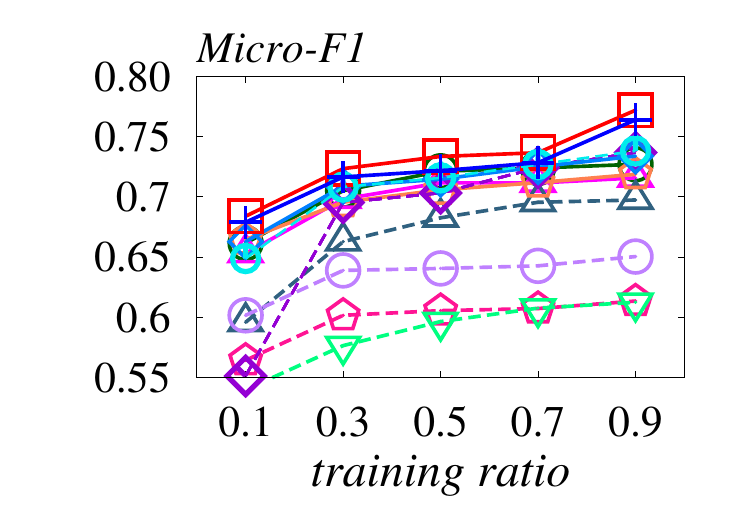}\label{fig:acc-class-citeseer}} &
\hspace{-5mm}\subfloat[{\em Pubmed}]{\includegraphics[width=0.26\linewidth]{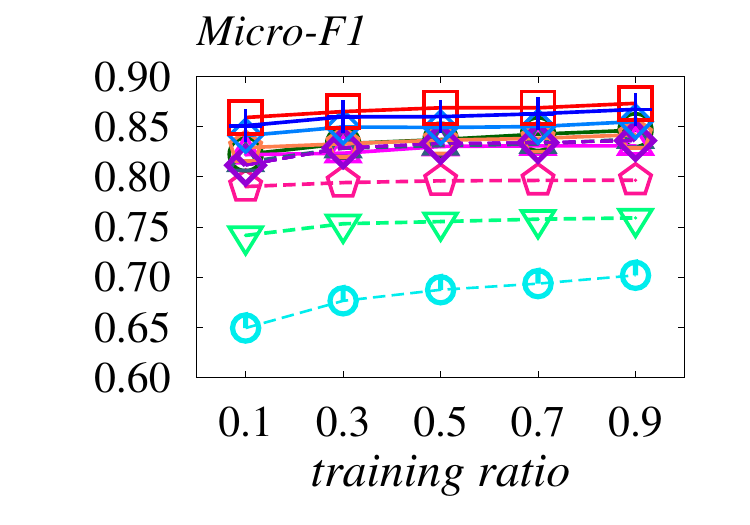}\label{fig:acc-class-pubmed}} &
\hspace{-5mm}\subfloat[{\em Facebook}]{\includegraphics[width=0.26\linewidth]{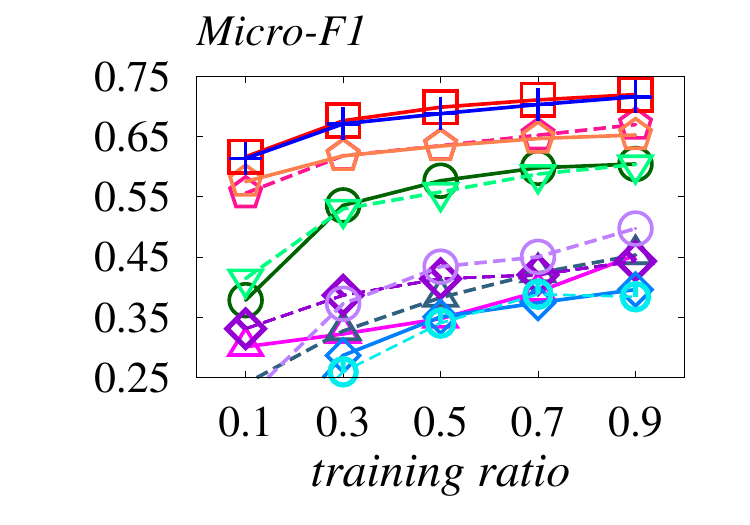}\label{fig:acc-class-facebook}}&
\hspace{-5mm}\subfloat[{\em Flickr}]{\includegraphics[width=0.26\linewidth]{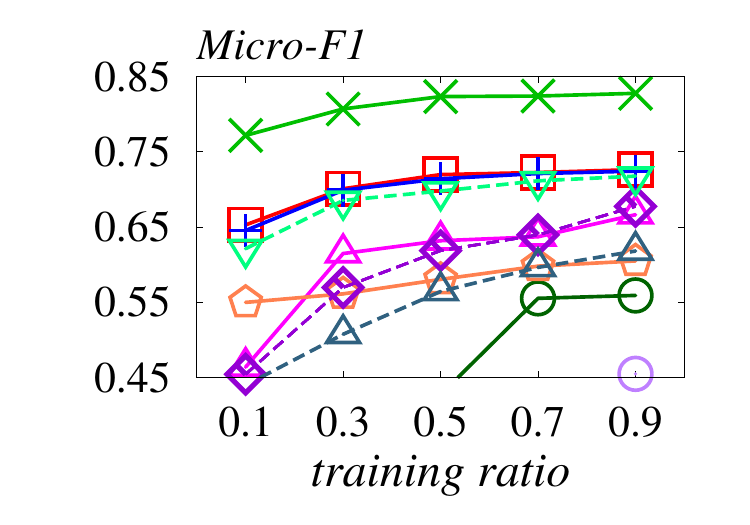}\label{fig:acc-class-flickr}} \\
\hspace{-5mm}\subfloat[{\em Google+}]{\includegraphics[width=0.26\linewidth]{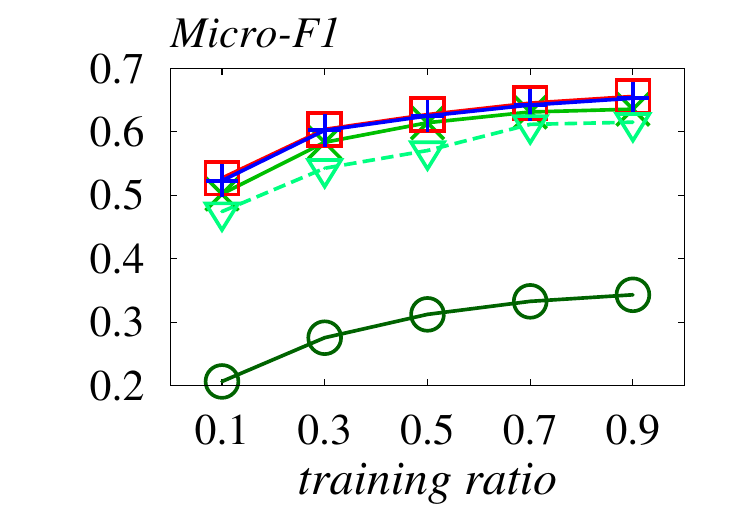}\label{fig:acc-class-google}} &
\hspace{-5mm}\subfloat[{\em TWeibo}]{\includegraphics[width=0.26\linewidth]{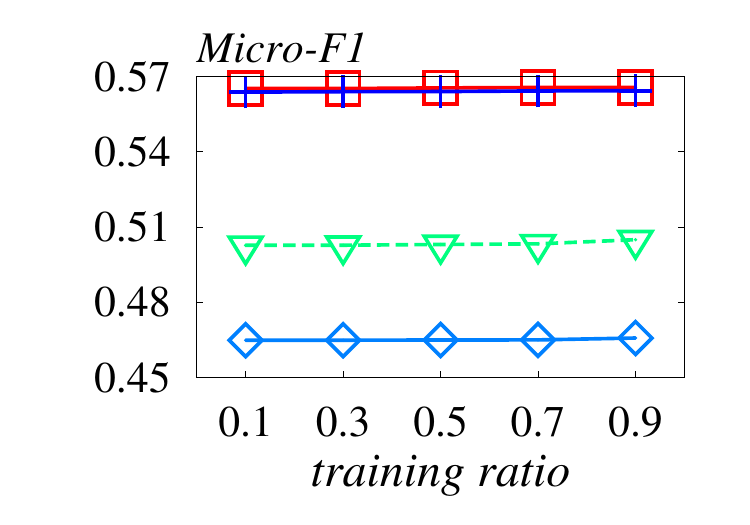}\label{fig:acc-class-tweibo}} &
\hspace{-5mm}\subfloat[{\em MAG}]{\includegraphics[width=0.26\linewidth]{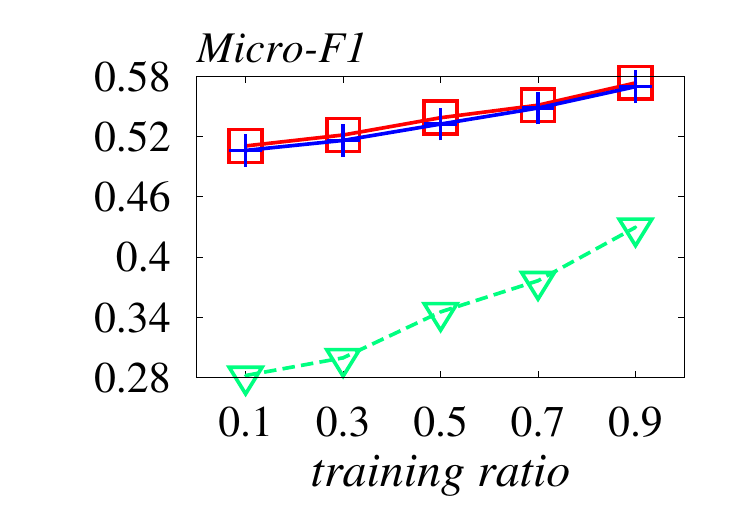}\label{fig:acc-class-mag}} &
\hspace{-5mm}\subfloat[{\em MAG-SC}]{\includegraphics[width=0.26\linewidth]{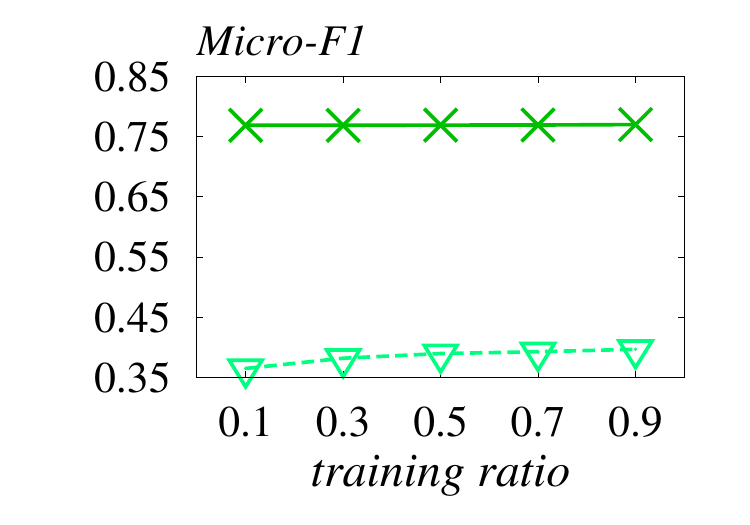}\label{fig:acc-class-magsc}} 
\end{tabular}
\caption{Node classification results on graphs (best viewed in color).} \label{fig:acc-class}
\end{small}
\vspace{-2mm}
\end{figure*}

% \begin{figure*}[!t]
% 	\centering
% 	\captionsetup[subfloat]{captionskip=-0.5mm}
% 	\begin{small}
% 		\begin{tabular}{ccc}
% 			\multicolumn{3}{c}{\hspace{-4mm} \includegraphics[height=6mm]{./figure/algo-legend-6.pdf}}\vspace{-3mm}  \\
% 			\hspace{-4mm}\subfloat[{\em Flickr}]{\includegraphics[width=0.32\linewidth]{./figure/link/flickr-eps-converted-to.pdf}\label{fig:link-flickr}} &
% 			\hspace{-4mm}\subfloat[{\em Google+}]{\includegraphics[width=0.32\linewidth]{./figure/link/google-eps-converted-to.pdf}\label{fig:link-google}}\vspace{-2mm} &
% 			\hspace{-4mm}\subfloat[{\em MAG-SC}]{\includegraphics[width=0.32\linewidth]{./figure/link/mag-sc-eps-converted-to.pdf}\label{fig:link-magsc}}
% 		\end{tabular}
% 		\caption{Link prediction results on graphs with large $d$ (best viewed in color).} \label{fig:link-ld}
% 	\end{small}
% 	\vspace{-2mm}
% \end{figure*}

\subsection{Link Prediction}\label{sec:link-pred}

Recall from Section~\ref{sec:ane-app} that the link prediction task aims to predict the edges that are most likely to form between nodes. In this set of experiments, we first randomly remove a certain number of edges $E_{rm}$ (ranging from $10\%$ to $90\%$) in input graph $G$, obtaining a residual graph $G'$. On undirected graphs, we then randomly sample the same amount of pairs of nodes without edges connecting each other as negative edges $E_{neg}$ (non-existing edges). On directed graphs, a node pair $(u, v)$ is ordered, and link prediction predicts whether there is a directed edge from $u$ to $v$. Hence, for directed graphs, $E_{neg}$ contains $|E_{rm}|/2$ non-existing edges obtained by reversing $|E_{rm}|/2$ edges picked from $E_{rm}$ and  $|E_{rm}|/2$ non-existing edges that are randomly sampled. The test set $E_{test}$ contains both the removed edges $E_{rm}$ and the negative edges $E_{neg}$.

We run all methods on the residual graph $G'$ to produce embedding vectors, and then evaluate the link prediction performance with $E_{test}$ as follows. 
Recall that our methods produce a forward embedding $\XMf[v_i]$ and a backward embedding $\XMb[v_i]$ for each node $v_i\in V$, as well as an attribute embedding $\YM[r_l]$ for each attribute $r_l\in R$.
As explained, given a node pair $(v_i,v_j)$, we use  $p(v_i, v_j)$ in \equref{eq:link-pred} for our methods to predict links on directed graphs, and use $p(v_i, v_j) + p(v_j, v_i)$ on undirected graphs.
% As explained, $\XMf[v_i]\cdot \YM[r_l]^{\top}$ preserves $\FM[v_i,r_l]$, and $\XMb[v_j]\cdot \YM[r_l]^{\top}$ preserves $\BM[v_j,r_l]$. Recall that  $\FM[v_i,r_l]$ measures the affinity from $v_i$ to $r_l$ over the attributed network; similarly given node $v_j$ and attribute $r_l$, $\BM[v_j,r_l]$ measures the affinity from $r_l$ to $v_j$ over the  network. 
% Intuitively, $\FM[v_i,r_l]\cdot\BM[v_j,r_l]$ represents the affinity from node $v_i$ to node $v_j$ based on attribute $r_l$. The affinity between nodes $v_i$ and $v_j$, denoted as $p(v_i,v_j)$, can be evaluated by summing up the affinity between the two nodes over all attributes in $R$, which can be computed according to \equref{eq:link-pred} and indicates the possibility of forming an edge from $v_i$ to $v_j$. 
% Therefore, for \algopt and \newalg, we can calculate $p(v_i,v_j)$ as the prediction score of the directed edge $(v_i,v_j)$. 
% To perform link prediction, we need to evaluate the affinity between nodes, considering both graph topology and attributes. The affinity between nodes can be obtained by concatenating forward affinity and backward affinity. Given node $v_i$ and attribute $r_l$, 
% we calculate $p(v_i,v_j)$ by \equref{eq:link-pred} described in \secref{sec:attr-app} as the prediction score of the directed edge $(v_i,v_j)$. 
Competitor \nrp generates a forward embedding $\XMf[v_i]$ and a backward embedding $\XMb[v_i]$ for each node $v_i$ and uses $\XMf[v_i]\cdot\XMb[v_j]^{\top}$ as the prediction score for the directed edge $(v_i,v_j)$ \cite{yang13homogeneous}, and $\XMf[v_i]\cdot\XMb[v_j]^{\top}+\XMf[v_j]\cdot\XMb[v_i]^{\top}$ for undirected edges. 
In terms of the remaining competitors that only work for undirected graphs, they learn one embedding $\XM[v_i]$ for each node $v_i$.
In literature, there are four ways to calculate the link prediction score $p(v_i,v_j)$, including {\em\ inner product} method used in \can and \arga, {\em cosine similarity} method used in \prre and \anrl, {\em Hamming distance} method used in \bane, as well as {\em edge feature} method used in \cite{node2vec2016,ma2018hierarchical}. 
%In particular, edge feature method fuses two $k$-dimensional node embedding vectors of each node pair $(v_i,v_j)$ into a length-$2k$ edge feature vector, and then trains a binary classifier to perform link prediction. 
We use all these four prediction methods for each method, and report the method's best performance. Following previous work \cite{meng2019co,pan2018adversarially}, we use {\em Area under the ROC Curve} (AUC) 
% and {\em Average Precision} (AP) 
to evaluate link prediction accuracy.
% Note that we report the best results achieved by the above four prediction methods for competiting methods on each dataset. 

\figref{fig:link} reports the AUC scores of each method on each dataset. 
% The highest scores are highlighted in blue, while the best results achieved by competitors are underlined. 
On undirected graphs including {\em Facebook} and {\em Flickr} in Figures \ref{fig:link-facebook} and \ref{fig:link-flickr}, our methods \algopt, \algoptp and \newalg achieve superior or comparable performance to the best competitors. Moreover, \algopt, \algoptp and \newalg consistently outperform all competitors over all directed graphs except \nrp on {\em Google+}, by a large margin 
% of up to $15.8\%$ for AUC
in terms of AUC.

For large attributed networks including {\em Google+, TWeibo}, {\em MAG}, and {\em MAG-SC}, most existing solutions cannot finish processing within a week and thus are not reported. 
% In particular, on the {\em MAG-SC} dataset, even \algopt and \algopt are unable to finish and \newalg is the only viable ANE solution. 
The superiority of our methods over competitors is achieved by (i) learning a forward embedding vector and a backward embedding vector for each node to capture the asymmetric transitivity (\ie edge direction) in directed graphs, and (ii) combining both node embedding vectors and attribute embedding vectors together for link prediction in \equref{eq:link-pred}, with the consideration of both topological and attribute features. On {\em Google+}, \nrp is slightly better 
% ($0.51\%$ absolute improvement at most) 
than \algopt since {\em Google+} has more than $15$ thousand attributes (see \tblref{tbl:exp-data}) leading to some accuracy loss when factorizing forward and backward affinity matrices into low dimensionality $k=128$ by \algopt. As shown in \figref{fig:link}, our \algoptp also outperforms all competitors significantly except \nrp on {\em Google+}. \algoptp also has comparable performance with \algopt  over all datasets.
%, which is due to the parallel approximation technique introduced in \secref{sec:split-merge} for efficiency. 
As reported in \secref{sec:exp-effi}, \algoptp is significantly faster than \algopt  by up to 9 times, with almost the same accuracy performance for link prediction.
For datasets with a large $d$, \ie{} {\em Flickr}, {\em Google+}, and {\em MAG-SC}, we can observe that the extended version of \algopt, \ie \newalg also yields competitive performance. In particular, \newalg is the only viable ANE solution and achieves a considerable gain 
% of up to $1.3\%$ 
over the best competitor \nrp.

\begin{figure*}[!t]
\centering
\captionsetup[subfloat]{captionskip=-0.5mm}
\begin{small}
\begin{tabular}{cccc}
\multicolumn{4}{c}{\hspace{-4mm} 
\includegraphics[height=2.7mm]{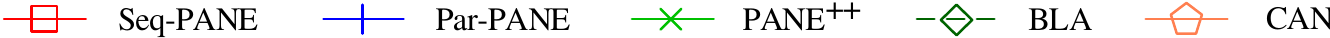}
}\vspace{-3mm}  \\
% \hspace{-4mm}\subfloat[{\em Wiki}]{\includegraphics[width=0.26\linewidth]{./figure/class/wiki-eps-converted-to.pdf}\label{fig:acc-class-wiki}} &
% \hspace{-4mm}\subfloat[{\em Cora}]{\includegraphics[width=0.32\linewidth]{./figure/attr/cora-eps-converted-to.pdf}\label{fig:attr-cora}} &
\hspace{-6mm}\subfloat[{\em Citeseer}]{\includegraphics[width=0.26\linewidth]{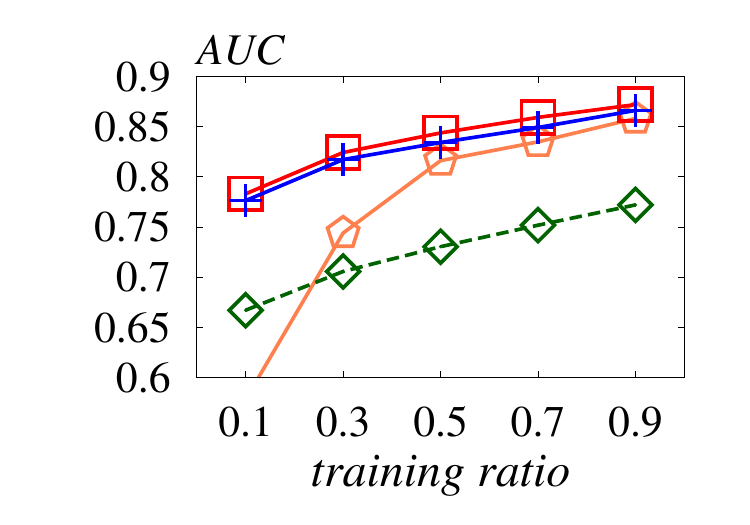}\label{fig:attr-citeseer}} &
\hspace{-5mm}\subfloat[{\em Pubmed}]{\includegraphics[width=0.26\linewidth]{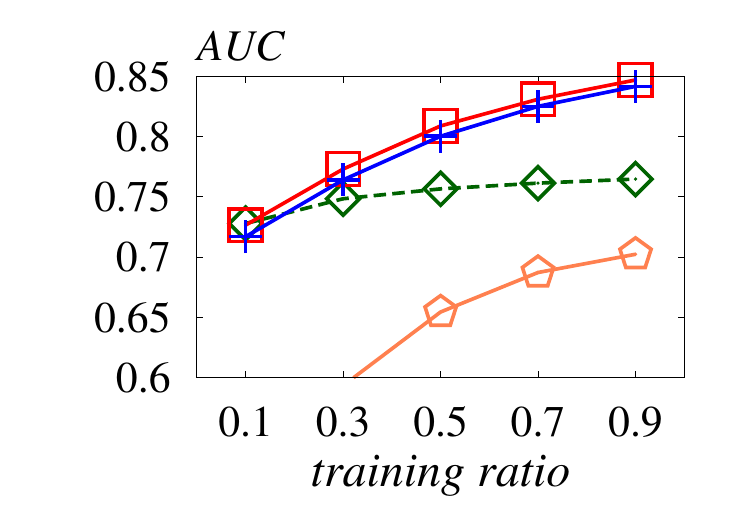}\label{fig:attr-pubmed}} &
\hspace{-5mm} \subfloat[{\em Facebook}]{\includegraphics[width=0.26\linewidth]{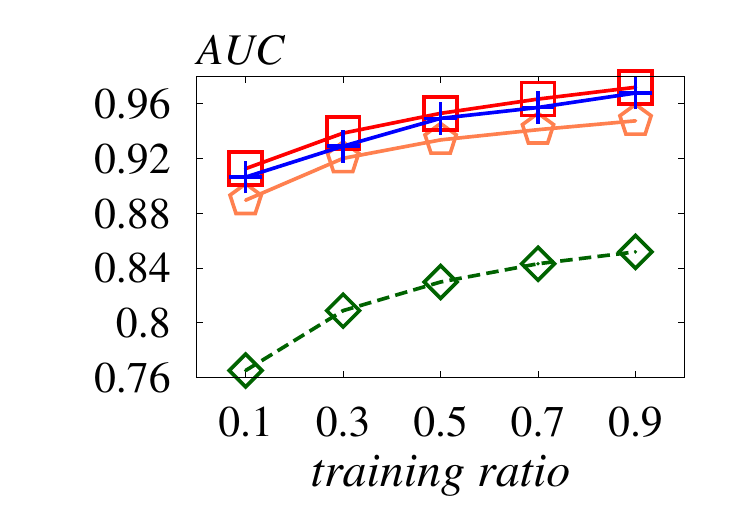}\label{fig:attr-facebook}}&
\hspace{-5mm}\subfloat[{\em Flickr}]{\includegraphics[width=0.26\linewidth]{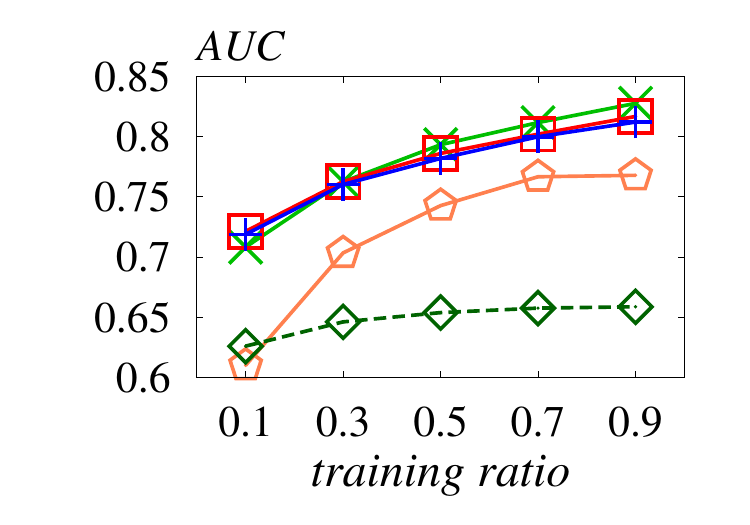}\label{fig:attr-flickr}} \\
\hspace{-6mm}\subfloat[{\em Google+}]{\includegraphics[width=0.26\linewidth]{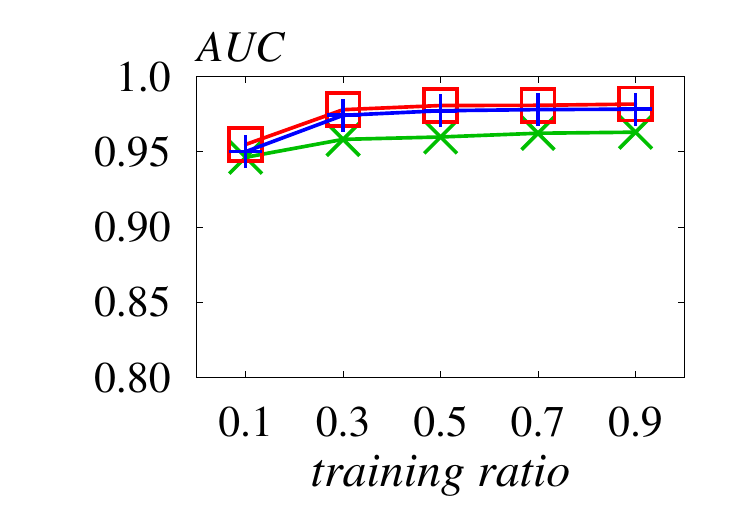}\label{fig:attr-google}}\vspace{-2mm} &
\hspace{-5mm}\subfloat[{\em TWeibo}]{\includegraphics[width=0.26\linewidth]{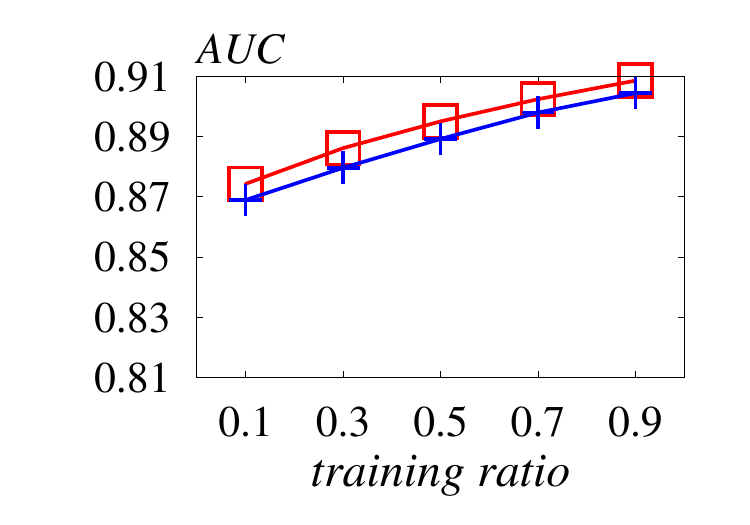}\label{fig:attr-tweibo}} &
\hspace{-5mm}\subfloat[{\em MAG}]{\includegraphics[width=0.26\linewidth]{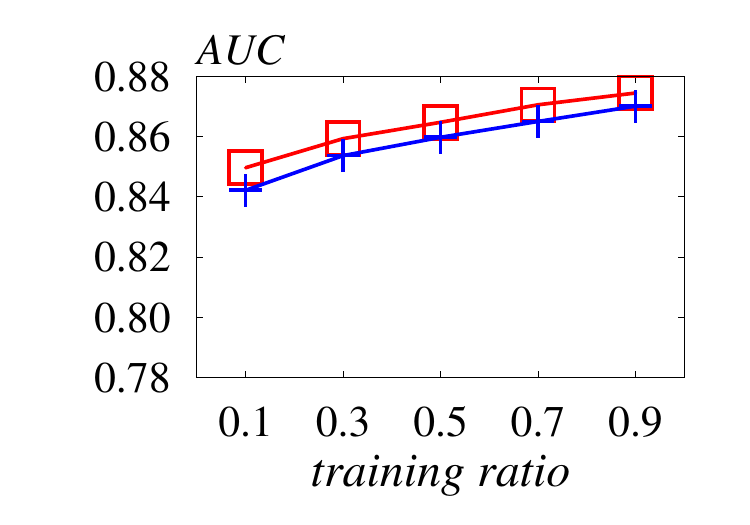}\label{fig:attr-mag}} &
\hspace{-5mm}\subfloat[{\em MAG-SC}]{\includegraphics[width=0.26\linewidth]{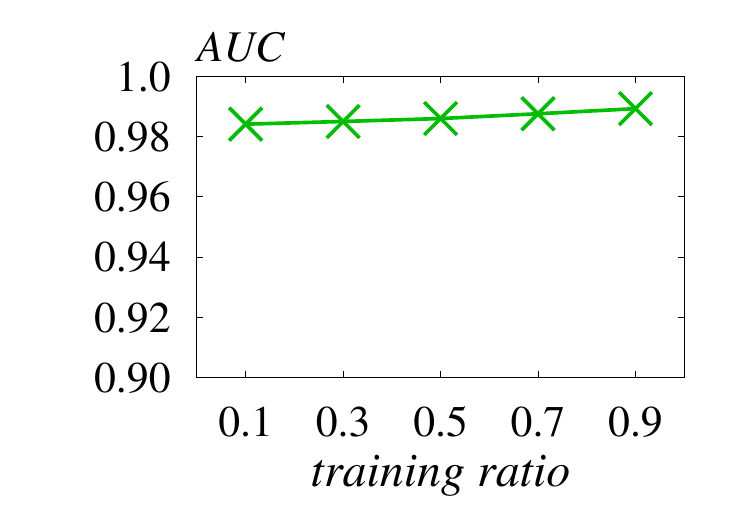}\label{fig:attr-magsc}}
\end{tabular}
% \vspace{-2mm}
\caption{Attribute inference results on graphs (best viewed in color).} \label{fig:attr}
\end{small}
\vspace{-2mm}
\end{figure*}

\subsection{Node Classification}\label{sec:node-class}
Node classification predicts the node labels. Note that {\em Facebook}, {\em Google+} and {\em MAG} are multi-labelled. That is, each node can have multiple labels. We first run \algopt , \algoptp, \newalg, and the competitors on the input attributed network $G$ to obtain their embeddings. Then we randomly sample a certain number of labelled nodes (ranging from 10\% to 90\%) to train a linear support-vector machine (SVM)  classifier \cite{cortes1995support} and use the rest for testing. \nrp, \algopt , \algoptp, and \newalg generate a forward embedding vector $\XMf[v_i]$ and a backward embedding vector $\XMb[v_i]$ for each node $v_i\in V$. 
As explained in Section \ref{sec:ane-app}, we normalize the forward and backward embeddings of each node $v_i$, and then concatenate them as the feature representation of $v_i$ to be fed into the classifier. Akin to prior work \cite{meng2019co,hou2019representation,ijcai2019-low}, we use Micro-F1 and Macro-F1  to measure node classification performance. We repeat for 5 times and report the average performance.

\figref{fig:acc-class} shows the Micro-F1 results when varying the percentage of nodes used for training from 10\% to 90\% (\ie 0.1 to 0.9). The results of  Macro-F1 are  similar and thus omitted for brevity.
For graphs with small $d$, including \textit{Citeseer, Pubmed, Facebook, TWeibo}, and \textit{MAG} in Figure \ref{fig:acc-class}, our methods \algopt, \algoptp and \newalg consistently outperform all competitors, which demonstrates that our proposed solutions effectively capture the topology and attribute information of the input attributed networks. 
%We omit the results of some methods on {\em Google+}, {\em TWeibo} and {\em MAG} due to their inability of handling large graphs. 
Specifically, compared with the competitors, \algopt  achieves a remarkable improvement
% , up to $3.4\%$ 
on {\em Citeseer}, {\em Pubmed}, and {\em Facebook}. On the large graphs {\em TWeibo} and {\em MAG}, most existing solutions fail to finish within a week and thus their results are omitted. Furthermore, \algopt  outperforms \nrp by a notable margin
% at least $6\%$ 
on {\em TWeibo} as displayed in  \figref{fig:acc-class-tweibo}. In addition, \algopt  and \algoptp are superior to \nrp with a significant gain 
% up to $17.2\%$ 
on {\em MAG}. Over all datasets, \algoptp has similar performance to that of \algopt , while as shown in \secref{sec:exp-effi}, \algoptp is significantly faster than \algopt .

As for the three datasets with large $d$, \ie{}  {\em Flickr}, {\em Google+}, and {\em MAG-SC} in Figure \ref{fig:acc-class}, \algopt , \algoptp, and \newalg still outperform all competitors. Further, we can observe that \newalg is significantly better than \algopt and \algoptp on \textit{Flickr}, and comparable to them on \textit{Google+}. In particular, on {\em Flickr}, \newalg outperforms \algopt by a significant margin of at least $10\%$ in terms of Micro-F1. On \textit{MAG-SC} with millions of attributes in Figure \ref{fig:acc-class}, \newalg can efficiently obtain effective embeddings for classification, while \algopt and \algoptp run out of memory, as reported in Section \ref{sec:exp-effi}. On \textit{MAG-SC}, \newalg is far better than the  only competitor \nrp, which is designed for homogeneous networks without considering the attributes in \textit{MAG-SC}.
The superior performance of \newalg over \algopt, \algoptp, and all competitors on graphs with a large number of attributes validates the effectiveness of the proposed techniques in Section \ref{sec:cluster-idea}, which boosts efficiency and also improves effectiveness in obtaining high-quality embeddings.

Another observation we can make from \figref{fig:acc-class} is that on {\em TWeibo} and {\em MAG-SC} datasets, the performance of all methods remain stable when increasing the training ratio from $0.1$ to $0.9$ whereas their performance on other datasets goes up notably. The reason is as follows. Note that both {\em TWeibo} and {\em MAG-SC} datasets have millions of nodes and most nodes are associated with only 2 to 4 dominant labels. As such, even with $10\%$ training data, we can learn a classifier with accuracy comparable to that with $90\%$ training data. In contrast, other datasets either have a small number of nodes or balanced label distributions, making the classifiers learned on them more sensitive to the training ratio.

%on {\em Flickr} is due to our effective attribute clustering method introduced in Section \ref{sec:cluster-idea}, which can accurately preserve attribute information in {\em Flickr} in a dense super attribute matrix $\widetilde{\RM}$. 
%In contrast, on {\em Google+}, \algopt and \newalg achieve similar performance, and outperform \nrp by up to $4.4\%$ absolute improvement. 
%As for {\em MAG-SC}, \newalg outperforms \nrp by a significant gain up to $40\%$. 
%Note that, as reported later in \secref{sec:exp-effi}, \newalg is much faster than \algopt. 
%This demonstrates the effectiveness of attribute clustering employed in \newalg.

\begin{figure*}[!t]
\centering
\begin{minipage}{0.8\textwidth}
	\captionsetup[subfloat]{captionskip=-0.5mm}
	\begin{small}
		\begin{tabular}{ccc}
% 			\multicolumn{3}{c}{			\hspace{-8mm} \includegraphics[height=2.6mm]{./figure/algo-legend3-new.pdf}}\vspace{-2mm}  \\
			\hspace{-4mm}\subfloat[{\em Varying $k$}]{\includegraphics[width=0.32\linewidth]{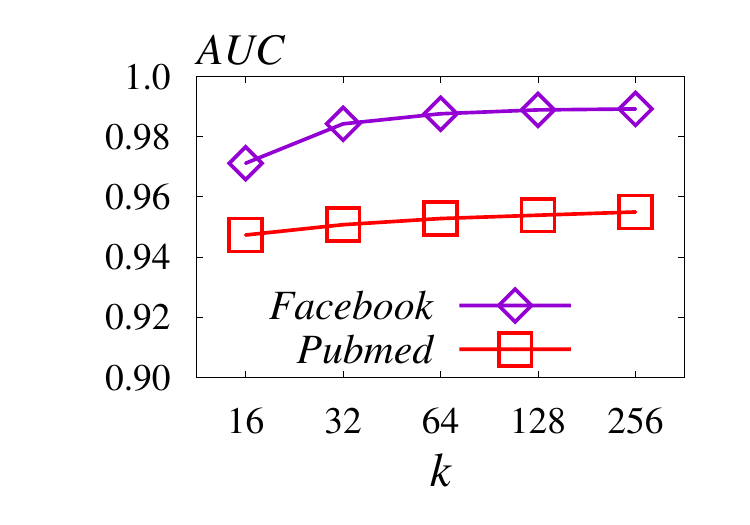}\label{fig:acc-link-k}} &
% 			\hspace{-8mm}\subfloat[{\em Varying $n_b$}]{\includegraphics[width=0.28\linewidth]{./figure/link-nb-eps-converted-to.pdf}\label{fig:acc-link-nb}} &
			\hspace{-6mm}\subfloat[{\em Varying $\epsilon$}]{\includegraphics[width=0.32\linewidth]{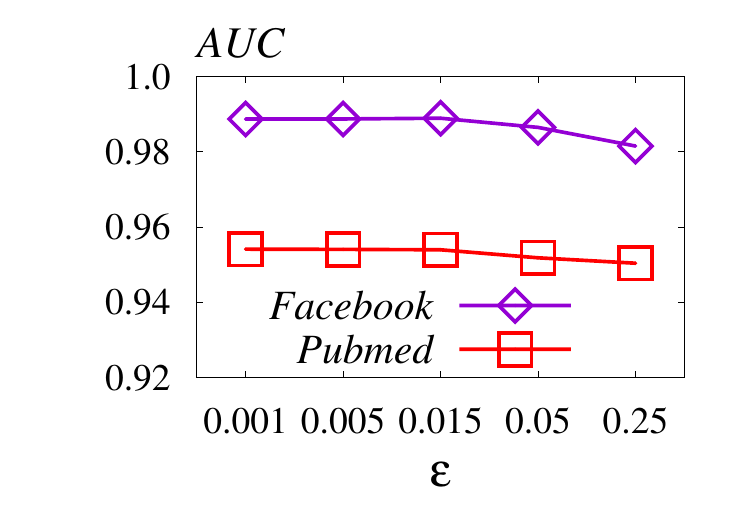}\label{fig:acc-link-eps}} &
			\hspace{-6mm}\subfloat[{\em Varying $\alpha$}]{\includegraphics[width=0.32\linewidth]{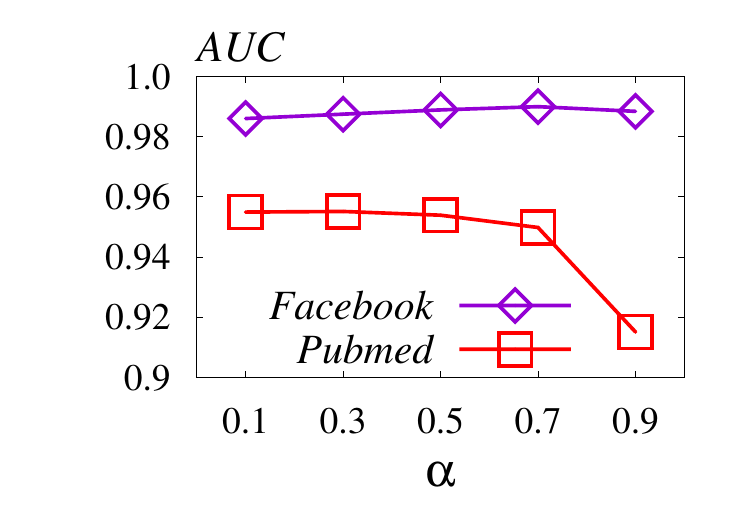}\label{fig:acc-link-alpha}}
		\end{tabular}
		\vspace{-2mm}
		\caption{Varying parameters in \algopt.} \label{fig:link-param}
	\end{small}
	\vspace{-2mm}
\end{minipage}
\hspace{-29mm}
\begin{minipage}{0.35\textwidth}
	\vspace{5mm}
	%\centering
\flushright
\captionsetup{type=figure,justification=raggedright}
\includegraphics[width=0.74\linewidth]{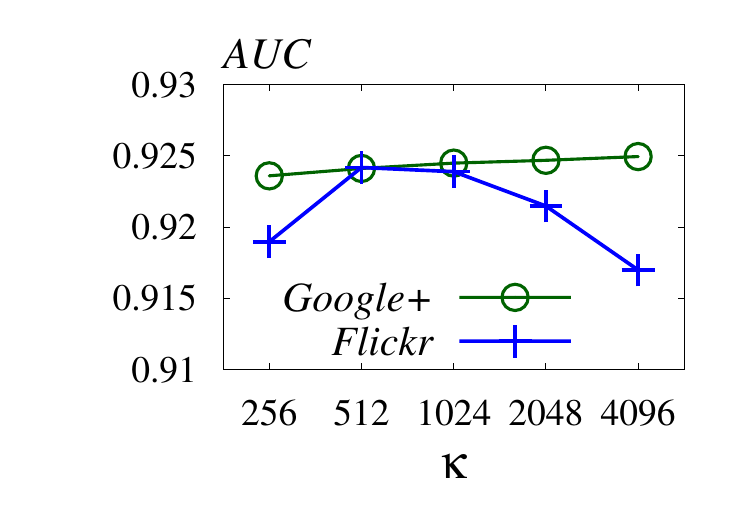}\label{fig:acc-link-kappa}
% \vspace{-3mm}
\vspace{0mm}
\caption{Varying $\kappa$ in \newalg.} \label{fig:link-kappa}
\end{minipage}
\end{figure*}

\begin{figure*}[!t]
\centering
\begin{minipage}{0.8\textwidth}
\begin{tabular}{ccc}
\hspace{-6mm}\subfloat[{ {speedups vs. $n_b$}}.]{\includegraphics[width=0.32\linewidth]{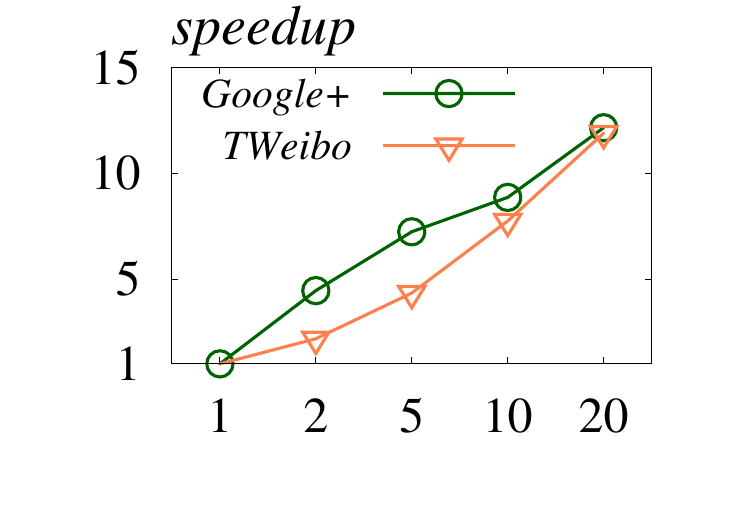}\label{fig:time-nb}} &
\hspace{-6mm}\subfloat[{ time vs. $k$}.]{\includegraphics[width=0.32\linewidth]{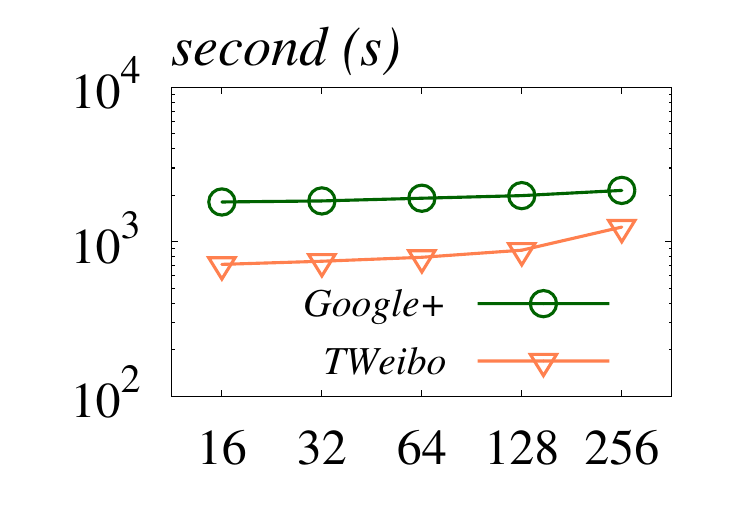}\label{fig:time-k}} &
\hspace{-6mm}\subfloat[{ time vs. $\epsilon$}.]{\includegraphics[width=0.32\linewidth]{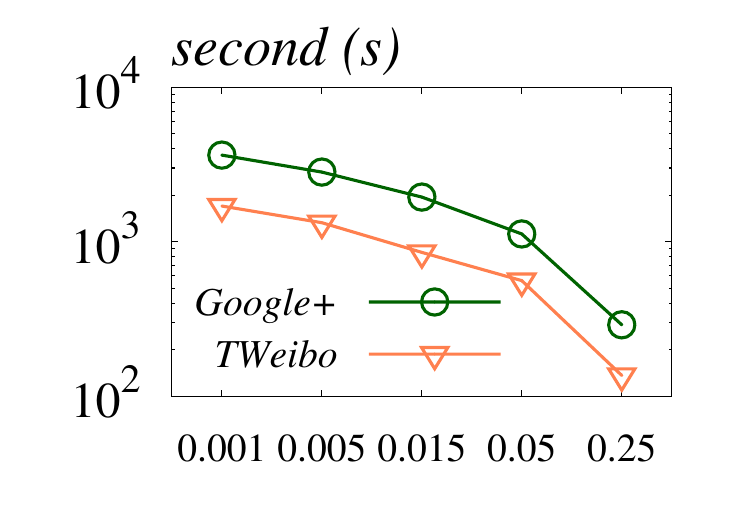}\label{fig:time-eps}} 
\end{tabular}
\vspace{-2mm}
\captionsetup{justification=raggedleft}
\caption[]{Varying parameters in \algopt/\algoptp} \label{fig:time-param}
\vspace{0mm}
\end{minipage}
\hspace{-29mm}
\begin{minipage}{0.35\textwidth}
	\vspace{4mm}
	%\centering
\flushright
	\captionsetup{type=figure,justification=raggedright}
	\includegraphics[width=0.74\linewidth]{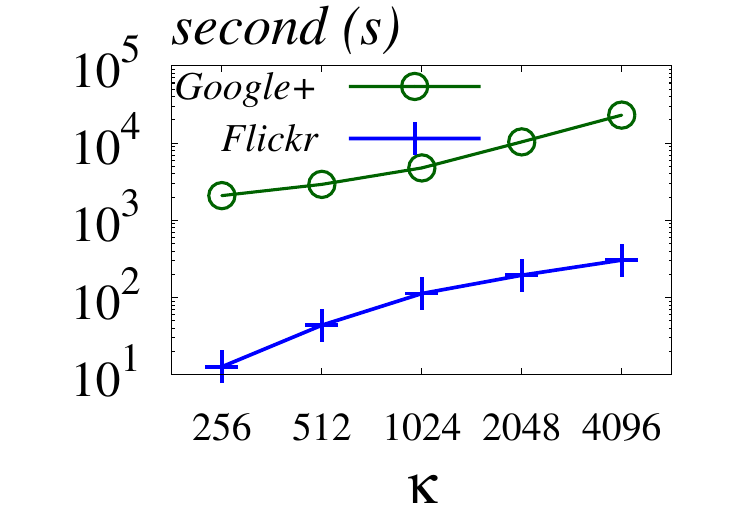}\label{fig:time-kappa}
\vspace{0mm}
\caption{Varying $\kappa$ in \newalg.} \label{fig:time-param-new}
\end{minipage}
\end{figure*}

\subsection{Attribute Inference}\label{sec:attr-infer} 

Attribute inference aims to predict the existence of an attribute in a node. Note that, except for \can \cite{meng2019co}, none of the other competitors is capable of performing attribute inference since they only generate embedding vectors for nodes and not attributes. Hence, we compare our solutions against \can for attribute inference. Further, we compare against \bla, the state-of-the-art attribute inference algorithm \cite{yang2017bi}. Note that \bla is not an ANE solution. 

We split the node-attribute associations $E_R$, and regard a certain number of these associations (ranging from $10\%$ to $90\%$) as the test set $E^{te}_R$ and the remaining part as the training set $E^{tr}_R$. In $E^{te}_R$, we also added an equal number of node-attribute pairs that are not in the original $E_R$ as negative samples. \can runs over $E^{tr}_R$ to generate node embedding vector $\XM[v_i]$ for each node $v_i\in V$ and attribute embedding vector $\YM[r_j]$ for each attribute $r_j\in R$, and uses the inner product of $\XM[v_i]$ and $\YM[r_j]$ as the predicted score of attribute $r_j$ with respect to node $v_i$. 
For our methods, we use \equref{eq:attr-infer} to calculate the predicted score of an attribute $r_j$ to a node $v_i$.
% Note that \algoptp generates a forward embedding vector $\XMf[v_i]$ and a backward embedding vector $\XMb[v_i]$ for each node $v_i\in V$, and also an attribute embedding vector $\YM[r_j]$ for each attribute $r_j\in R$.
% Based on objective function in \equref{eq:obj1}, $\XMf[v_i]\cdot \YM[r_j]^{\top}$ is expected to preserve forward affinity value $\FM[v_i,r_j]$, and $\XMb[v_i]\cdot \YM[r_j]^{\top}$ is expected to preserve backward affinity value $\BM[v_i,r_j]$. Thus, we predict the score between $v_i$ and $r_j$ through the affinity between node $v_i$ and attribute $r_j$, including both forward affinity and backward affinity, denoted as $p(v_i,r_j)$, by utilizing their embedding vectors as in \equref{eq:attr-infer}.
Following prior work \cite{meng2019co}, we adopt the {\em Area under the ROC Curve} (AUC) to measure the performance. 

%\header
%{\bf Results.} 
In \figref{fig:attr}, we present the AUC scores of \algopt, \algoptp, \newalg,  \can, and  \bla in terms of attribute inference when varying the percentage of node-attribute associations (\ie the entries in $\RM$) used for training from 10\% to 90\% (\ie 0.1 to 0.9).
% Particularly, in \figref{fig:attr-ld}, we also show the AUC results of \newalg on the datasets with large $d$, \ie{} {\em Flickr}, {\em Google+}, and {\em MAG-SC}. 
Observe from \figref{fig:attr} that our methods consistently outperform existing solutions often by a large margin, demonstrating the power of the learned embedding vectors $\XMf,\XMb$ and $\YM$, which capture the affinity between nodes and attributes in attributed networks.
In particular, on graphs with small $d$, including \textit{Citeseer, Pubmed, Facebook, TWeibo}, and \textit{MAG} in Figure \ref{fig:attr}, \algopt always obtains the highest AUC scores among all methods, while \algoptp and \newalg have comparable performance. 
% For instance, on {\em Pubmed}, \algopt has the highest AUC 
% up to $0.847$
% , while that of the state-of-the-art competitor $\mathtt{BLA}$ is far inferior.
% $0.765$.
On {\em Pubmed}, the difference of AUC between \algopt  and \algoptp is very small.
% just $0.5\%$. 
This negligible difference is introduced by the split-merge-based parallel SVD technique \smsvd for matrix decomposition.
As shown in \secref{sec:exp-effi}, parallel \algoptp is considerably faster than \algopt  by up to 9 times, while obtaining almost the same accuracy performance.
Moreover, on graphs with large $d$  (\textit{Flickr, Google+,} and \textit{MAG-SC}), \newalg is the only method able to handle \textit{MAG-SC} with millions of attributes (Figure \ref{fig:attr-magsc}) and also is slightly better than \algopt and \algoptp on \textit{Flickr} (Figure \ref{fig:attr-flickr}). It also has comparable performance on \textit{Google+} as depicted in Figure \ref{fig:attr-google}.
Moreover, on \textit{Flickr}, all our methods outperform competitors \can and $\mathtt{BLA}$ by a remarkable
% at least 3.6\% 
absolute improvement in terms of AUC.
In addition, \can and \bla fail to process large attributed networks, including {\em TWeibo}, {\em Google+}, {\em MAG} and {\em MAG-SC} in one week. Thus, their results are omitted in Figure \ref{fig:attr}.
Considering the high efficiency of \newalg demonstrated in in Section \ref{sec:exp-effi}, we conclude that the techniques in Section \ref{sec:cluster-idea} are effective and efficient to learn ANE embeddings on graphs with many attributes.

\subsection{Parameter Analysis}\label{sec:exp-param}
In this section, we evaluate the performance of our methods with varying parameter values, to study the effects of these parameters. First, we vary the embedding dimensionality $k$, error threshold $\epsilon$ and random walk stopping probability $\alpha$ in \algopt for link prediction on {\em Facebook} and {\em Pubmed}. The AUC results are reported in \figref{fig:link-param}.
Specifically, \figref{fig:acc-link-k} displays the link prediction AUC scores of \algopt on {\em Facebook} and {\em Pubmed}, with varying values of the embedding size $k$ in $\{16,32,64,128,256\}$. 
Observe that the AUC scores grow notably when $k$ increases from $16$ to $256$, indicating that a large embedding dimensionality generally leads to more effective % benefits to producing more accurate 
embedding vectors.
\figref{fig:acc-link-eps} reports the AUC scores of \algopt when varying $\epsilon$ from $0.001$ to $0.25$. Observe that the link prediction performance remains relatively stable when increasing $\epsilon$ from 0.001 to 0.015, and decreases slightly when $\epsilon$ increases from 0.015 to 0.25.
Note that when $\alpha=0.5$, varying $\epsilon$ from $0.001$ to $0.25$ is equivalent to varying the number of iterations $t$ from $9$ down to $1$.
We then vary $\alpha$ from 0.1 to 0.9, and report the AUC scores of \algopt on link prediction in \figref{fig:acc-link-alpha}.
Observe that when $\alpha$ increases, the performance of \algopt on \textit{Facebook} is relatively stable, while that on \textit{Pubmed} decreases significantly when $\alpha>0.5$. Therefore, we choose to set $\alpha=0.5$ by default. 
We speculate that the different behaviors of \algopt are due to the different underlying graph properties of \textit{Facebook} and \textit{Pubmed}. 

Next, we vary $\kappa$ from $256$ to $4096$ in \newalg on {\em Flickr} and {\em Google+}, and report the link prediction performance in \figref{fig:link-kappa}.
When $\kappa$ increases from $256$ to $4096$, the AUC performance on {\em Google+} goes up slightly, while the performance on {\em Flickr} increases first and then decreases after $\kappa>1024$. Therefore, we choose to set $\kappa=1024$ by default.
The different behaviors of \newalg on {\em Flickr} and {\em Google+} are probably due to their different graph properties.

{ \figref{fig:time-nb} displays the speedups of \algoptp over \algopt on {\em Google+} and {\em TWeibo} when varying the number of threads $n_b$ from $1$ to $20$.
When $n_b$ increases, \algoptp becomes much faster than \algopt, demonstrating the parallel scalability of \algoptp with respect to $n_b$. \figref{fig:time-k} and \figref{fig:time-eps} illustrate the running time of \algopt with varying space budget $k$ from $16$ to $256$ and error threshold $\epsilon$ from $0.001$ to $0.25$, respectively.}
%When one of these parameters varies, we keep others as default settings described in \secref{sec:exp-set}.
In \figref{fig:time-k},  when $k$ is increased from $16$ to $256$, the running time is quite stable and goes up slowly, indicating the robust efficiency of our solution.
In \figref{fig:time-eps}, the running time of \algopt decreases considerably when error threshold $\epsilon$ is increased in $\{0.001,0.005,0.015,0.05,0.25\}$. When $\epsilon$ increases  from $0.001$ to $0.25$, the running time on {\em Google+} and {\em TWeibo} reduces by about $10$ times, which is consistent with our analysis that \algopt runs in  linear to $\log\left({1}/{\epsilon}\right)$ in \secref{sec:parallel}. 
\figref{fig:time-param-new} presents the running time of \newalg when varying the number of attribute clusters $\kappa$ from $256$ to $4096$ on {\em Flickr} and {\em Google+}. As $\kappa$ increases, the running time of \newalg increases, which is consistent with the time complexity analysis of \newalg in Section \ref{sec:cluster-als}.

%% file: rw.tex
% \vspace{-1mm}
\section{Related Work}\label{sec:rw}
% \vspace{-1mm}

The work reported in this paper is an extended version of \cite{yang2020scaling,yang2022scaling}. It differs from these earlier versions in the following ways. First, this work introduces the \newalg algorithm for handling an input network with a large attribute set. Second, it presents a detailed description on how the obtained embeddings are exploited for downstream machine learning tasks, in particular, node classification, attribute inference, and link prediction. Lastly,  
% (ii) a detailed description on how to use our new prediction functions with the obtained embeddings for downstream machine learning tasks including attribute inference and link prediction,
our experimental study is extended by incorporating a new dataset, \textit{MAG-SC}, with millions of attributes and billions of node-attribute associations, as well as an extensive parameter analysis.

In the following, we review related work in the literature.

\subsection{Network Embedding}
Network embedding (NE)~\cite{perozzi2014deepwalk} is to learn low-dimensional, fixed-length vector representations of network nodes such that the similarity in the embedding space reflects the similarity in the network. A pioneering effort is $\mathtt{DeepWalk}$ \cite{perozzi2014deepwalk}, which adopts the SkipGram model and random walks to capture the graph structure surrounding a node and map it into a low-dimensional embedding vector. Several studies \cite{tang2015line,node2vec2016,zhou2017scalable,tsitsulin2018verse} aim to improve the performance over $\mathtt{DeepWalk}$ by exploiting different random walk schemes. These random-walk-based solutions suffer from severe efficiency issues as they need to sample a large number of random walks and conduct expensive training processes.
To address these challenges, massively parallel network embedding systems, including $\mathtt{PBG}$ \cite{pbg2019}, $\mathtt{Graphy}$ \cite{zhu2019graphvite} and $\mathtt{LightNE}$ \cite{qiu2021lightne}, are developed to utilize a large system with multiple processing units, including CPUs and GPUs. However, these systems consume immense amounts of computational resources that are financially expensive.
Qiu {\it et al.} proved that the aforementioned random-walk-based methods have their equivalent matrix factorization forms and proposed an efficient factorization-based UNE solution \cite{qiu2018network}.
In the literature, there are many factorization-based UNE solutions exhibiting superior efficiency and effectiveness, such as $\mathtt{RandNE}$ \cite{zhang2018billion}, $\mathtt{AROPE}$ \cite{zhang2018arbitrary}, $\mathtt{STRAP}$ \cite{yin2019scalable}, \nrp \cite{yang13homogeneous}, and $\mathtt{FREDE}$ \cite{tsitsulin2021frede}.
% and $\mathtt{Lemane}$ \cite{zhang2021learning}.
\revise{ In addition to preserving the affinities between nodes, a number of studies \cite{DBLP:journals/kbs/YeJJWH22,DBLP:journals/tetci/GongCXW20,DBLP:conf/aaai/WangCWP0Y17,DBLP:journals/isci/DuanSZCZT21} propose to incorporate community structures into network embedding.}
% for enhanced result utility
However, all NE solutions ignore attributes associated with nodes, limiting their utility in real-world attributed networks.

\subsection{Attributed Network Embedding}
% Unlike HNE, {\em Attributed Network Embedding} (ANE) considers not only graph structure but aslo attributes of each node. 
% ANE is also a popular topic and there exists a large body of literature. Here we mainly category them into three types based the methods they adopt.

\header
{\bf Factorization-based methods.} Given an attributed network $G$ with $n$ nodes, existing factorization-based methods mainly involve two stages: (i) building a proximity matrix $\MM\in \mathbb{R}^{n\times n}$ that models the proximity between nodes based on graph topology or attribute information; (ii) factorizing $\MM$ via techniques such as SGD \cite{bottou2010large}, ALS \cite{comon2009tensor}, and coordinate descent \cite{wright2015coordinate}.
Specifically, \tadw \cite{yang2015network} constructs a second-order  proximity matrix $\MM$ based on the adjacency matrix of $G$, and aims to reconstruct $\MM$ by the product of the learned embedding matrix and the attribute matrix. \hsca \cite{zhang2016homophily}  ensures that the learned embeddings of connected nodes are close in the embedding space. \aane \cite{huang2017accelerated} constructs a proximity matrix $\MM$ using the cosine similarities between the attribute vectors of nodes. \bane \cite{yang2018binarized} learns a binary embedding vector per node, \ie $\{-1,1\}^{k}$, by minimizing the reconstruction loss of a unified matrix that incorporates both graph topology and attribute information. \bane reduces space overheads at the cost of accuracy. To further balance the trade-off between space cost and representation accuracy, \lqanr \cite{ijcai2019-low} learns embeddings $\in \{-2^{b},\cdots,-1,0,1,\cdots,2^b\}^k$, where $b$ is the bit-width. $\mathtt{GAGE}$ \cite{kanatsoulis2022gage} formulates the ANE problem based on multi-dimensional scaling \cite{davison1983introduction} and employs the tensor factorization over distance matrices to produce node embeddings.
\revise{$\mathtt{ANEM}$ \cite{DBLP:journals/tkdd/LiHWHLC21} utilizes {\em nonnegative matrix factorization} \cite{arora2012computing} to jointly decompose the low-order proximity matrix and community membership strength matrix to obtain node embeddings.}
All these factorization-based methods incur immense overheads in building and factorizing the $n\times n$ proximity matrix. Further, these methods are designed for undirected graphs only.

%A classic methodology for ANE is matrix-factorization-based, which involves two stages: (i) building a proximity matrix $\MM\in \mathbb{R}^{n\times n}$ where $n$ is the number of nodes in the input graph $G$, by modelling the proximity between nodes in the graph or the similarities between nodes' attributes; (ii) factorizing $\MM$ via  dimensionality reduction techniques \cite{golub1971singular,wold1987principal} or optimization techniques \cite{bottou2010large,comon2009tensor,wright2015coordinate}.
%\tadw \cite{yang2015network} constructs a second-order topological proximity matrix $\MM$ based the adjacency matrix of $G$, and then aims to reconstruct $\MM$ by the product of embedding matrix and the attribute matrix. Similar to \tadw, \hsca \cite{zhang2016homophily} ensures the embedding vectors of the connected nodes are close in the vector space with a regularization term. \aane \cite{huang2017accelerated} improve \hsca by constructing $\MM$ based on the cosine similarity of every two nodes' attribute vectors. \bane \cite{yang2018binarized} proposes to incorporate both topology and attribute information into a unified matrix, minimizes the reconstruction loss and meanwhile ensures that each embedding vector is binary, \ie $\{-1,1\}^{k}$, which reduces data redundancy (\ie space overheads) but leads a loss of representation accuracy. To achieve a trade-off between data redundancy and representation accuracy, \lqanr \cite{ijcai2019-low} ensures the embedding vector $\in \{-2^{b},\cdots,-1,0,1,\cdots,2^b\}^k$, where $b$ determines the bit-width.

\header
{\bf Auto-encoder-based methods.}
An auto-encoder \cite{goodfellow2016deep} is a neural network model consisting of an encoder that compresses the input data to obtain embeddings and a decoder that reconstructs the input data from the embeddings, with the goal of minimizing the reconstruction loss. 
%The basic idea of auto-encoder-based ANE methods is to feed a proximity matrix and/or attribute matrix to an auto-encoder and obtain the compressed representation of nodes as the embedding vectors.
Existing methods  either use different proximity matrices as inputs or design various neural network structures for the auto-encoder. Specifically, \anrl \cite{zhang2018anrl} combines auto-encoder with the SkipGram model \cite{mikolov2013distributed} to learn embeddings. \dane \cite{gao2018deep} designs two auto-encoders to reconstruct the high-order proximity matrix and the attribute matrix respectively. \arga \cite{pan2018adversarially} integrates  auto-encoder with graph convolutional networks \cite{kipf2016semi} and generative adversarial networks \cite{goodfellow2014generative} together. \stne \cite{liu2018content} samples nodes via random walks and feeds the attribute vectors of the sampled nodes into a LSTM-based auto-encoder \cite{hochreiter1997long}. %In reality, graph structures and node attributes are often from different sources and orthogonal, but are typically treated in the same way by many methods. To address this issue, 
\netvae \cite{ijcai2019-370} compresses the graph structures and node attributes with a shared encoder for transfer learning and information integration. 
%Unlike prior work focusing on node embedding vectors only, 
\can \cite{meng2019co} embeds both nodes and attributes into two Gaussian distributions using a graph convolutional network and a dense encoder. None of these methods based on auto-encoders considers edge directions. Further, they suffer from severe efficiency issues due to the expensive training process of auto-encoders.

{$\mathtt{SAGE2VEC}$ \cite{sheikh2019simple} proposes an enhanced auto-encoder model that preserves global graph structure and meanwhile handles the non-linearity and sparsity of both graph structures and attributes. $\mathtt{AdONE}$ \cite{bandyopadhyay2020outlier} designs an auto-encoder model for detecting and minimizing the effect of community outliers while generating embeddings.} $\mathtt{SAGES}$ \cite{wang2022sages} first samples subgraphs containing highly relevant nodes with the consideration of node connections and attributes, and then learn the node embeddings by applying an unbiased graph autoencoder on the sampled subgraphs  with the guide of structure, content and community loss.

\header
{\bf Other methods.}
%There also exist some ANE techniques without factorizations and auto-encoders.
%trains an aggregator function (\eg LSTM, pooling) that aggregates the attribute of the node's neighborhood, using random walks. such that the embedding vectors of nodes on random walks are close in the embedding space. 
%Unlike prior work that directly preserves proximities between nodes, 
\prre \cite{zhou2018prre} categorizes node relationships into positive, ambiguous, and negative types, according to the graph and attribute proximities between nodes, and then employs  Expectation Maximization  \cite{dempster1977maximum} to learn embeddings.
\graphsage \cite{hamilton2017inductive} samples and aggregates features from a node’s local neighborhood and learns embeddings by LSTM and pooling.
\nethash \cite{wu2018efficient} builds a rooted tree for each node by expanding along the neighborhood of the node, and then recursively sketches the rooted tree to get a summarized attribute list as the embedding vector of the node.
In contrast to learning-based algorithms, \nethash \cite{wu2018efficient} expands each node along with its neighboring nodes into a rooted tree and then recursively sketches the rooted tree to get a summarized attribute list as the embedding vector. 
\pge \cite{hou2019representation} groups nodes into clusters based on their attributes, and then trains neural networks with biased neighborhood samples in clusters to generate embeddings. \progan \cite{gao2019progan} adopts generative adversarial networks to generate node proximities, followed by neural networks to learn node embeddings from the generated node proximities. \dgi \cite{velickovic2018deep} derives embeddings via graph convolutional networks, such that the mutual information between the embeddings for nodes and the embedding vector for the whole graph is maximized.
% All these methods are not designed for attributed network co-embedding problem studied in this paper.
\cite{liao2018attributed} proposes  a generic
framework \asne for embedding social networks, which captures the structural proximity and attribute proximity using a deep neural network architecture model.
$\mathtt{MUSAE}$ \cite{rozemberczki2021multi} extends the SkipGram model with negative sampling used in homogeneous network embedding \cite{perozzi2014deepwalk,node2vec2016} for attributed networks and show their method implicitly factorizes a matrix of pointwise mutual information.
\sane \cite{wang2018united} trains embeddings via a united approach which combines the attention network with CBOW model \cite{mikolov2013efficient} to  learn the similarity of the graph structure and attributes simultaneously.
{ $\mathtt{MARINE}$ \cite{wu2019scalable} preserves the long-range spatial dependencies between nodes into embeddings by minimizing the information discrepancy in a Reproducing Kernel Hilbert Space.}
$\mathtt{BiANE}$ \cite{huang2020biane} jointly models the attribute proximity and the structure proximity through latent correlation training to embed bipartite attributed networks. $\mathtt{MTSN}$ \cite{liu2021motif} learns dynamic node embeddings by simultaneously modeling both local high-order structural proximities and temporal dynamics for dynamic attributed networks.
\revise{Inspired by \cite{hamilton2017inductive}, $\mathtt{InfomaxANE}$ \cite{DBLP:conf/cikm/LiangLM20} leverages feature aggregation for the  combination of topological features and node attributes, and then trains global and local embeddings based on mutual information estimation.}
\revise{$\mathtt{ANGM}$ \cite{DBLP:journals/www/LiuYSMZCY21} focuses on embedding networks with multipartite, hubs, and hybrid structures by combining neural networks and the {\em stochastic block model} \cite{holland1983stochastic}.}

\revise{{
\subsection{Other Related Work}
Heterogeneous networks contain nodes and edges of different types. A series of studies focus on 
%In real-world scenarios, graph data is often heterogeneous, i.e., nodes and edges are multi-modal and multi-typed, which motivates a series of studies on 
embedding heterogeneous networks as surveyed in \cite{DBLP:journals/tkde/YangXZSH22,DBLP:journals/pr/XieYLZWG21,DBLP:conf/ijcai/DongHWS020}. We review several representative studies as follows. 
Inspired by $\mathtt{LINE}$ \cite{tang2015line} for NE, \cite{DBLP:conf/kdd/TangQM15} and \cite{DBLP:conf/kdd/ShiZGZ018} preserve first/second-order proximities into the embeddings with the consideration of edge types.
To incorporate high-order proximities, $\mathtt{metapath2ec}$ \cite{DBLP:conf/kdd/DongCS17}, $\mathtt{HIN2Vec}$ \cite{DBLP:conf/cikm/FuLL17}, and $\mathtt{JUST}$ \cite{DBLP:conf/cikm/HusseinYC18} exploit different random walk models that are guided by pre-defined meta-paths to sample node context for representation learning. Li et al. \cite{DBLP:conf/icde/0002ZLZWWJ0YM20} propose  a biased correlated random walk model to capture node proximity inside each view (i.e., node type) without user-specified meta-paths and, further, a cross-view algorithm to transfer information across views. Without the assumption that different meta-paths share the same semantic space, $\mathtt{SAHE}$ \cite{DBLP:journals/www/ZhengGY22} measures the relative proximities on each meta-path in its own semantic space and then aggregates them to obtain the final node proximity for embedding generation.
% \cite{DBLP:conf/pakdd/ZhangYZZ18},
% \cite{DBLP:conf/kdd/ChangHTQAH15}
}}

Recently, there are substantial embedding studies \cite{chang2015heterogeneous,wang2019heterogeneous,zhang2019shne,yang2020heterogeneous} on attributed heterogeneous networks that consist of not only graph topology and node attributes, but also node types and edge types. When there are only one type of node and one type of edge, these  methods effectively work on attributed networks. For instance, Alibaba proposed \gatne \cite{cen2019representation}, to process attributed heterogeneous network embedding. 
For each node on every edge type, it learns an embedding vector, by using the SkipGram model and random walks over the attributed heterogeneous network. Then it obtains the overall embedding vector for each node by concatenating the embeddings of the node over all edge types. \gatne incurs expensive training overheads and highly relies on the power of distributed systems.

\revise{
To cope with dynamic graphs that evolve over time,
%with node/edge insertions and deletions,  
increasing research efforts \cite{DBLP:journals/ijon/XueZLCZK22} have been invested in {\em dynamic network embedding} (DNE) in recent years. For instance, \cite{DBLP:conf/ijcai/DuWSLW18} generalizes the Skip-gram model to DNE through a decomposable objective equivalent to that of $\mathtt{LINE}$ \cite{tang2015line} and a carefully-designed mechanism to select the greatly affected nodes that need to be updated. $\mathtt{DynGEM}$ \cite{DBLP:journals/corr/abs-1805-11273} represents dynamic graphs as a collection of snapshots and incrementally updates the embeddings based on the ones from the previous snapshot via deep auto-encoders. Building on $\mathtt{node2vec}$ \cite{node2vec2016}, $\mathtt{dynnode2vec}$ \cite{DBLP:conf/bigdataconf/MahdaviKA18} keeps updating the list of random walks for evolving nodes and utilizes the dynamic Skip-gram model to generate updated embeddings. In lieu of embedding the entire graph, \cite{DBLP:conf/kdd/GuoZS21} learns embeddings for a subset of interesting nodes in large graphs with a dynamic algorithm for PPR computation. Bielak et al. \cite{DBLP:journals/kbs/BielakTFKC22} develop a general framework $\mathtt{FILDNE}$ for DNE, which integrates embeddings from any existing NE methods by an incremental updating scheme with batched data and an alignment mechanism. 

% Nguyen et al. \cite{DBLP:conf/www/NguyenLRAKK18}

% In \cite{DBLP:conf/aaai/ZhouYR0Z18}, the authors  
}

%% file: conclude.tex
\section{Conclusions}\label{sec:ccl}
This paper presents \algo, an effective solution for ANE computation that scales to massive graphs with tens of millions of nodes, while obtaining state-of-the art result utility. The high scalability and effectiveness of \algo are mainly due to a novel problem formulation based on a random walk model, a highly efficient and sophisticated solver, and non-trivial parallelization. Further, we extend \algo to \newalg with an effective attribute clustering algorithm to efficiently handle large attributed networks with 
numerous attributes. Extensive experiments show that \algo and \newalg achieve substantial performance advantages over the previous state-of-the-art in terms of both efficiency and result utility. Regarding future work, we plan to further develop GPU / multi-GPU versions of \algo and extend \algo to heterogeneous networks and dynamic graphs.

\begin{acknowledgements}
This work is supported by the National University of Singapore SUG grant R-252-000-686-133, Singapore Government AcRF Tier-2 Grant MOE2019-T2-1-029, NPRP grant \#NPRP10-0208-170408 from the Qatar National Research Fund (Qatar Foundation), and the financial support of Hong Kong RGC ECS (No. 25201221) and Start-up Fund (P0033898) by PolyU. The findings herein reflect the work, and are solely the responsibility, of the authors.
\end{acknowledgements}